\documentclass[preprint,nofootinbib,aps,superscriptaddress,eqsecnum]{revtex4-1}
\pdfoutput=1
\usepackage[colorlinks=true,citecolor=blue,linkcolor=blue,breaklinks=true]{hyperref}
\usepackage{graphicx}
\usepackage{mathrsfs}
\usepackage{soul}
\usepackage{natbib}
\usepackage[utf8]{inputenc}
\usepackage{enumerate}
\usepackage{amsmath,amssymb}
\usepackage{slashed}
\usepackage{amssymb,amsmath,subfigure,xcolor}

\newcommand{\lapprox}{%
\mathrel{%
\setbox0=\hbox{$<$}
\raise0.6ex\copy0\kern-\wd0
\lower0.65ex\hbox{$\sim$}
}}
\newcommand{\gapprox}{%
\mathrel{%
\setbox0=\hbox{$>$}
\raise0.6ex\copy0\kern-\wd0
\lower0.65ex\hbox{$\sim$}
}}

\newcommand{\ba}{\begin{array}}
\newcommand{\ea}{\end{array}}
\newcommand{\bd}{\begin{displaymath}}
\newcommand{\ed}{\end{displaymath}}
\newcommand{\beq}{\begin{equation}}
\newcommand{\eeq}{\end{equation}}
\newcommand{\bea}{\begin{eqnarray}}
\newcommand{\eea}{\end{eqnarray}}

\newcommand{\nn}{\nonumber}

\def\ie{ {\em i.e.,\ }}
\def\viz{ {\em viz.\ }}

\def\a{\alpha}

\def\b{\beta}

\def\g{\gamma}

\def\l{\lambda}
\def\m{\mu}
\def\n{\nu}

\def\q2 {q^2}

\def\bt{\begin{table}}
\def\et{\end{table}}

\catcode`@=11 
\def \gsim{\mathrel{\mathpalette\@versim>}}
\def \lsim{\mathrel{\mathpalette\@versim<}}
\def \@versim#1#2{\lower0.4ex\vbox{\baselineskip\z@skip\lineskip\z@skip
     \lineskiplimit\z@\ialign{$\m@th#1\hfil##\hfil$%
     \crcr#2\crcr\sim\crcr}}}
\catcode`@=12 

\begin{document}

\title{Relaxed constraints on the heavy scalar masses in 2HDM}

\author{Siddhartha Karmakar}
\email{ phd1401251010@iiti.ac.in}
\author{Subhendu Rakshit}
\email{rakshit@iiti.ac.in}
\affiliation{\em Discipline of Physics, Indian Institute of Technology Indore,\\
 Khandwa Road, Simrol, Indore - 453\,552, India}





\begin{abstract} 
\vskip 20pt
In the wake of new scalar searches at LHC in various channels, it is interesting to investigate the sacrosanctity of the constraints on the masses and couplings of the heavier scalars in a two-Higgs-doublet model~(2HDM). 
We consider the effects of new physics beyond a 2HDM encoded in terms of bosonic dim-6 operators. 
Although these constraints are mostly immune to such new physics, we demonstrate for a specific class of bosonic operators, the constraints on the masses of the exotic scalars from cascade decays can get substantially relaxed.
We present such effects for both degenerate and hierarchical mass spectra of the heavier scalars in 2HDM.
Some decay channels of the new scalars vanish at the alignment limit in the tree-level 2HDM. 
But the inclusion of dim-6 terms can lead to significant cross-sections for such processes. 
It is also pointed out that observation of such processes can no longer rule out the alignment limit if such dim-6 operators are present. 
\end{abstract}
 \vskip 1 true cm
 \pacs {}
\maketitle
\section{Introduction}
\label{intro}
Even after the discovery of a Higgs boson~\cite{Aad:2012tfa,Chatrchyan:2012xdj} whose characteristics resemble that of the standard model~(SM) Higgs, the dynamics of electroweak symmetry breaking and the structure of the scalar sector remains an open question. 
The non-cancellation of quadratic divergence of Higgs mass under the framework of SM has motivated plethora of beyond-standard model~(BSM) theories for decades. 
The two-Higgs-doublet model~(2HDM) is an archetype of an extended scalar sector, theoretically well-motivated from the viewpoint of supersymmetry, composite Higgs models, etc. 
For example, in supersymmetric models~\cite{Fayet:1976et} the motivation behind a second Higgs doublet is twofold: firstly to cancel chiral anomalies created by the superpartners of such scalars, and secondly from the requirement of the superpotential to be holomorphic. 2HDMs arising in the framework of composite Higgs~\cite{Mrazek:2011iu,DeCurtis:2016tsm}, Little Higgs~\cite{Schmaltz:2010ac}, Twin Higgs~\cite{Yu:2016swa} have also been studied in the literature. 
Even keeping the hierarchy problem aside, it is often deployed to explain issues of electroweak baryogenesis~\cite{Trodden:1998ym,Cline:2011mm}, flavour anomalies~\cite{Crivellin:2012ye,Crivellin:2019dun}, neutrino mass~\cite{Ma:2006km,Campos:2017dgc}, dark matter~\cite{Barbieri:2006dq}, etc.

In light of measurements of the signal strengths of the observed Higgs, any model with a scalar sector beyond the SM must contain a CP-even neutral scalar whose couplings are aligned to that of the SM Higgs boson. 
 Such an alignment can be realised when the new scalars which mix with the SM-like Higgs, are decoupled from the mass spectrum of SM \textit{$\grave{a}\,\,la$} Applequist-Carrazone~\cite{Georgi:1978ri,Gunion:2002zf}. 
The `alignment without decoupling' scenario becomes viable only for models with additional scalar doublets~\cite{Gunion:2002zf,Delgado:2013zfa,Carena:2013ooa,Bernon:2015qea,Bernon:2015wef,Haber:2013mia,Haber:2018ltt}.
 In such cases, the scalars can have masses below a TeV,\ie well within the reach of  LHC.  
Thus, along with the signal strengths of the SM-like Higgs, the direct bounds on the masses of exotic scalars also play a pivotal role in constraining the parameter space of 2HDM. Such bounds also depend on the specifications of the Yukawa sector of the models. 
Non-observation of such new scalars rule out a significant region of parameter space in the `alignment without decoupling' scenarios.
Also, some decay channels involving exotic scalars remain absent at the alignment limit in a CP-conserving 2HDM~\cite{Grzadkowski:2018ohf}.
If the LHC discovers any new scalar state in one of these channels, the interpretation involving a CP-conserving 2HDM would readily imply a deviation from the alignment limit.

If new physics beyond 2HDM exists as a decoupled sector from the mass scale of 2HDM, the effects of such new physics can be encoded in the higher-dimensional operators in an effective theory where the fields of 2HDM constitute the low-energy spectrum~\cite{DiazCruz:2001tn,Crivellin:2016ihg,Karmakar:2017yek}.
 Such an effective theory is dubbed as two-Higgs-doublet model effective field theory~(2HDMEFT) in the literature.
Several aspects of such effective theories for various extended scalar sectors have been addressed in the literature~\cite{Chala:2017sjk,Kikuta:2011ew,Kikuta:2015pya,Fonseca:2015gva,Banerjee:2019gmr}.
A complete basis of the 6-dim operators in 2HDMEFT has been introduced only recently~\cite{Karmakar:2017yek}. It has also been shown that such 6-dim operators are capable of masking the true alignment limit in a 2HDM, by modifying various decay channels of the SM-like Higgs boson~\cite{Karmakar:2018scg}. 
In the present paper, we have investigated the role of such 6-dim terms while extracting the LHC constraints on the masses of the new scalars in a 2HDM.
 We consider different mass spectra of these new scalars allowed from the theoretical constraints and  measurements of the oblique parameters. 
The constraints ensuing from different searches for the heavy scalars at the LHC and possible deviations in the presence of 6-dim terms have been illustrated.

In Sec.~\ref{review} we briefly review the theoretical framework of a general 2HDM. In Sec.~\ref{bounds} we discuss the theoretical as well as phenomenological constraints on the parameter space of a 2HDM relevant to this work. In Sec.~\ref{2hdmeft} we introduce the 6-dim terms that have been considered in this work, along with the modified couplings of the scalars. In Sec.~\ref{bp} we present the benchmark scenarios to illustrate the effect of such 6-dim terms on the parameter space of the 2HDM and eventually conclude in Sec.~\ref{summary}.  
\section{2HDM: A review}
\label{review}
The two scalar doublets are defined as: 
\begin{equation} 
\varphi_{I} =  \left(
\begin{array}{c}
\phi_{I}^{+}\\
\frac{1}{\sqrt{2}}(v_{I} + \rho_{I}) + i\, \eta_{I}\\
\end{array}
\right),
\end{equation}
with $I=1,2$. Here $\phi^{\pm}_I$, $\rho_I$, $\eta_I$, and $v_I$ denote the charged, neutral CP-even and neutral CP-odd degrees of freedom~(d.o.f.) and the vacuum expectation value~(vev) of the $I$-th doublet respectively.  

Before spontaneous symmetry breaking~(SSB), the tree-level 2HDM Lagrangian, augmented with 6-dim operators, assumes the form 
\bea
\label{lagrangian}
&&\mathcal{L} = \mathcal{L}_{kin} + \mathcal{L}_{Yuk} - V(\varphi_{1},\varphi_{2}) + \mathcal{L}_{6},
\eea
where, 
\bea
\label{lagrangianterms}
\mathcal{L}_{kin} &=&  -\frac{1}{4} \sum_{X = {G^{a}},W^{i},B} X_{\m\n} X^{\m\n} + \sum_{I = 1,2} |D_{\m} \varphi_{I}|^2 + \sum_{\psi = Q,L,u,d,l} \bar{\psi} i \slashed{D} \psi,\nn\\
\mathcal{L}_{Yuk} &=& \sum_{I=1,2} Y^{e}_I \, \bar{l} \,e \varphi_{I} + \sum_{I=1,2} Y^{d}_I \, \bar{q} \, d \varphi_{I} + \sum_{I=1,2} Y^{u}_I \, \bar{q} \, u \tilde{\varphi}_{I}, \nn\\
V(\varphi_{1},\varphi_{2}) &=& m_{11}^2 |\varphi_{1}|^2
 + m_{22}^2 |\varphi_{2}|^2 - ( \m^2 \varphi_{1}^{\dagger} \varphi_{2} +  h.c.) + \lambda_1 |\varphi_{1}|^4 + \lambda_2 |\varphi_{2}|^4 + \lambda_{3} |\varphi_{1}|^2 |\varphi_{2}|^2 \nn \\
  &&+  \lambda_4 |\varphi_{1}^{\dagger} \varphi_{2}|^2 + \Big[ \Big( \frac{\l_5}{2} \varphi_{1}^{\dagger} \varphi_{2} + \l_6 |\varphi_{1}|^2 + \l_7 |\varphi_{2}|^2 \Big) \varphi_{1}^{\dagger} \varphi_{2}  + h.c.\Big], \nn\\
  \mathcal{L}_{6} &=& \sum_{i} c_i O_{i}/f^2. 
\eea
Here, $c_i$ is the Wilson coefficient of the 6-dim operator $O_{i}$ and $f$ is the scale of new physics beyond the tree-level 2HDM. 
The terms proportional to $\l_{6,7}$ are called `hard-$Z_2$ violating' because they lead to a quadratically divergent amplitude for $\varphi_1 \leftrightarrow \varphi_2$ transition~\cite{Ginzburg:2004vp} and they also lead to CP-violation in the scalar sector when they attain complex values~\cite{ElKaffas:2007rq}.
But it is possible to realise the CP-conserving limit with non-zero values of $\lambda_{6,7}$ as well~\cite{Grzadkowski:2018ohf}. 
In this paper we contain our discussion to the CP-conserving 2HDM, and we take $\l_{6,7} = 0$.
The electroweak symmetry is broken by the vacuum expectation values~(vev), namely $v_1$ and $v_2$  corresponding to the the two doublets $\varphi_{1,2}$ respectively.
This leads to the mixing of similar types of degrees of freedom pertaining to $\varphi_{1,2}$. 
In the CP-conserving case, the mass matrices of the neutral CP-even and odd scalars and the charged scalars are diagonalised by the following field rotations: 
\bea
\left(
\begin{array}{c}
H\\
h\\
\end{array}\right)
= R(\a)
\left(
\begin{array}{c}
\rho_1\\
\rho_2\\
\end{array}\right), 
\hspace{15pt}
\left(
\begin{array}{c}
W_L^{\pm}\\
H^{\pm}\\
\end{array}\right)
= R(\b)
\left(
\begin{array}{c}
\phi_1^{\pm}\\
\phi_2^{\pm}\\
\end{array}\right),
\hspace{15pt}
\left(
\begin{array}{c}
Z_L\\
A\\
\end{array}\right)
= R(\b)
\left(
\begin{array}{c}
\eta_1\\
\eta_2\\
\end{array}\right).
\label{dofdef}
\eea  
Here, 
\bea
 R(\theta) =
 \begin{bmatrix}
    \cos \theta        & \sin \theta \,\, \\
    -\sin \theta      & \cos \theta   \,\, 
\end{bmatrix}.
\eea
$h,H$ are the neutral CP-even physical d.o.f., whereas $A$ and $H^{\pm}$ are the neutral CP-odd and charged d.o.f respectively. As it can be seen from eq.~(\ref{dofdef}), $\beta$ is the mixing angle of the charged and CP-odd sectors and it is given by $\b = \tan^{-1} (v_2/v_1)$. $\a$ is the mixing angle of the CP-even neutral scalars and can be expressed as
\bea
\alpha &=& \sin^{-1} \Bigg[\frac{\mathcal{M}_{\rho 12}^2}{\sqrt{(\mathcal{M}_{\rho 12}^2)^2 + (\mathcal{M}_{\rho 11}^2 - m_{h}^2)^2}}\Bigg],
\eea 
with $\mathcal{M}_{\rho}^2$ being the mass-squared matrix in the neutral CP-even sector. 
In this paper, we assume $h$ to be the SM-like Higgs with a mass of $m_h \sim 125.09$~GeV and $m_H > m_h$.
The case of an additional  CP-even scalar in 2HDM with mass lower than the SM-like Higgs has been explored in the literature as well~\cite{Bernon:2015wef}. 
It was shown that the tree-level Higgs-mediated flavour-changing neutral currents~(FCNC) appear in models where more than one scalar doublet give mass to the same kind of SM fermions~\cite{Glashow:1976nt,Paschos:1976ay}.
Such a situation can be avoided under the framework of various discrete symmetries, for example, a $Z_2$ symmetry~\cite{Glashow:1976nt,Paschos:1976ay}.
There are four possible ways in which such a $Z_2$ charge assignment of the SM fermions can be done, namely the type-I~($Y^{u}_1 = Y^{d}_1 = Y^{e}_1 = 0$), type-II~($Y^{u}_1 = Y^{d}_2 = Y^{e}_2 = 0$), type-III~($Y^{u}_1 = Y^{d}_2 = Y^{e}_1 = 0$) and type-IV~($Y^{u}_1 = Y^{d}_1 = Y^{e}_2 = 0$) cases. 
Type-II scenario is also dubbed as the MSSM-like case due to similarity in the Yukawa sectors. 
Type-III and -IV are sometimes also referred to as flipped and lepton-specific scenarios respectively.
Due to the rotation in the scalar sector following eq.~(\ref{dofdef}), the couplings of the SM gauge bosons and fermions to the SM-like Higgs boson are rescaled compared to the corresponding SM values. After SSB the Yukawa sector of the 2HDM can be written as,
\bea
- \mathcal{L}_{Yuk}& = &\frac{1}{\sqrt{2}} (\kappa_D s_{\b-\a} + \rho_D c_{\b-\a}) \bar{D} D h + \frac{1}{\sqrt{2}} (\kappa_D c_{\b-\a} - \rho_D s_{\b-\a}) \bar{D} D H \nn\\
&&+ \frac{1}{\sqrt{2}} (\kappa_U s_{\b-\a} + \rho_U c_{\b-\a}) \bar{U} U h + \frac{1}{\sqrt{2}} (\kappa_U c_{\b-\a} - \rho_U s_{\b-\a}) \bar{U} U H \nn\\
&&+ \frac{1}{\sqrt{2}} (\kappa_L s_{\b-\a} - \rho_L c_{\b-\a}) \bar{L} L h + \frac{1}{\sqrt{2}} (\kappa_L c_{\b-\a} - \rho_L s_{\b-\a}) \bar{L} L H \nn\\
&&- \frac{i}{\sqrt{2}} \bar{U} \g_5 \rho_U U A + \frac{i}{\sqrt{2}} \bar{D} \g_5 \rho_D D A + \frac{i}{\sqrt{2}} \bar{L} \g_5 \rho_L L A \nn\\
&&+\Big(\bar{U}(V_{\text{CKM}}\, \rho_D P_R - \rho_U V_{\text{CKM}} P_L) D H^{+} + \bar{\n} \rho_L P_R L H^{+} + \text{h.c.} \Big),
\label{coupmult}
\eea

with, $\kappa_f = \sqrt{2} M_f/v$ for $f = U,D,L$ and,
\begin{center}
\begin{tabular}{c c c c c}
& Type-I & Type-II & Type-III & Type-IV \\
$\rho_D$ & $\kappa_D \cot \b$  & $-\kappa_D \tan \b$ & $-\kappa_D \tan \b$ & $\kappa_D \cot \b$ \\
$\rho_U$ & $\kappa_U \cot \b$  & $\kappa_U \cot \b$ & $\kappa_U \cot \b$ & $\kappa_U \cot \b$ \\
$\rho_L$ & $\kappa_L \cot \b$  & $-\kappa_L \tan \b$ & $\kappa_L \cot \b$ & $-\kappa_L \tan \b.$ \\
\end{tabular}
\end{center} 
$U,D,L$ and $\nu$ represent the up-type, down-type quarks, charged leptons and neutrinos in their mass bases respectively. 
The generation indices of the fermionic fields have been suppressed in eq.~\eqref{coupmult}.
As mentioned earlier, the measurement of the signal strengths of the SM-like Higgs at LHC demands that the properties of one of the neutral CP-even neutral scalars, here $h$, should closely resemble that of the SM Higgs. As eq.~(\ref{coupmult}) indicates, this is satisfied at the vicinity of the so-called `alignment limit',\ie $\cos (\b-\a) \rightarrow 0$. Thus, the current measurement of Higgs signal strengths have pushed the 2HDMs close to the alignment limit~\cite{Bernon:2015wef,Bernon:2015qea,Gu:2017ckc,Dorsch:2016tab}. 
 The measurement of the Higgs signal strengths dictate that for type-II 2HDM, at $\tan \b \sim 1$, the constraint on $\cos (\b-\a)$ is given by $-0.05 \lesssim \cos (\b-\a) \lesssim 0.15$ at $95\%$~CL. The allowed region becomes even smaller for higher values of $\tan \beta$.
 The situation for type-III and -IV are quite similar to that of type-II 2HDM.
This constraint is comparably relaxed in type-I 2HDM, where the allowed range is $|\cos (\b-\a)| \lesssim 0.4$.
As we are working under the assumption of a CP-even vacuum of the 2HDM potential, vertices like $AWW$ and $AZZ$ are not present at the tree-level. 
Among the tree-level scalar-gauge couplings which are important for the cascade decays of the new scalars, $AZh$ and $H^{\pm}hW^{\mp}$ are proportional to $\cos (\b-\a)$, whereas $AZH$ and $H^{\pm}HW^{\mp}$ are proportional to $\sin (\b-\a)$. 
It is possible to realise an exact alignment in  the multi-Higgs-doublet models in the framework of certain additional symmetries of the 2HDM potential~\cite{Dev:2014yca,Pilaftsis:2016erj,Das:2017zrm,Benakli:2018vjk,Pramanick:2017wry}.

It is evident from eqs.~(\ref{coupmult})  when $\tan \beta \gtrsim 1$, the $hbb$ coupling multiplier deviates significantly from unity in type-II 2HDM. 
For such values of $\tan \beta$ the branching ratio Br($h \rightarrow b\bar{b}$) as well as the production of $h$ in both $gg$ and $b\bar{b}$-fusion substantially increase. 
It is mainly due to the measurement of the processes like $gg \rightarrow \gamma\gamma, b\bar{b}, VV^{*}$ and $Vh \rightarrow b\bar{b}$, that the parameter space of a type-II 2HDM in the $\cos (\b-\a)-\tan \b$ plane is quite strongly constrained. 
The impact of the measurement of the Higgs signal strengths in each individual search channels on the $\cos (\b-\a) - \tan \b$ plane has been discussed in ref.~\cite{Craig:2013hca}.
It should be mentioned that the coupling multipliers of the SM-like Higgs also becomes close to unity when $\sin (\beta + \a) = 1$ \ie at the so-called `wrong-sign Yukawa' limit~\cite{Ferreira:2014qda} for type-II, -III, and -IV 2HDM.
 Though, with better measurement of the processes like $Vh \rightarrow b\bar{b}$, $h \rightarrow \gamma\gamma, \Upsilon \gamma$~\cite{Modak:2016cdm,Ferreira:2014naa} the fate of the wrong-sign Yukawa region will be decided in near future. 

It is also clear from eq.~(\ref{coupmult}) that, though the coupling multipliers of the SM-like Higgs become unity at the alignment limit, the couplings of the exotic Higgses with SM fermions can be non-zero.    
The $HVV$ coupling becomes identically zero at the alignment limit, protecting the alignment limit against measurements like $gg \rightarrow H \rightarrow WW,ZZ$.
Though the couplings of $H$ and $A$ with SM fermions do not vanish at $\cos (\b-\a) = 0$.    
Thus it is possible to constrain the parameter space of the type-II 2HDM even at the alignment limit from the non-observation of the heavier scalars~\cite{Grzadkowski:2018ohf} in processes like $gg/b\bar{b} \rightarrow H/A \rightarrow \tau\bar{\tau}, \gamma\gamma$, $gg \rightarrow H \rightarrow hh$, etc. 
Both ATLAS and CMS are involved in numerous dedicated searches of these kinds, for instance ref.~\cite{ATLAS:2017nxi,Sirunyan:2018iwt,TheATLAScollaboration:2016loc,ATLAS:2017mpg,CMS:2015mca}, resulting in significant constraints on the 2HDM parameter space.

Different kinds of mass spectra of the new scalars in a 2HDM can lead to quite a rich phenomenology at the LHC.
 In addition to the regular search channels consisting of a pair of SM particles as mentioned earlier, exotic states decaying into one another can also provide strong constraints on 2HDM parameter space. 
 At this point, we define $m_A = m_H = m_{H^\pm}$ as the `degenerate' case and the case when any of the three exotic scalars is more massive than the rest two, as the `hierarchical' case.
The constraints from the decay of the new scalars into SM particles are significantly relaxed in the hierarchical scenario compared to the degenerate case~~\cite{Dorsch:2016tab}.
But for the hierarchical spectrum of new scalars, the channels like $H(A) \rightarrow ZA (H)$ dominate the total decay width of such states, leading to new bounds on the parameter space which are not applicable for the degenerate case.
A hierarchical spectrum such as $m_A > m_H \sim m_{H^{\pm}} \sim v$ can lead to a first-order electroweak phase transition providing an explanation for the matter-antimatter asymmetry, with $A \rightarrow ZH$ being its smoking gun signature at LHC~\cite{Dorsch:2013wja,Dorsch:2014qja}. 
In general, the importance of Higgs cascade decays as the possible probes of an extended scalar sector have been discussed in the literature~\cite{Gao:2016ued,Bauer:2015fxa,Kling:2015uba,Coleppa:2014cca,Kling:2016opi,Adhikary:2018ise} and $A \rightarrow ZH$ decay is dubbed as a `golden channel' in this context~\cite{Kling:2018xud}.

\section{Constraints on 2HDM parameter space}
\label{bounds}
We work with the 2HDM parameters in the physical basis which consists of \{$m_h$, $m_H$, $m_A$, $m_{H^{\pm}}$, $\tan \beta$, $\cos (\b -\a), m_{12}^2, \lambda_6, \lambda_7, v $\}. 
Along with $m_h = 125.09$~GeV, $v = 246$~GeV and $\l_{6,7} = 0$, we are left with six free parameters. The conversion between the generic and physical basis can be found in, for instance, ref.~\cite{Gunion:2002zf,Kling:2016opi}.
The theoretical constraints are discussed below.

\noindent $\bullet$ {\it Vacuum stability}  \hspace{5pt} The stability of the EW vacuum in a 2HDM at the tree-level is ensured if~\cite{Gunion:2002zf}, 
\bea
\l_1 > 0, \,\,\; \l_2 > 0, \,\,\; \l_3 > - \sqrt{\l_1 \l_2}, \;\,\, \l_3 + \l_4 - |\l_5| > - \sqrt{\l_1 \l_2}.
\eea
It can be shown that at the alignment limit, the first two conditions are satisfied if $m_{12}^2 = m_H^2 s_{\b} c_{\b}$. Along with that, the last two criteria are satisfied if $m_{h}^2 + m_{H^{\pm}}^2 - m_{H}^2 > 0$ and  $m_{h}^2 + m_{A}^2 - m_{H}^2  > 0$ respectively. This means for degenerate masses of the new scalars the last two criteria are automatically satisfied  if the first two are satisfied. For hierarchical mass spectrum, the mass of the exotic scalars cannot be arbitrarily different. 

\noindent $\bullet$ {\it Perturbativity} \hspace{5pt} The perturbativity of the quartic couplings is satisfied if $|\l_{i}| \lesssim 4\pi$. At the alignment limit, this implies for $t_{\b} \gtrsim 1$, $|m_{12}^2 - m_H^2 s_{\b} c_{\b}| \lesssim v^2$.

\noindent $\bullet$ {\it Unitarity} \hspace{5pt} Tree-level unitarity of the S-matrix requires the eigenvalues of the $2 \rightarrow 2$ scattering matrix to be less than $8\pi$. At the alignment limit for $m_{12}^2 \sim m_H^2 s_{\b} c_{\b}$, this implies that the differences between the masses of the new scalars have to be $\lesssim v$.  

\noindent $\bullet$ {\it Oblique parameters} The new scalars in 2HDM contributes to the oblique parameters through their couplings to the massive gauge bosons~\cite{Gerard:2007kn,deVisscher:2009zb,Grimus:2007if,Grimus:2008nb,Haber:2010bw,He:2001tp}. At the alignment limit such contributions to the $T$-parameter assume the form, 
\bea
\Delta T = \frac{g^2}{64 \pi^2 m_W^2} \Big(F(m^2_{H^{\pm}},m_A^2) + F(m^2_{H^{\pm}},m_H^2) - F(m^2_A,m_H^2)\Big),
\label{tparam}
\eea
with,
\bea
 F(a,b) &&= \frac{a+b}{2} -  \frac{ab}{a-b}  \ln \Big(\frac{a}{b}\Big) \hspace{15pt}(a \neq b) \nn\\
 &&= 0 \hspace{117pt}(a = b)\nn. 
\eea
As eq.~(\ref{tparam}) suggests, this anomalous contribution to the $T$-parameter vanishes at the limit $m_{H^{\pm}} = m_A$ or  $m_{H^{\pm}} = m_H$. However, substantially away from the alignment limit this does not hold for the entire range of $m_A$. As mentioned earlier, the measurements of the Higgs signal strengths imply that the maximum values of $\cos (\b-\a)$ can be attained in type-I 2HDM, $|\cos (\b-\a)| \lesssim 0.4$. For the rest three types of Yukawa structure, $|\cos (\b-\a)| \lesssim 0.1$. We have checked that the limits $m_H = m_{H^{\pm}}$ and $m_A = m_{H^{\pm}}$ ensure that the contribution to the $T$-parameter remains in the experimentally allowed range at $95\%$~CL even when we consider small deviations from alignment limit, $\cos (\b-\a) \lesssim 0.1$.
For $\cos (\b-\a) \sim 0.4$, even for $m_H = m_{H^{\pm}} = 300$~GeV, $m_A \gtrsim 480$~GeV is ruled out from the measurement of $T$-parameter.

LEP searches put a constraint on the mass of the charged scalar as $m_{H^{\pm}} \gtrsim 72$~GeV (80~GeV) for type-I~(II) 2HDM~\cite{Abbiendi:2013hk}. 
Also the searches for $Z \rightarrow AH \rightarrow \tau\bar{\tau}\tau\bar{\tau}$ lead to the constraint $m_H + m_A \gtrsim 208$~GeV~\cite{Schael:2006cr}.
The  charged scalar mediates flavour-violating processes such as $B_d \rightarrow X_s \gamma$, $B_{s} \rightarrow \mu^{+} \mu^{-}$, $B_d^{+} \rightarrow \tau^{+} \nu$, etc., which in turn lead to constraints on $m_H^{\pm}$~\cite{Mahmoudi:2009zx,Eberhardt:2013uba,Misiak:2017bgg}.
 The measurement of the width of $B_d \rightarrow X_s \gamma$ leads to the most stringent constraint on the charged scalar mass for type-II 2HDM, $m_{H^{+}} \gtrsim 580$~GeV, almost independent of the value of $\tan \b$~\cite{Misiak:2017bgg}. 
For type-I 2HDM, the constraint from meson decays is comparatively less stringent and depends rather strongly on $\tan \b$. For $\tan \b \sim 1.5$ the constraint is, $m_{H^{+}} \gtrsim 200$~GeV~\cite{Misiak:2017bgg}.
Based on the similarity in couplings of the scalars to the quarks, the constraints on charged scalar mass for type-I and -II 2HDM can also be used for type-IV and -III cases respectively. 
Though we do not consider this as a hard bound for our purpose as it can be ameliorated in several extensions of 2HDM~\cite{Han:2013mga}.
Moreover, the mass differences of the three exotic scalars are constrained from the measurement of the oblique parameters~\cite{Gerard:2007kn}. For $\cos (\b -\a) = 0$, the contributions  of the new scalars to the $S$ and $T$ parameters do not depend upon $m_{12}^2$ or $\tan \b$. For a fixed value of $m_H$, the constraints from precision tests are satisfied only if $m_H \sim m_A$ or $m_{H} \sim m_{H^{\pm}}$.
As mentioned earlier, the measurements of the signal strengths of the SM-like Higgs boson constrain the value of $\cos (\b-\a)$ to be close to zero, especially for type-II, -III and -IV 2HDM. 
Thus, while working with the hierarchical spectrum of the new scalars, we take $\cos (\b-\a)$ to be close to zero.  
But for the degenerate spectrum of exotic scalars, we consider possible large deviation from the alignment limit while we present the excluded region on the $\cos (\b-\a) - \tan \b$ plane. 
In this paper, we have considered the constraints due to the non-observation of the processes: $gg \rightarrow H \rightarrow ZZ$~\cite{ATLAS:2017nxi},  $gg \rightarrow H \rightarrow hh$~\cite{Sirunyan:2018iwt}, $gg \rightarrow H(A) \rightarrow A (H/h) Z$~\cite{TheATLAScollaboration:2016loc} and $gg/b\bar{b} \rightarrow H/A \rightarrow \tau\bar{\tau}$~\cite{ATLAS:2017mpg,CMS:2015mca}.

\section{Couplings of the heavier scalars in 2HDMEFT}
\label{2hdmeft} 
We contain our discussion only to the bosonic operators of 2HDMEFT for simplicity. 
Phenomenology of the fermionic dim-6 terms will be reported elsewhere.
As discussed in ref.~\cite{Karmakar:2018scg}, the measurement of EW oblique parameters, triple gauge boson vertices, and Higgs signal strengths, constrain the bosonic operators other than type $\varphi^4 D^2$ at $\mathcal{O}(10^{-3})$.
Moreover, some of the $\varphi^4 D^2$-type of operators violate the $T$-parameter at the tree-level and the corresponding Wilson coefficients are rather small. 
Thus we have considered only the operator of type $\varphi^4 D^2$ which do not contribute to  the $T$-parameter in the tree-level~\cite{Karmakar:2017yek}:
\bea
\label{ops}
O_{H1} &=& (\partial_{\m}|\varphi_1|^2)^2, \hspace{10pt} O_{H2} = (\partial_{\m}|\varphi_2|^2)^2,\hspace{10pt} O_{H12} = (\partial_{\m}(\varphi_1^{\dagger} \varphi_2 + h.c.))^2, \\
O_{H1H2} &=& \partial_{\m}|\varphi_1|^2 \partial^{\m}|\varphi_2|^2, O_{H1H12} = \partial_{\m}|\varphi_1|^2\partial^{\m}(\varphi_1^{\dagger} \varphi_2 + h.c.), O_{H2H12} = \partial_{\m}|\varphi_2|^2\partial^{\m}(\varphi_1^{\dagger} \varphi_2 + h.c.). \nn
\eea
In presence of such operators, the non-diagonal kinetic terms arise after SSB~\cite{Karmakar:2017yek}. 
In order to get rid of such terms, one needs to rescale the neutral CP-even d.o.f.\ie $\rho_1$ and $\rho_2$. This implies that the physical neutral CP-even scalars in presence of these operators are rescaled compared to the tree-level 2HDM. 
\bea
\label{hfieldred}
h & \rightarrow & (1-x_1)h+  y H,\nn\\
H & \rightarrow & (1-x_2)H+  y h.
\eea  
$x_1$, $x_2$ and $y$ can be written in terms of the Wilson coefficients of the operators appearing in eq.~(\ref{ops}) and the scale of new physics beyond 2HDM. The analytical forms of $x_1$, $x_2$ and $y$ can be found in Appendix~\ref{xydef}. 
In our convention, the Wilson coefficient of an operator $O_i$ is given as $c_i$, which include the symmetry factors.
Eq.~(\ref{hfieldred}) dictates that any coupling involving at least one $h$ or $H$ field are modified compared to 2HDM at the tree-level. For example, 
\bea
\label{scalemult1}
\kappa_{hff}^{\prime} &=& (1- x_1)\kappa_{hff}+ y \kappa_{Hff},\\
\label{scalemult2}
\kappa_{Hff}^{\prime} &=& (1- x_2)\kappa_{Hff}+ y \kappa_{hff},\\
\label{scalemult3}
\kappa_{hVV}^{\prime} &=& (1 - x_1) \sin (\b-\a) + y \cos (\b-\a), \\ 
\label{scalemult4}
\kappa_{HVV}^{\prime} &=& (1 - x_2) \cos (\b-\a) + y \sin (\b-\a), \\ 
\label{scalemult5}
\kappa_{AZh}^{\prime} &=& (1 - x_1)\kappa_{AZh}  + y \kappa_{AZH}, \\ 
\label{scalemult6}
\kappa_{AZH}^{\prime} &=& (1 - x_2)\kappa_{AZH} + y \kappa_{AZh}.
\eea
Eqs.~\eqref{scalemult2} and \eqref{scalemult4} affect decay processes like $H \rightarrow \tau \bar{\tau}$ and $ H \rightarrow ZZ$ which are particularly important in the degenerate case.  
Similarly, the decay widths $H(A) \rightarrow Z A(H)$, which become relevant in the hierarchical scenarios are changed according to eq.~\eqref{scalemult6}. 
What is more interesting that processes, which apparently look unaffected by the rescaling of fields, such as $gg \rightarrow A \rightarrow \tau \bar{\tau}$, are also changed. It is due to the fact that away from the alignment limit, for a large range of values of $m_A$ and $\tan \b$, Br($A \rightarrow Zh$) is quite significant. 
The change in Br($A \rightarrow Zh$) according to eq~\eqref{scalemult5} in turn modifies Br($A \rightarrow \tau \bar{\tau}$). 
Even the change in $h \rightarrow f\bar{f}$ can be important for bounds on the heavier scalars. 
For example, the search for $A \rightarrow Zh$ assumes that the Higgs in the final state further decays into a pair of $b$-tagged jets~\cite{TheATLAScollaboration:2016loc}.  
Moreover, the coupling multipliers of the SM-like Higgs also modify upon the inclusion of dim-6 operators compared to the tree-level 2HDM according to eqs.~\eqref{scalemult1} and \eqref{scalemult3}. Thus, the allowed parameter space changes in the $\cos (\b-\a) - \tan \b$ plane~\cite{Karmakar:2018scg}.

Many of the sum rules involving various gauge couplings, which hold in 2HDM at the tree-level, are no longer valid in the presence of 6-dim operators~\cite{Mrazek:2011iu}.
These sum rules can play an important role in deciphering new physics beyond 2HDM. 
For instance, in 2HDM at the tree-level, the sum rule $\kappa_{hVV}^2 + \kappa_{HVV}^2 = 1$ holds true, but in presence of the dim-6 terms mentioned in eq.~\eqref{ops}, $\kappa_{hVV}^2 + \kappa_{HVV}^2 = 1 - 2(x_1 s_{\b-\a}^2 + x_2 c_{\b-\a}^2 + 2 y c_{\b-\a} s_{\b-\a})$.
If another CP-even neutral scalar, $H$ is discovered after $h(125)$, the measurement of its decay width and Br($H \rightarrow WW$) will facilitate the verification of such a sum rule. 
A deviation from $\kappa_{hVV}^2 + \kappa_{HVV}^2 = 1$ will point to a departure from the CP-conserving 2HDM. 
If $\kappa_{hVV}^2 + \kappa_{HVV}^2 < 1$, then it may indicate towards CP-violating 2HDM or CP-conserving NHDM~($N>2$).
But such an interpretation does not hold if there are more than one neutral BSM scalars with degenerate masses, or masses same as the SM-like Higgs. 
Anyway, even the dim-6 operators in 2HDMEFT can lead to $\kappa_{hVV}^2 + \kappa_{HVV}^2 < 1$. 
On contrary, neither CP-violating 2HDM nor NHDM can lead to $\kappa_{hVV}^2 + \kappa_{HVV}^2 > 1$. 
Though, such a scenario can be interpreted in terms of the dim-6 terms of 2HDMEFT. 
At the CP-conserving limit with $\l_{6,7} \neq 0$ the similar argument for tree-level 2HDM is valid in this context, whereas in general $\l_{6,7} \neq 0$ will follow the argument for CP-violating 2HDM.

\section{Benchmark scenarios}
\label{bp}
Following the discussions in Section~\ref{bounds} in the context of oblique parameters, for the hierarchical mass spectrum, we consider either $m_A = m_{H^{\pm}}$ or $m_{H} = m_{H^{\pm}}$.
The limit $m_A = m_H$ is highly constrained from the measurement of $S,T$ parameters and the decays of $H$ and $A$ into each other are kinematically forbidden.  

So the mass spectra under scrutiny for the hierarchical case are~\cite{Kling:2016opi}:
\bea
&&{\bf C1}: m_{A}  = m_{H^{\pm}} > m_{H} \hspace{20pt}
{\bf C2}: m_{A} > m_{H} = m_{H^{\pm}} \nn\\
&&{\bf C3}: m_{A}  = m_{H^{\pm}} < m_{H} \hspace{20pt}
{\bf C4}: m_{A} < m_{H} = m_{H^{\pm}}.
\eea
For the hierarchical case, we have studied the bounds on the $m_A = m_H$ plane for $\tan \b = 1.5$ and $\cos (\b-\a) = 0$ with and without considering the dim-6 terms. 

On the other hand, for the degenerate mass spectrum~($ m_H = m_A = m_{H^{\pm}}$) new scalars can no longer decay into each other,
thus making the SM decay channels of these scalars $H \rightarrow ZZ, \tau\bar{\tau}, b\bar{b}$, $A \rightarrow Zh, \tau\bar{\tau}$ more important.
We have studied the change in the constraints due to inclusion of dim-6 operators on the $\cos (\b-\a) - m_A$ plane. We define the benchmark for  the degenerate case as,
\bea
{\bf C5}: m_H = m_A = m_{H^{\pm}}, \tan \b = 1.5.
\eea 
As mentioned earlier, $\l_{6,7} = 0$ for both degenerate and hierarchical cases.
Also the quadratic mass parameter is taken to be $ m_{12}^2 = m_H^2 s_{\b} c_{\b}$ following the discussions on the theoretical constraints in Section~\ref{bounds}.
For both the hierarchical and degenerate cases, the bounds on 2HDM at the tree-level are compared with a specific case of 2HDMEFT with the following values of the Wilson coefficients and the new physics scale~\cite{Karmakar:2018scg}, 
\bea 
{\bf BP1}: \;\;c_{H1} = -1, \, c_{H2} = 1.5, c_{H12} = c_{H1H2} = c_{H1H12} = c_{H2H12} = 0, \, f = 1~\text{TeV}.
\eea

The theoretical constraints such as perturbativity  and stability  do not change upon the inclusion of bosonic operators considered in this paper, as these operators do not modify the  2HDM scalar potential. However, they can lead to additional contributions in the S-matrix for $2 \rightarrow 2$ scattering of bosonic states. Implementing these changes in \texttt{2HDMC-1.7.0}~\cite{Eriksson:2009ws} we have checked that for the 2HDMEFT benchmark scenario {\bf BP1}, there are no significant modifications of the allowed parameter space for $\sqrt{\hat{s}}\sim $ few TeVs while $f = 1$~TeV. Similar conclusions were obtained  for a composite 2HDM based on $SO(6)/SO(4)\times SO(2)$~\cite{DeCurtis:2016scv}. 

The branching ratios of various scalars in 2HDM at the tree-level, as well as in the presence of the 6-dim operators, have been calculated with \texttt{2HDMC-1.7.0}~\cite{Eriksson:2009ws} after incorporating the modified couplings. 
The production cross-section of the neutral scalars have been computed up to NNLO in QCD using \texttt{SusHi-1.6.1}~\cite{Harlander:2012pb}. 
As mentioned earlier, the constraints on $\cos (\b-\a)- \tan \b$ plane from the measurement of signal strengths of $h(125)$ change in presence of the dim-6 operators. For instance, in type-II 2HDM for $\tan \beta = 1.5$, the allowed range of $\cos (\b-\a)$ in 2HDM is $[-0.05, 0.12]$  which changes in {\bf BP1} of 2HDMEFT to, $[-0.02, 0.11]$~\cite{Karmakar:2018scg}. The allowed range of $\cos (\beta - \alpha)$ changes for the other types of Yukawa couplings as well, but the exact tree-level alignment limit $\cos (\b - \a) = 0$ is allowed in all these cases for $\tan \b = 1.5$. 

\begin{figure}[h!]
 \begin{center}
\subfigure[]{
   \includegraphics[width=2.8in,height=2.8in, angle=0]{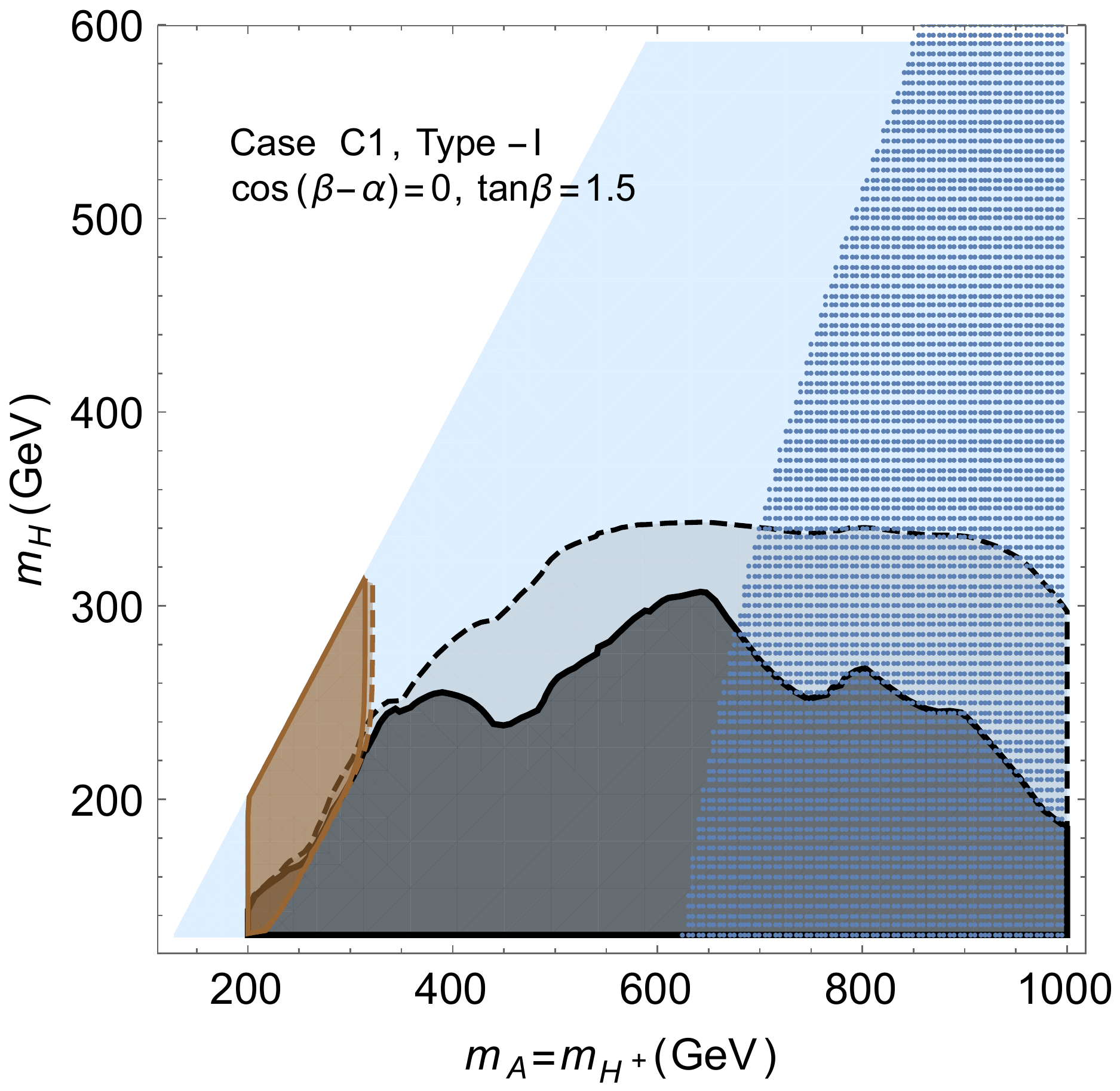}} 
 \hskip 15pt
 \subfigure[]{
\includegraphics[width=2.8in,height=2.8in, angle=0]{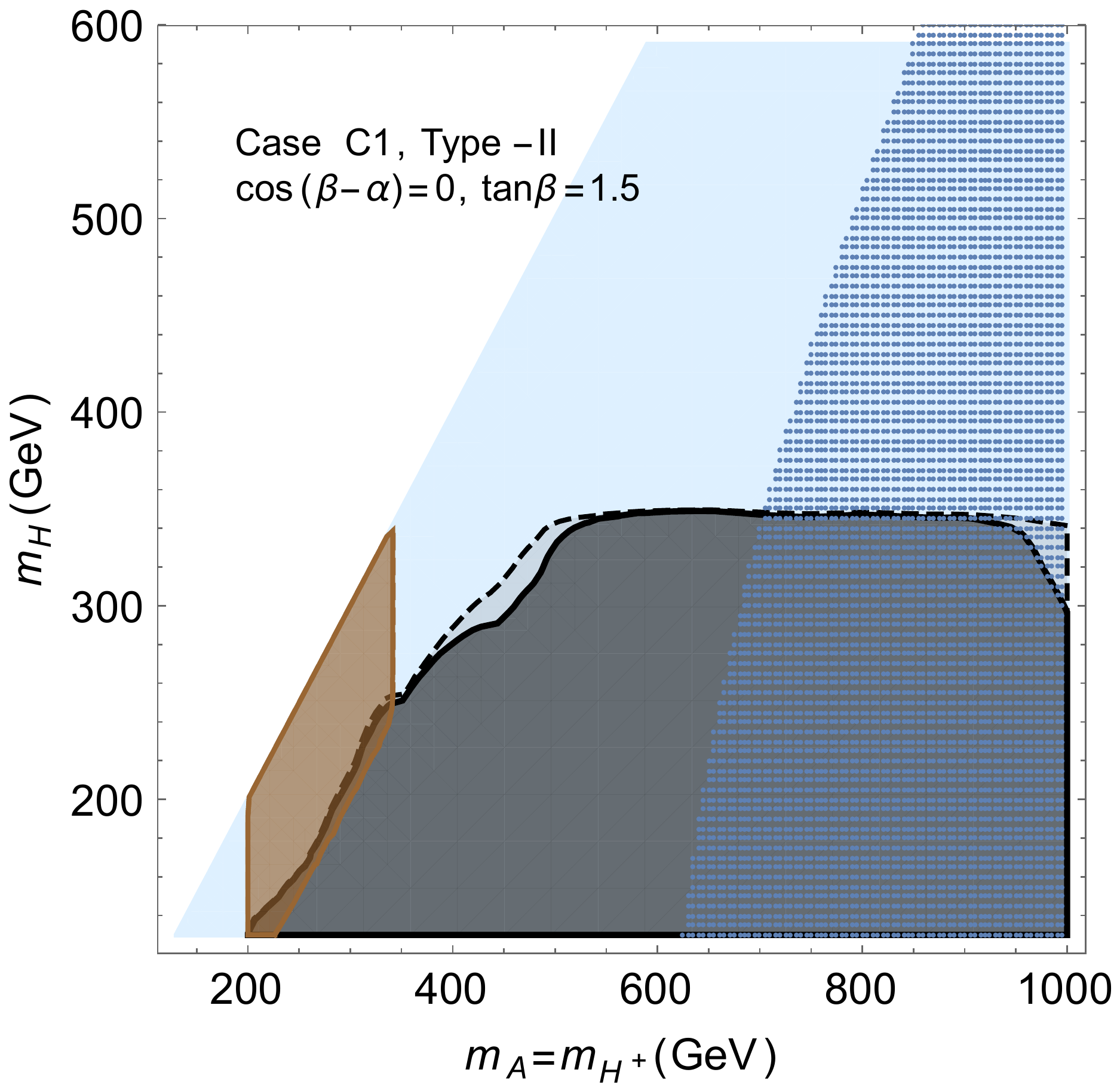}} 
 \hskip 15pt
 \subfigure[]{
 \includegraphics[width=2.8in,height=2.8in, angle=0]{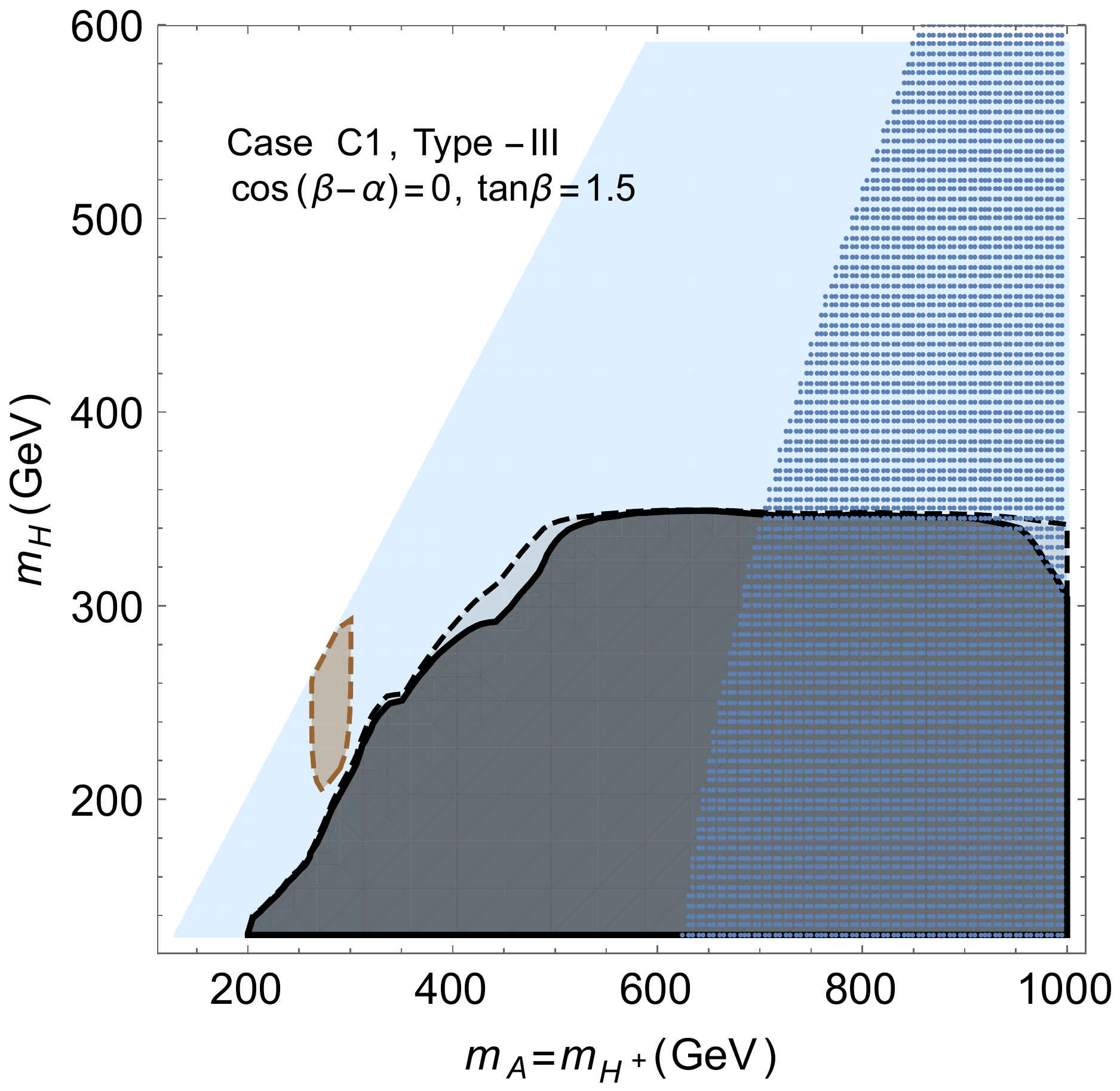}} 
  \hskip 15pt
 \subfigure[]{
 \includegraphics[width=2.8in,height=2.8in, angle=0]{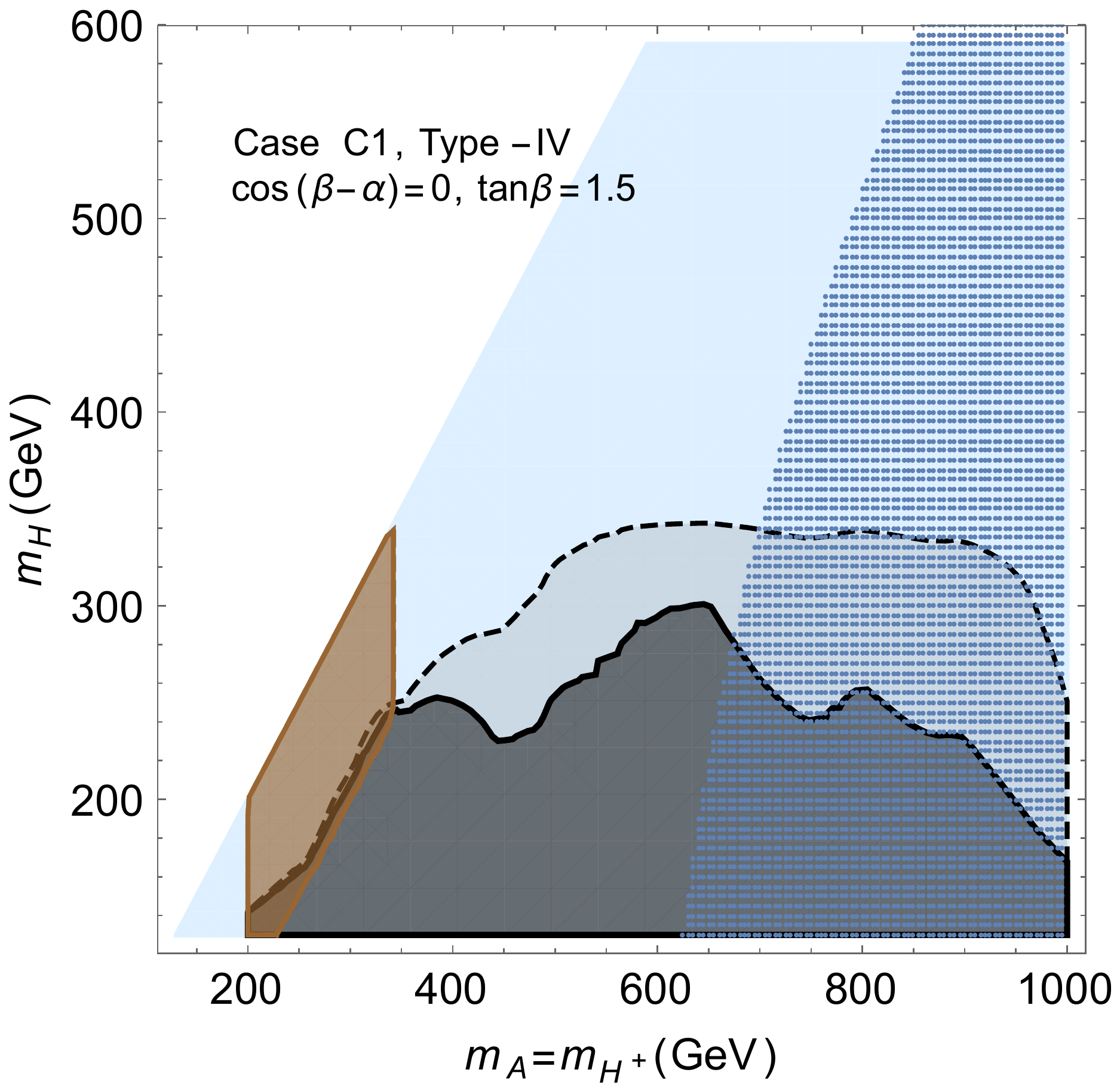}} 
 \caption{The effect of $\varphi^4 D^2$ type of operators for mass spectrum {\bf  C1}.
Only the coloured regions are kinematically viable. The grey regions with the dashed and solid boundaries are ruled out from $gg \rightarrow A \rightarrow ZH(b\bar{b})$~\cite{TheATLAScollaboration:2016loc} in 2HDM and {\bf BP1} of 2HDMEFT respectively. The brown regions with the dashed and solid boundaries are ruled out from $gg \rightarrow A \rightarrow \tau\bar{\tau}$~\cite{ATLAS:2017mpg} in 2HDM and {\bf BP1} of 2HDMEFT. The meshed blue region is disfavoured from the theoretical constraints,\viz stability, perturbativity and unitarity. }
 \label{fig:c1}
\end{center}
 \end{figure}

\begin{figure}[h!]
 \begin{center}
\subfigure[]{
   \includegraphics[width=2.8in,height=2.8in, angle=0]{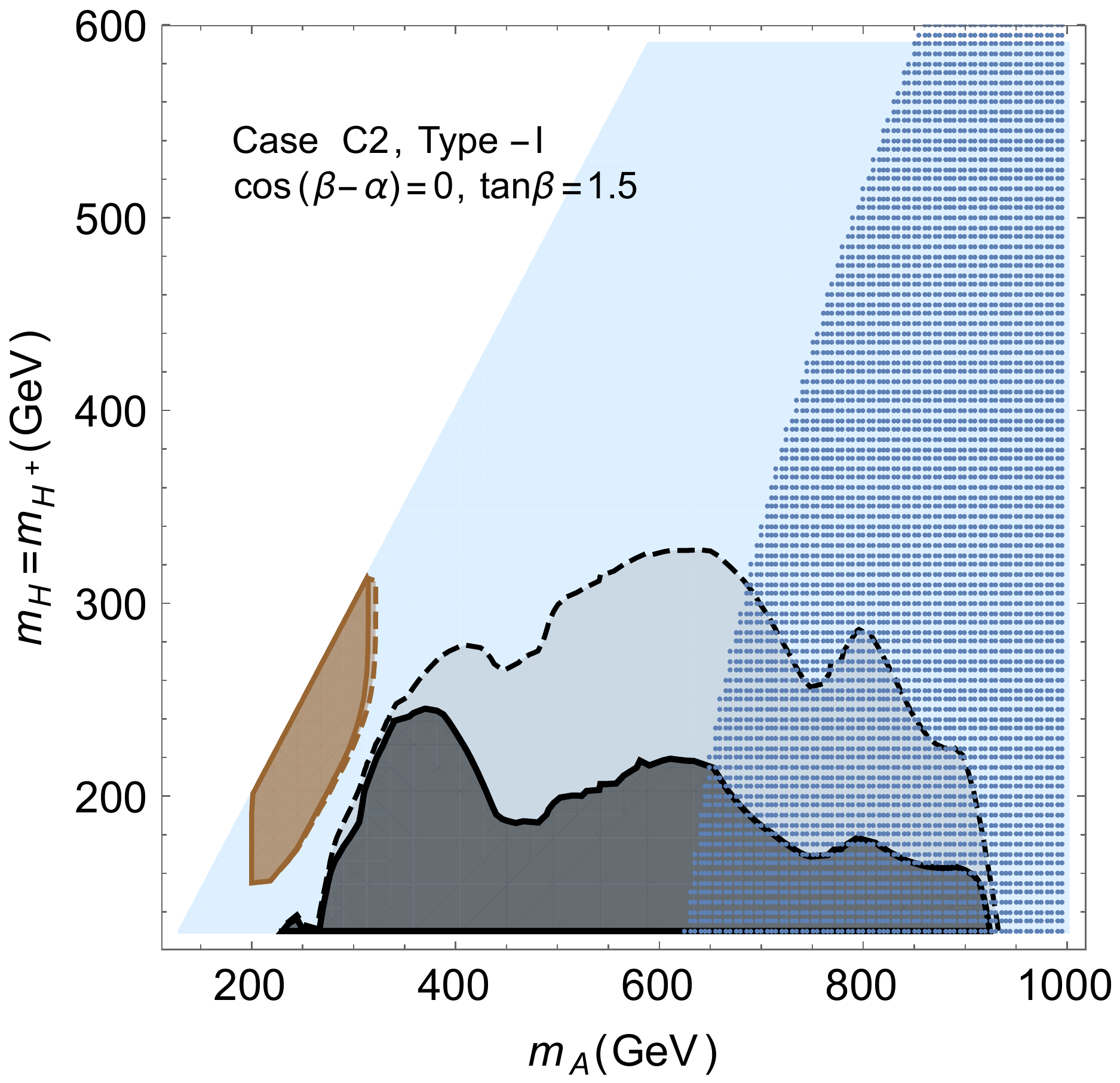}} 
 \hskip 15pt
 \subfigure[]{
\includegraphics[width=2.8in,height=2.8in, angle=0]{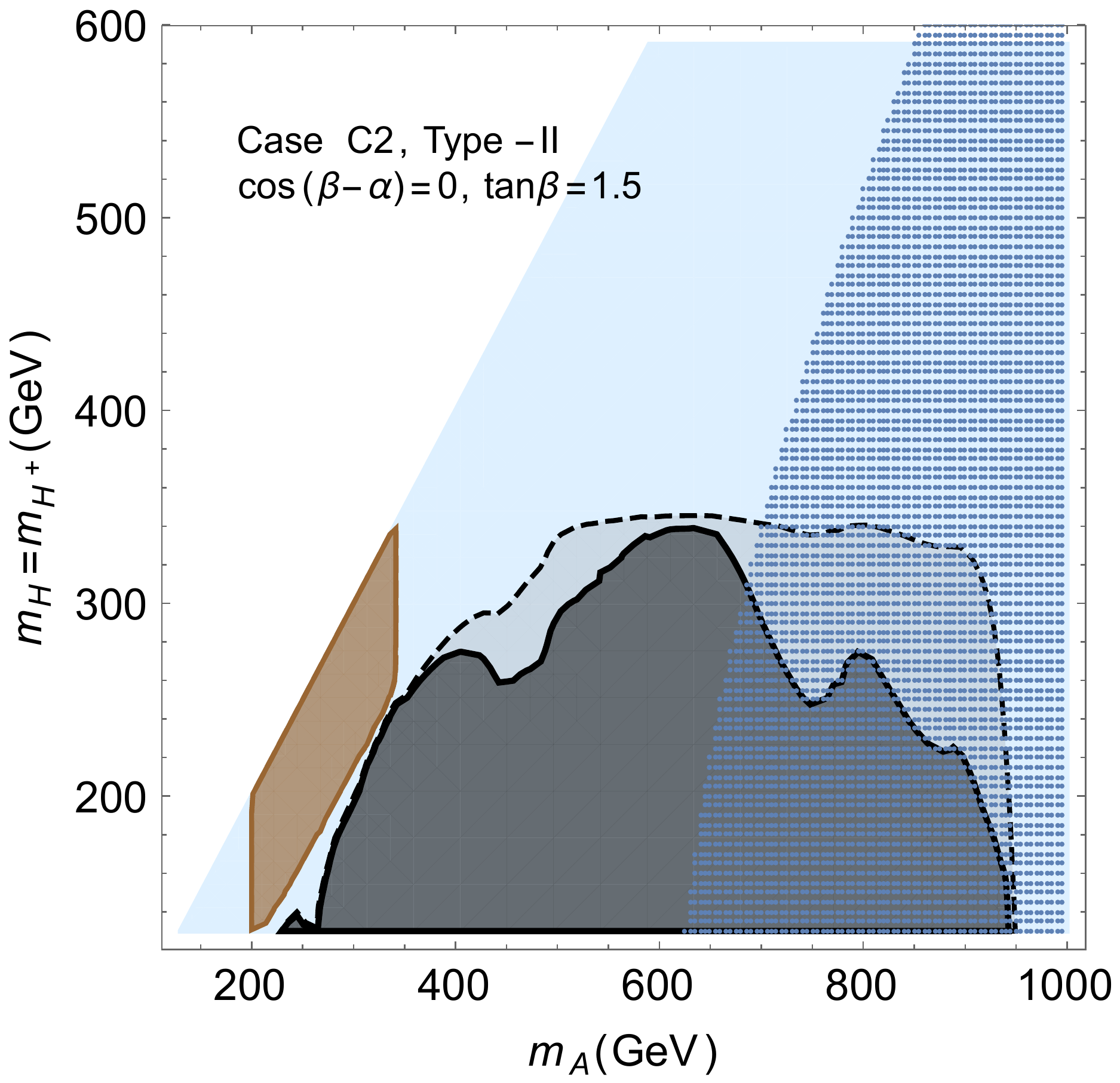}} 
 \hskip 15pt
 \subfigure[]{
 \includegraphics[width=2.8in,height=2.8in, angle=0]{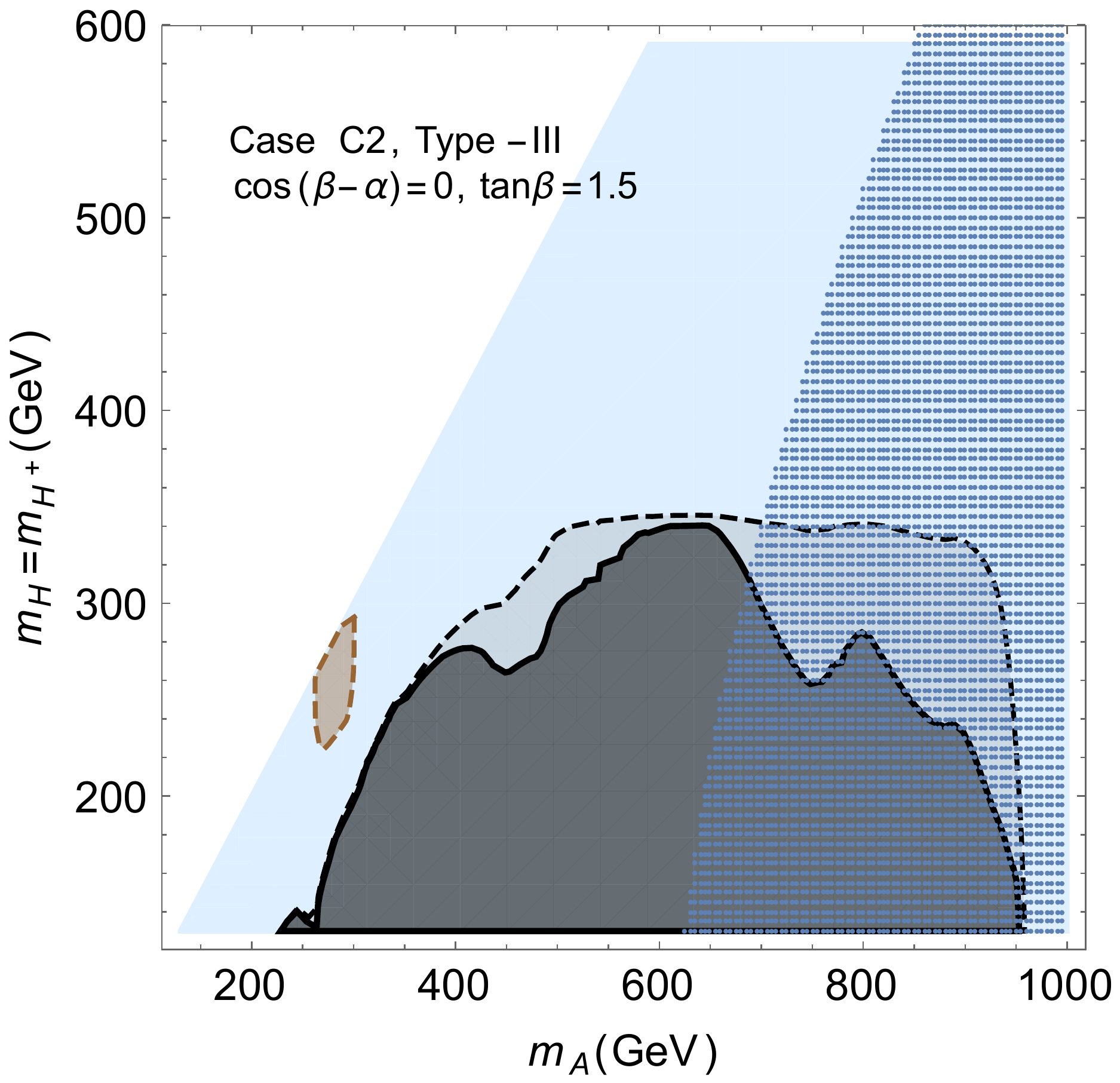}} 
  \hskip 15pt
 \subfigure[]{
 \includegraphics[width=2.8in,height=2.8in, angle=0]{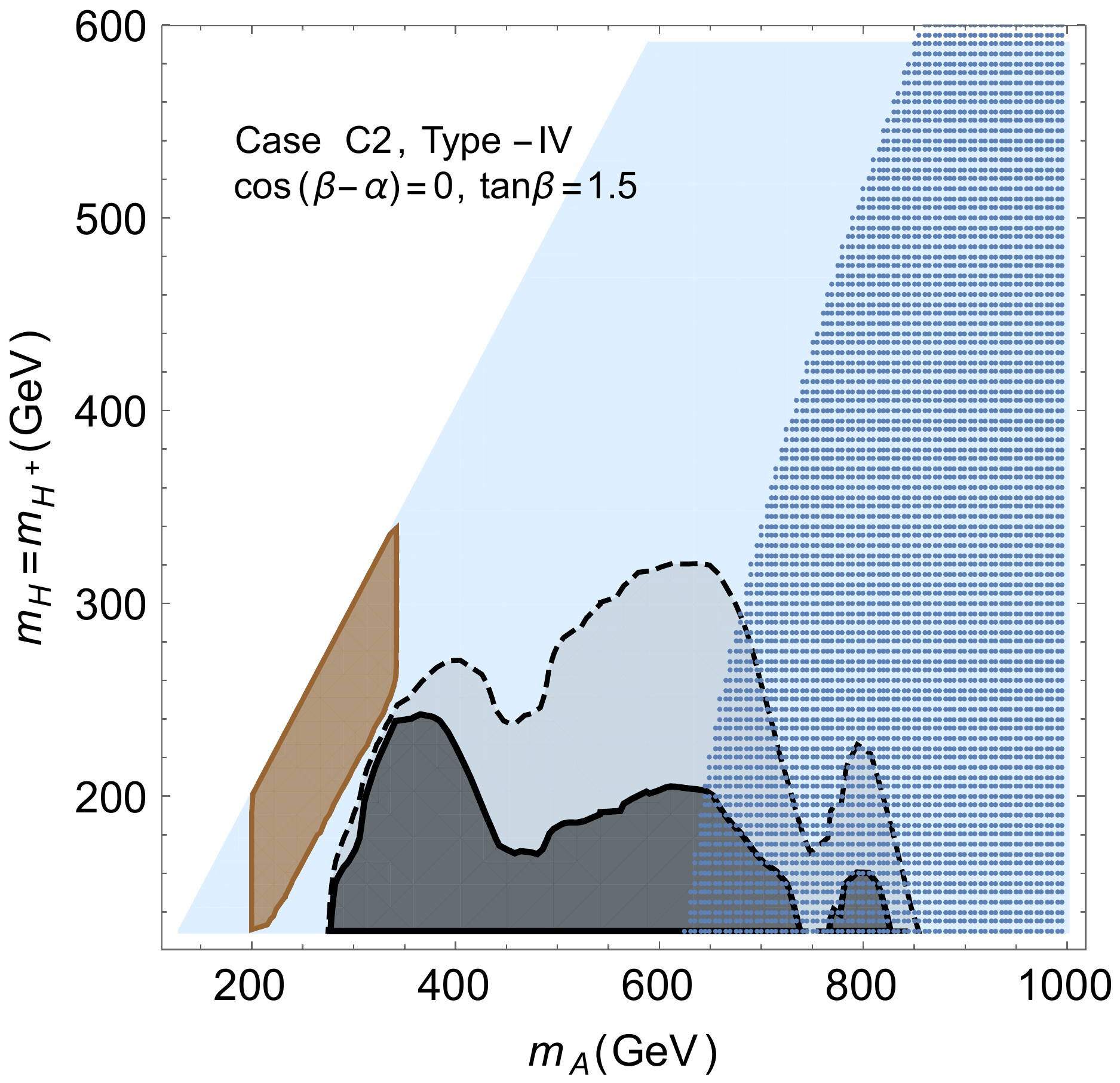}} 
 \caption{The effect of $\varphi^4 D^2$ type of operators for mass spectrum {\bf C2}.  Colour coding is the same as in fig.~\ref{fig:c1}.}
 \label{fig:c2}
\end{center}
 \end{figure}

\begin{figure}[h!]
 \begin{center}
\subfigure[]{
   \includegraphics[width=2.8in,height=2.8in, angle=0]{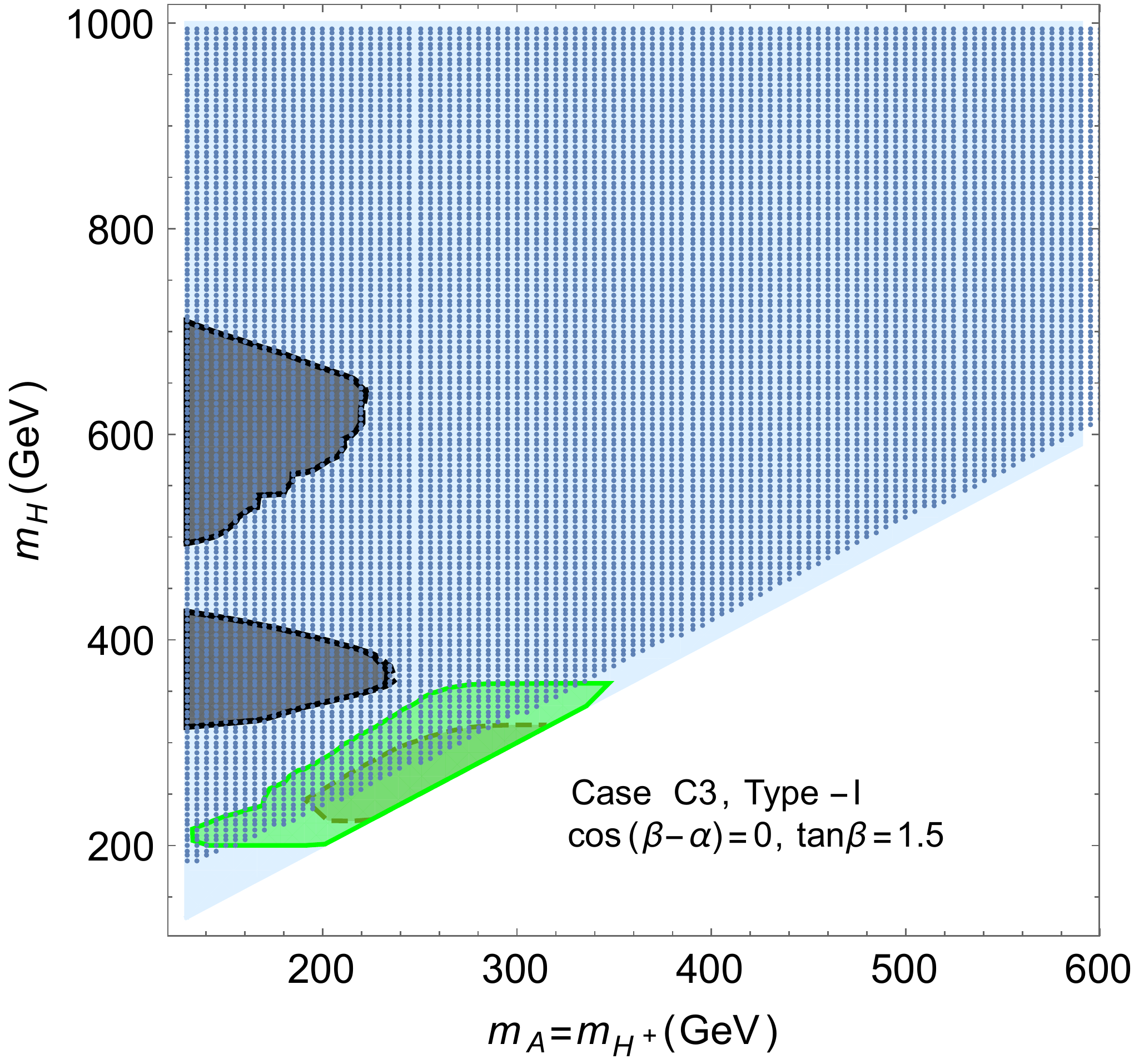}} 
 \hskip 15pt
 \subfigure[]{
\includegraphics[width=2.8in,height=2.8in, angle=0]{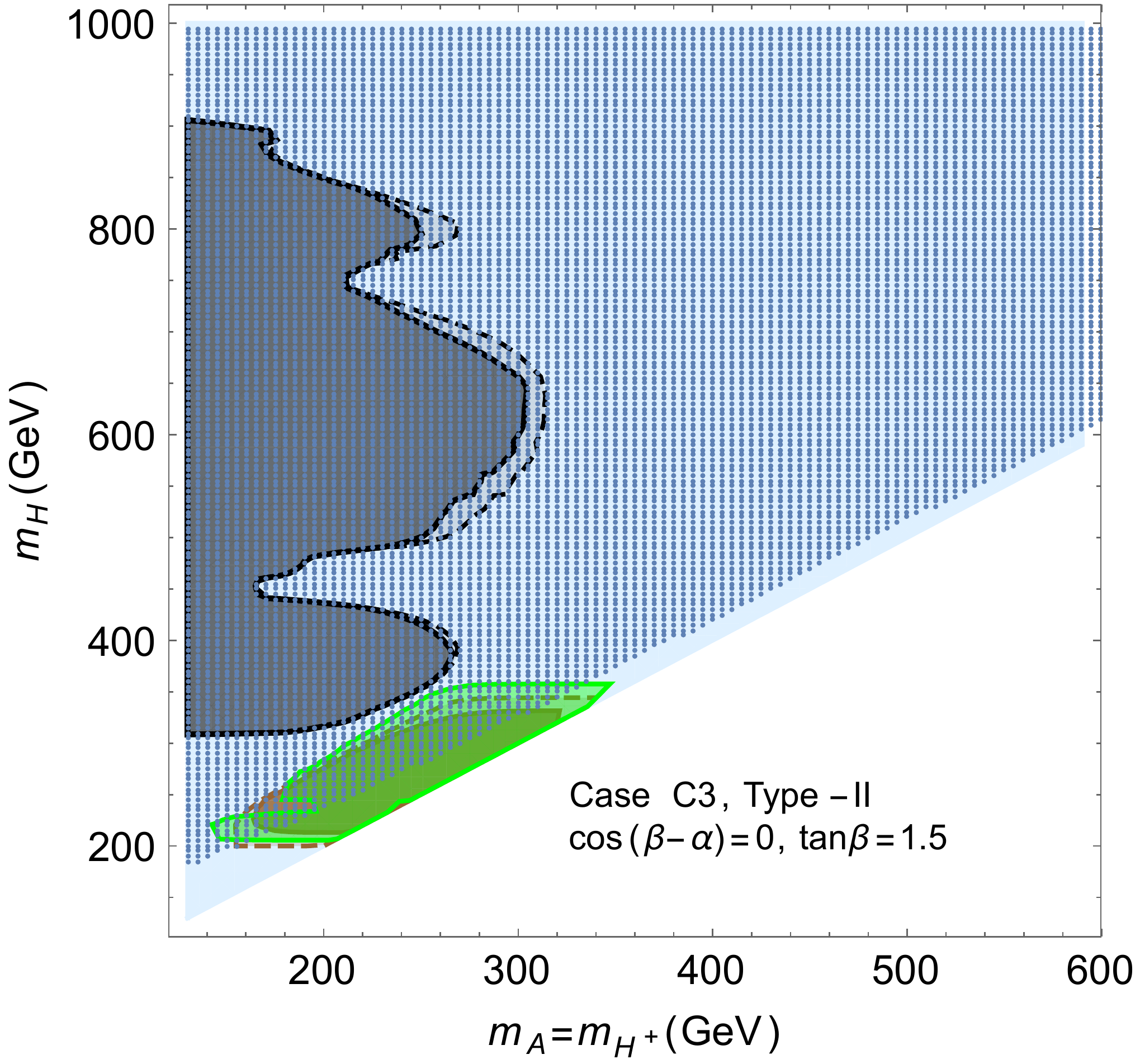}} 
 \hskip 15pt
 \subfigure[]{
 \includegraphics[width=2.8in,height=2.8in, angle=0]{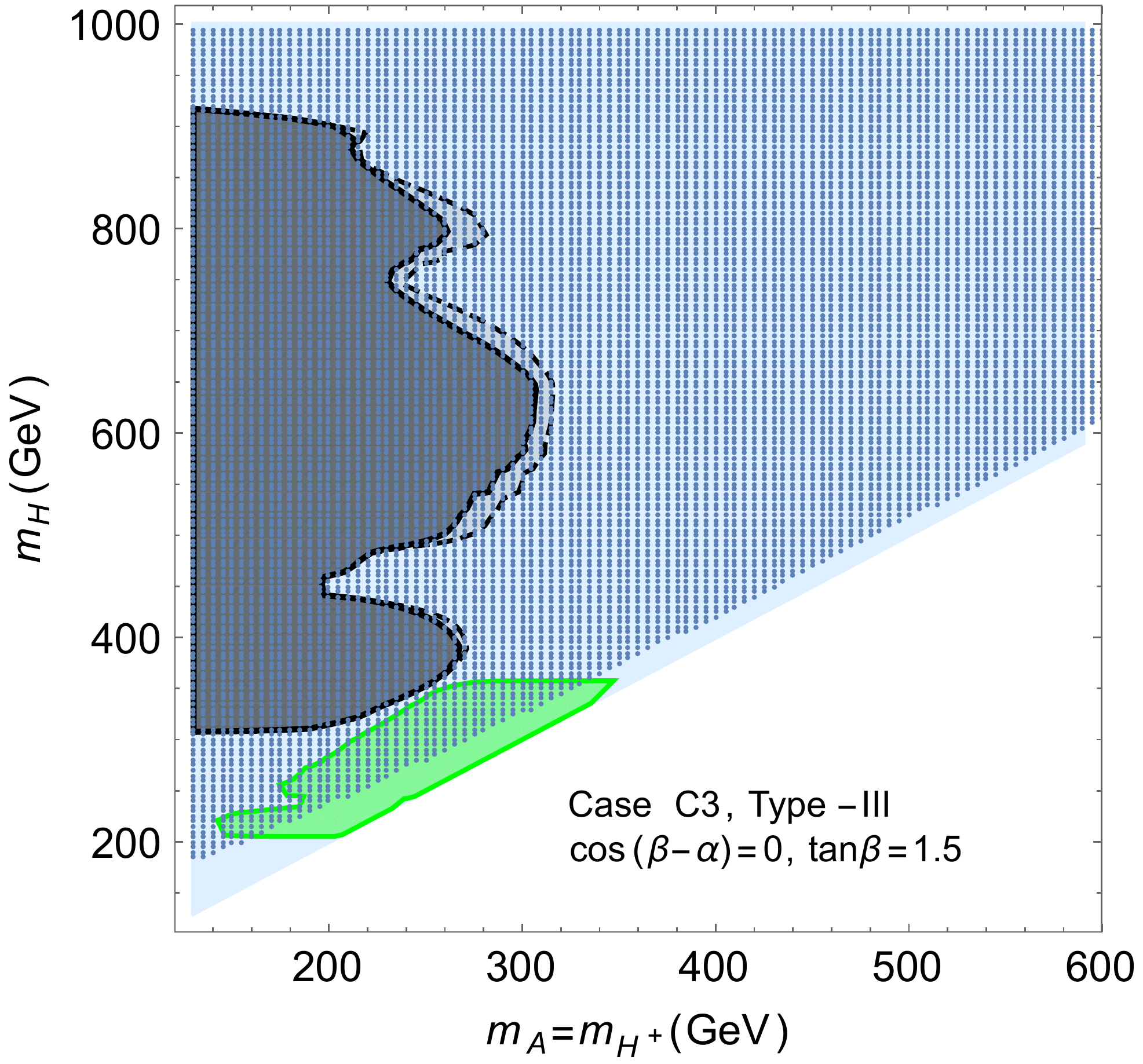}} 
  \hskip 15pt
 \subfigure[]{
 \includegraphics[width=2.8in,height=2.8in, angle=0]{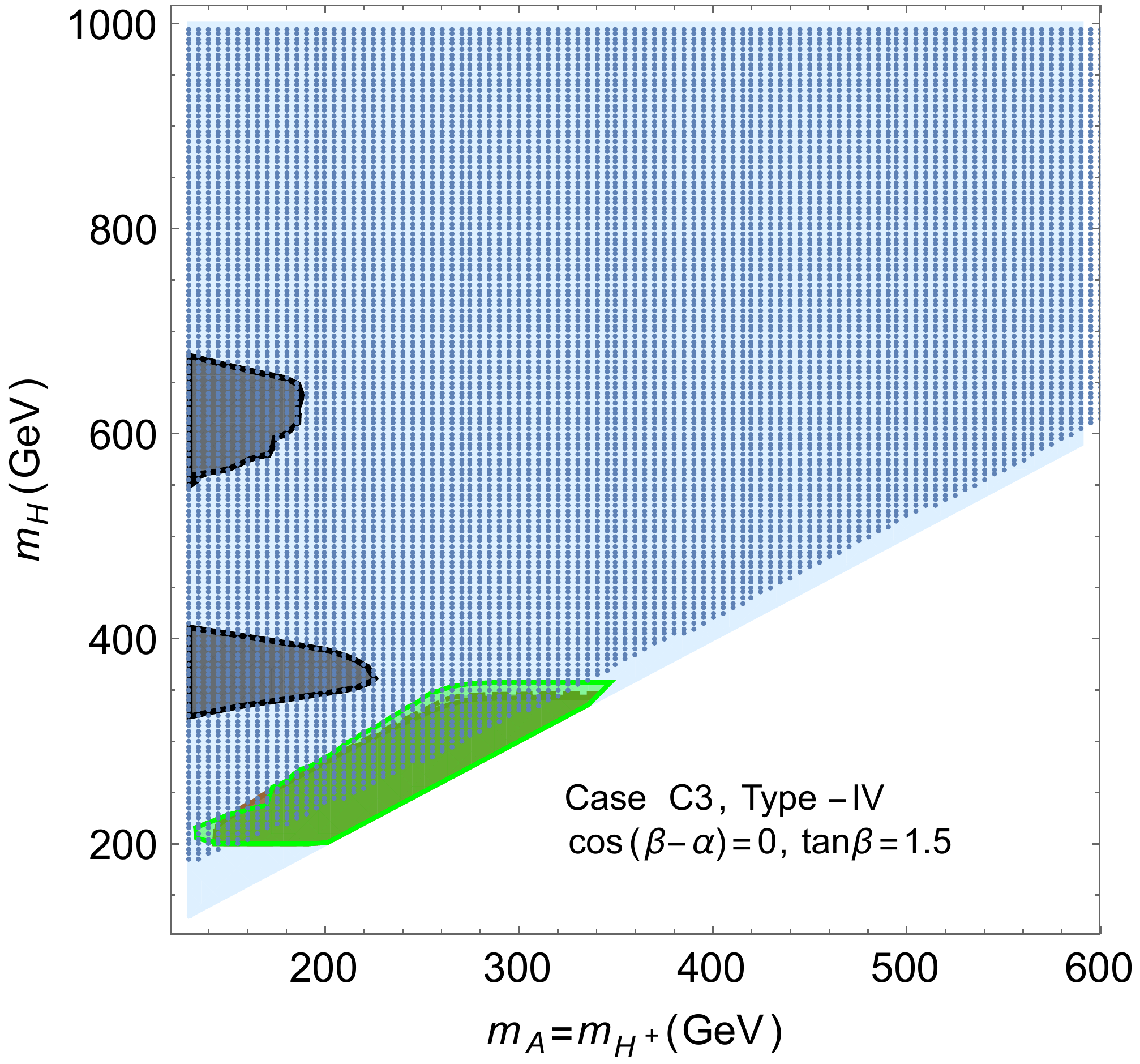}} 
 \caption{The effect of $\varphi^4 D^2$ type of operators for mass spectrum {\bf C3}.
 The grey regions with the dashed and solid boundaries are ruled out from $gg \rightarrow H \rightarrow ZA(b\bar{b})$~\cite{TheATLAScollaboration:2016loc} in 2HDM and {\bf BP1} of 2HDMEFT respectively. The brown regions with the dashed and solid boundaries are ruled out from $gg \rightarrow H \rightarrow \tau\bar{\tau}$~\cite{ATLAS:2017mpg} in 2HDM and {\bf BP1} of 2HDMEFT. 
The green region is disfavoured from the non-observation of $gg \rightarrow H \rightarrow ZZ$~\cite{ATLAS:2017nxi} in {\bf BP1} of 2HDMEFT, which is absent in 2HDM at the tree-level for $\cos (\b-\a) = 0$. 
 Rest of the colour coding is the same as in fig.~\ref{fig:c1}.}
 \label{fig:c3}
\end{center}
 \end{figure}

\begin{figure}[h!]
 \begin{center}
\subfigure[]{
   \includegraphics[width=2.8in,height=2.8in, angle=0]{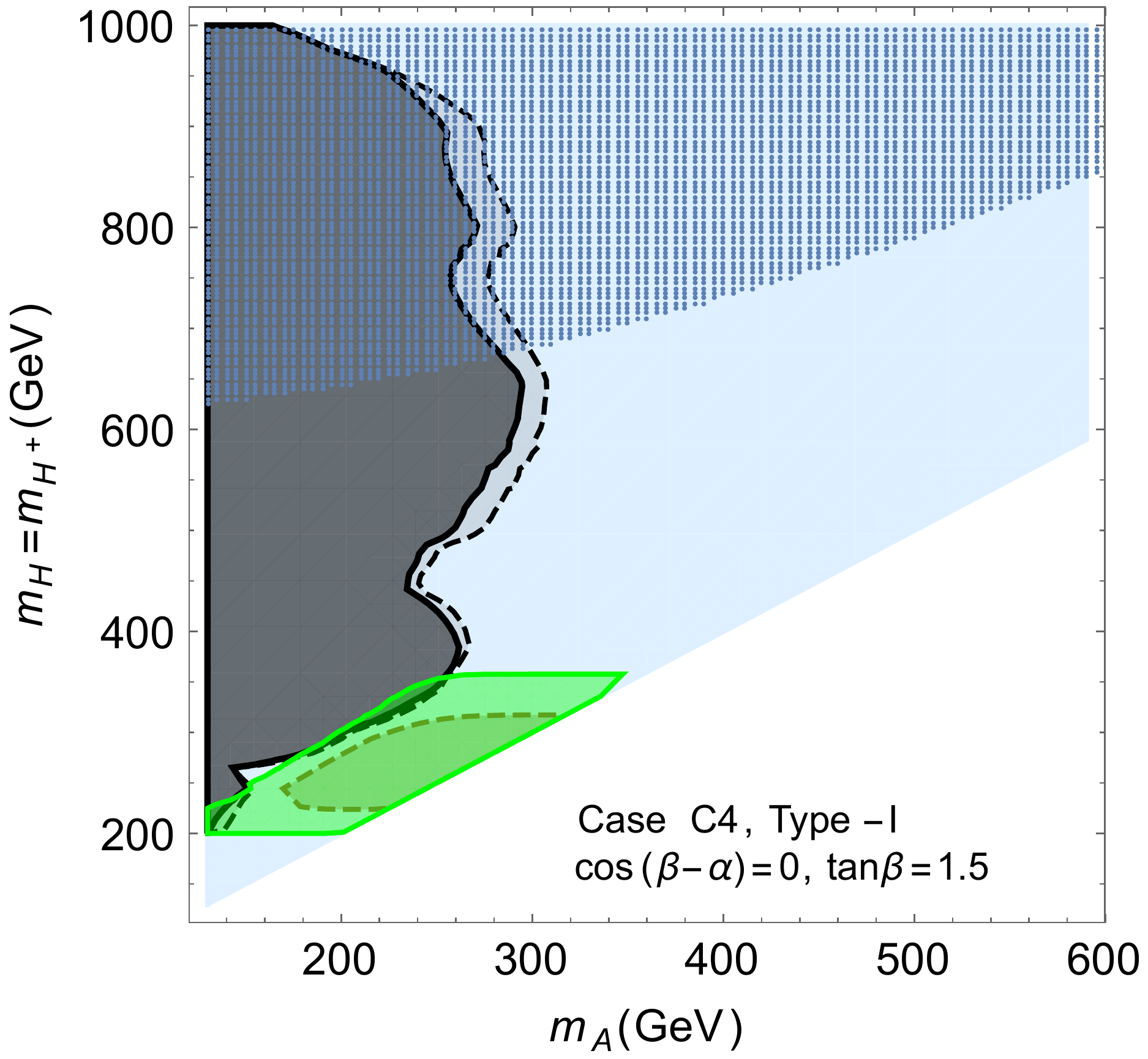}} 
 \hskip 15pt
 \subfigure[]{
\includegraphics[width=2.8in,height=2.8in, angle=0]{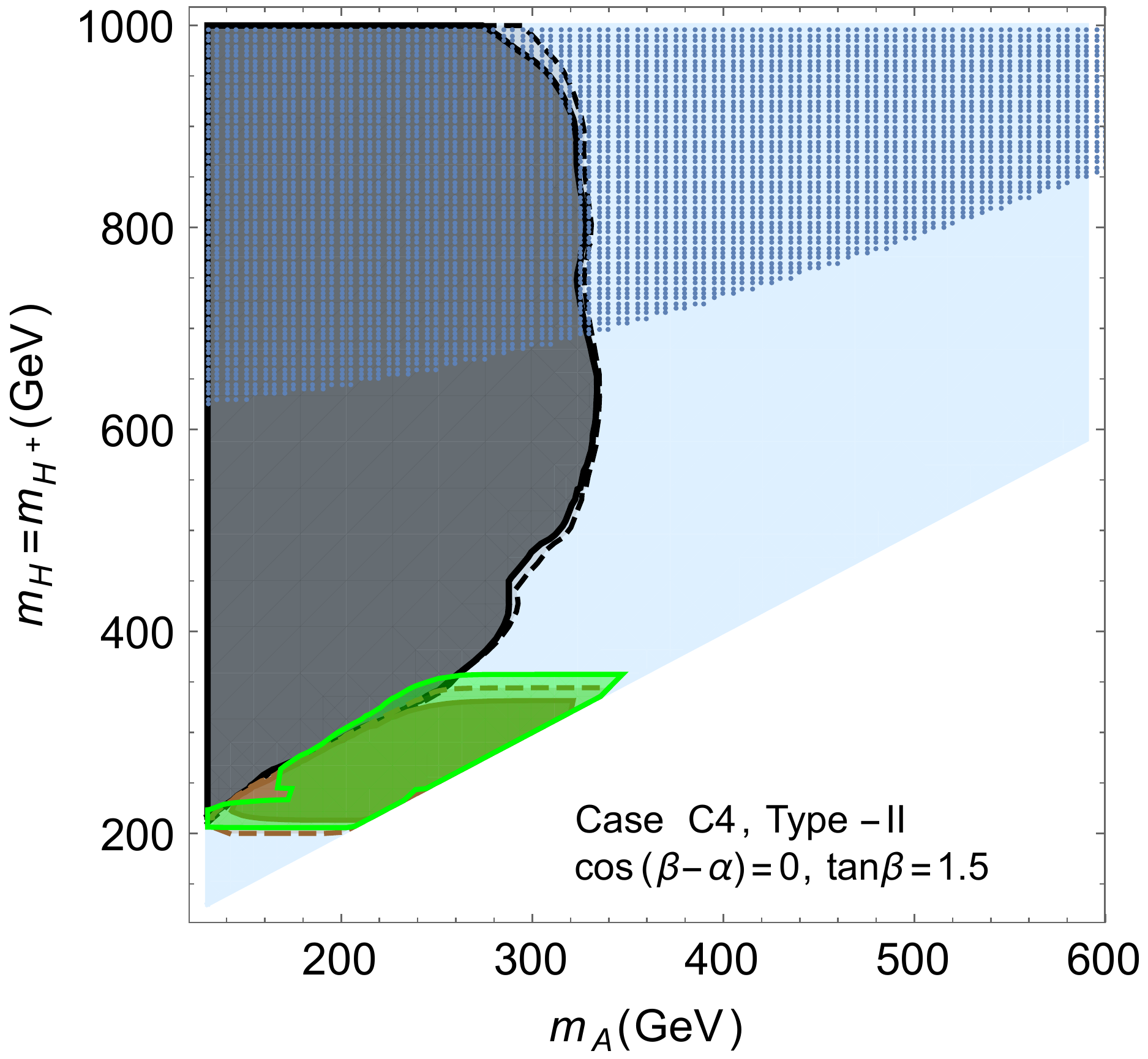}} 
 \hskip 15pt
 \subfigure[]{
 \includegraphics[width=2.8in,height=2.8in, angle=0]{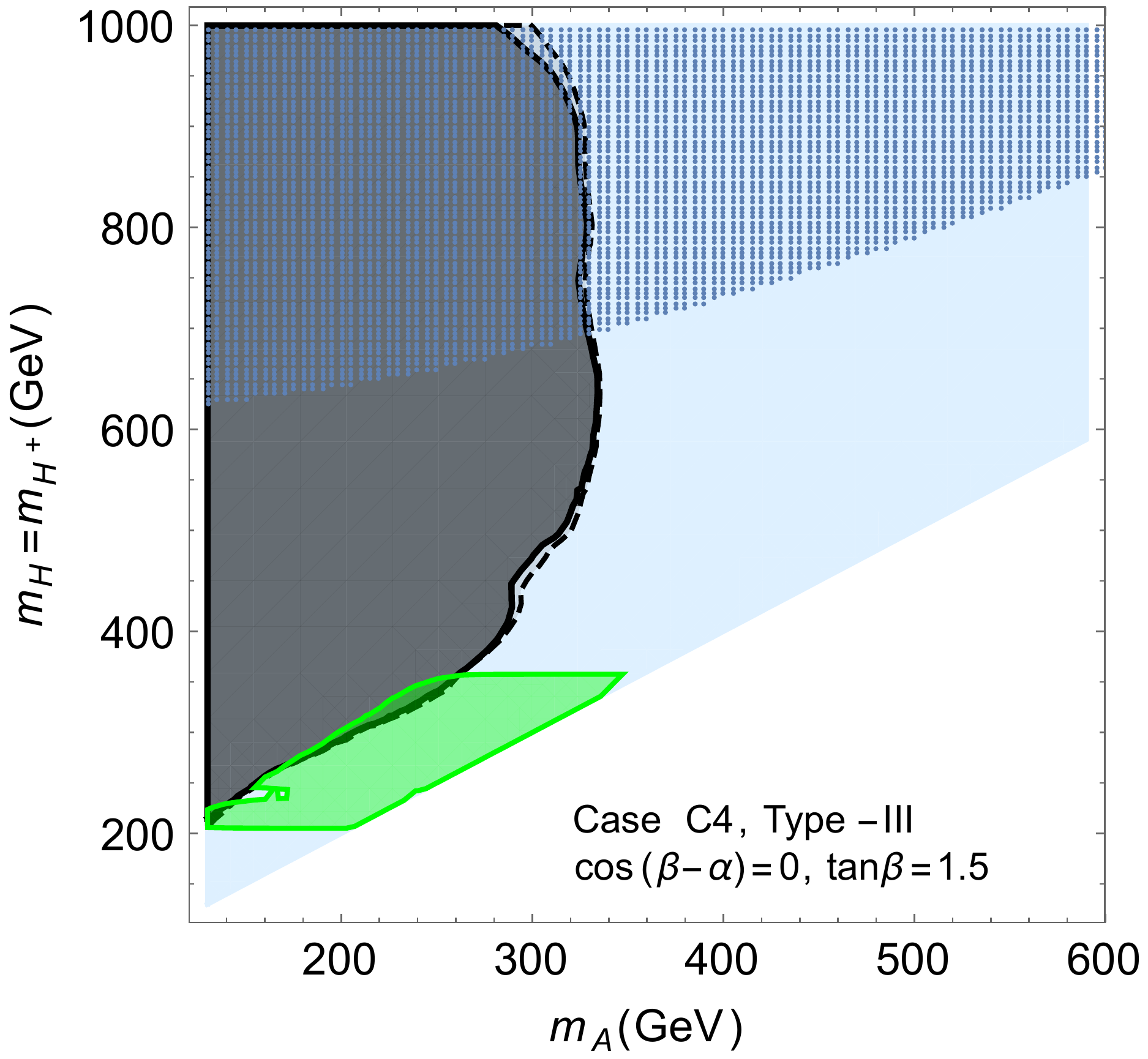}} 
  \hskip 15pt
 \subfigure[]{
 \includegraphics[width=2.8in,height=2.8in, angle=0]{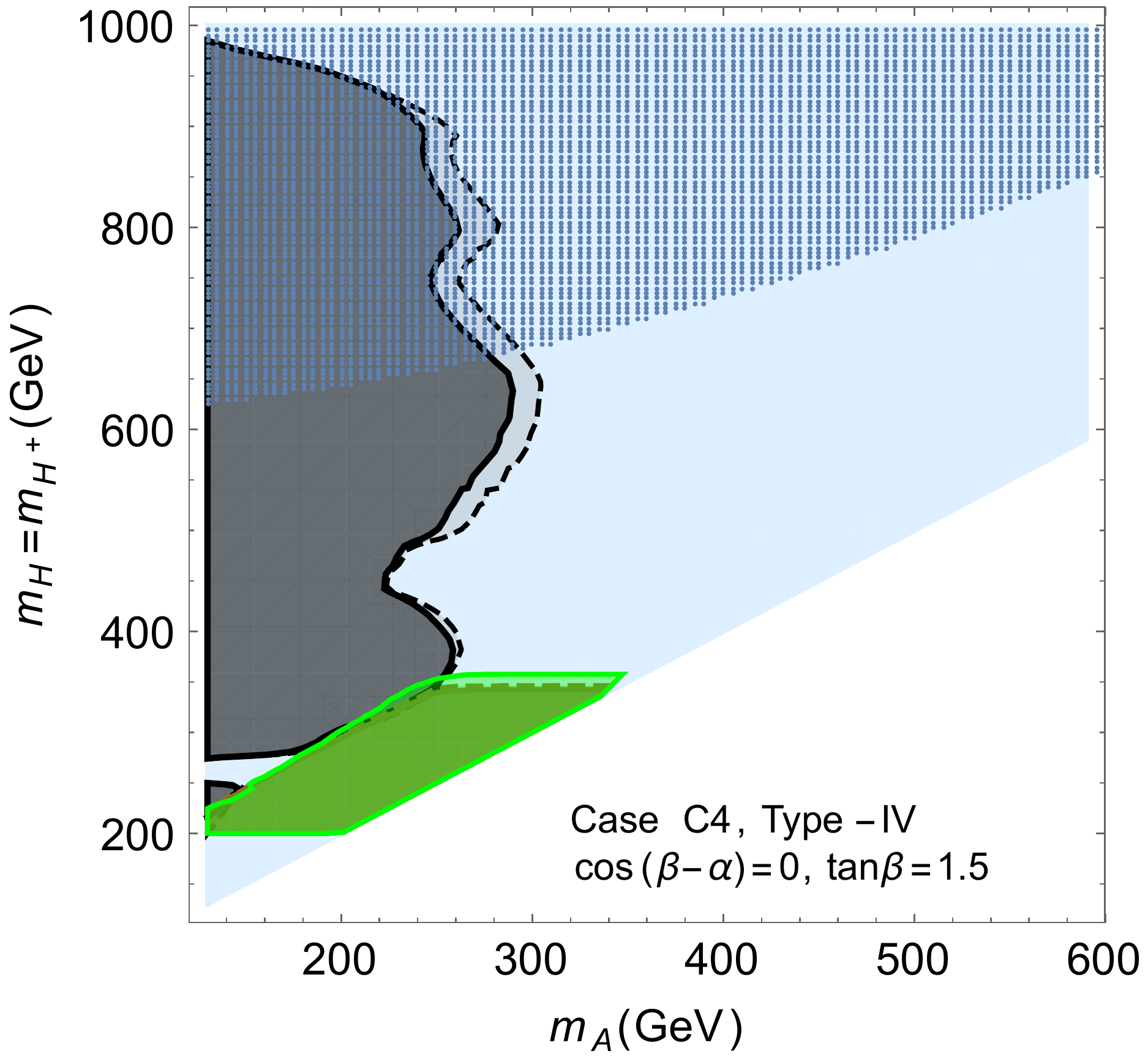}} 
 \caption{The effect of $\varphi^4 D^2$ type of operators for mass spectrum {\bf C4}. Colour coding is the same as in fig.~\ref{fig:c3}. 
 }
 \label{fig:c4}
\end{center}
 \end{figure}

The constraints on the $m_A - m_H$ plane from various exotic Higgs search channels are elucidated in figs.~\ref{fig:c1}, \ref{fig:c2}, \ref{fig:c3} and \ref{fig:c4} corresponding to the mass spectra {\bf C1}, {\bf C2}, {\bf C3} and {\bf C4} respectively.
In fig.~\ref{fig:c1}, the appreciable change from the tree-level 2HDM scenario takes place only for type-I and type-IV Yukawa couplings. 
Moreover, it can be seen that the characteristics of the type-I and -IV 2HDM are similar in this context, as is the case for type-II and type-III. 
This pattern can be attributed to the similarity in the couplings of the new scalars with the SM quarks, following eq.~\eqref{coupmult}, which dictate their production cross-sections at LHC.   
The most significant search channel in the context of the mass spectra {\bf C1} is $A \rightarrow ZH(b\bar{b})$. 
For all four Yukawa types, the change in Br($H \rightarrow b\bar{b}$) can be substantially different in 2HDMEFT compared to tree-level 2HDM, though Br($A \rightarrow ZH$)  does not change significantly.
This happens because a key decay channel of $H$,\ie $H \rightarrow WW$, becomes viable in {\bf BP1} of 2HDMEFT, which is absent in tree-level 2HDM at $\cos (\b-\a) = 0$. 
For example, with $m_A = 400$~GeV and $m_H = 300$~GeV, Br($H \rightarrow b\bar{b}$) becomes $\sim 0.45$ in {\bf BP1} of 2HDMEFT compared to its value $\sim 0.73$ in tree-level 2HDM.
Such a reduction in Br$(H \rightarrow b\bar{b})$ is compensated by the newly viable channel $H \rightarrow WW$.   
This leads to a reduced value of $A \rightarrow ZH(b\bar{b})$, and eventually, to a more relaxed constraint on $m_H$.
Compared to type-I and -IV 2HDM, Br($H \rightarrow b\bar{b}$) is larger for type-II and -III
 Yukawas even in the tree-level scenario.
Thus the percentage change in Br($H \rightarrow b\bar{b}$) is much lower for type-II and -III compared to type-I and -IV.
So the change in the excluded region from the non-observation of $A \rightarrow ZH$ is larger in type-I and -IV compared to type-II and -III.
$A \rightarrow \tau \bar{\tau}$ rules out a region of parameter space where the on-shell decay $A \rightarrow ZH$ is not allowed.
 As eq.~\eqref{coupmult} suggests, Br($A \rightarrow \tau \bar{\tau}$) attains the smallest value for type-III case among all the Yukawa types.
 In both type-I and -III the $A\tau\bar{\tau}$ coupling is proportional to $\cot \b$ as opposed to $\tan \b$ in type-II and -IV. 
Br($A \rightarrow \tau\bar{\tau}$) is even smaller in type-III 2HDM compared to type-I 2HDM because Br($A \rightarrow b\bar{b}$) becomes larger in the latter case. 
On inclusion of the dim-6 operators, the region excluded from $gg \rightarrow A \rightarrow \tau \bar{\tau}$ is not significantly altered for type-I, -II and -IV cases.  

In the context of Yukawa types, a similar pattern in the excluded region can be seen for mass spectra {\bf C2} as depicted in fig.~\ref{fig:c2}.
It can also be seen that the excluded region from $A \rightarrow ZH$ is small in the case {\bf C2} compared to {\bf C1}. 
For mass spectrum {\bf C2}, the decay channel $A \rightarrow W^{\pm}H^{\mp}$ becomes kinematically viable and has branching ratio almost similar to that of $A \rightarrow ZH$. 
However, the branching ratios of $H$ are the same in the cases {\bf C1} and {\bf C2}.
This, along with Br$(A \rightarrow ZH)|_{\textbf{C1}} \gtrsim 2\times \text{Br}(A \rightarrow ZH)|_{\textbf{C2}}$ dictate that the excluded region for {\bf C1} is larger than that in {\bf C2}.  
 As the branching ratios of $H$ in cases {\bf C1} and {\bf C2} are the same in both 2HDM and 2HDMEFT, the pattern of deviation in terms of different Yukawa couplings are similar for {\bf C1} and {\bf C2}. 
The changes in the branching ratios of have been illustrated in figs.~\ref{brratiosA} and \ref{brratiosH} in Appendix~\ref{subsection:brratio}.

$H \rightarrow Z A(b\bar{b})$ can rule out a significant area of parameter space on the $m_A -m_H$ plane for cases {\bf C3} and {\bf C4}.  
 The excluded region in the case {\bf C4} does not show a substantial change upon the inclusion of dim-6 terms as in {\bf C1} or {\bf C2}. 
This happens because Br($A \rightarrow b\bar{b}$) remains almost the same in 2HDM and 2HDMEFT. 
It changes only by $\lesssim 10\%$ due to a small non-zero value of Br($A \rightarrow Zh$) in {\bf BP1} of 2HDMEFT. 
 As opposed to 2HDMEFT, for $\cos (\b-\a) = 0$, $A \rightarrow Zh$ is absent in 2HDM at the tree-level. 
 So the constraints on $m_A$ from $H \rightarrow ZA$ can be relaxed at most by $\sim 19$~GeV for type-I and -IV for {\bf C4} around $m_H \sim 795$~GeV, a region which is already disfavoured from the criteria of stability and unitarity.   
 It can be seen that a significant area is ruled out from $H \rightarrow WW$ for $m_H \lesssim 350$~GeV and $m_H - m_A \lesssim 90$~GeV if the 6-dim operators are included. 
 $H \rightarrow WW$ is otherwise absent in the alignment limit irrespective of the mass spectrum of the new scalars.   
 $H \rightarrow t\bar{t}$ becomes dominant if $m_H \gtrsim 2 m_t \sim 350$~GeV. For $m_H \lesssim 350$~GeV, Br($H \rightarrow ZA$) becomes substantial if $m_H - m_A \gtrsim 90$~GeV. 
Thus the area disfavoured by $H \rightarrow WW$ is confined to a small strip close to the $m_H \sim m_A$ line as shown in fig.~\ref{fig:c4}. 
A similar effect can be seen in fig.~\ref{fig:c3} for the case {\bf C3}.
The appearance of such an exclusion region originates from the fact that in presence of the 6-dim terms, the coupling multiplier $\kappa_{HVV}$ does not vanish at $\cos (\b -\a) = 0$ as opposed to the tree-level 2HDM. 
The region ruled out from $H \rightarrow \tau\bar{\tau}$ overlap with that for $H \rightarrow WW$ in 2HDMEFT in most of the cases.
However, for type-I and -III 2HDM,  the constraint from $H \rightarrow \tau\bar{\tau}$ appears to be relaxed compared to the two other types. 
It can also be seen that the region excluded from $H \rightarrow ZA$ is smaller for {\bf C3} compared to {\bf C4} even in 2HDM at tree-level. 
This occurs because the decay channel $H \rightarrow W^{\pm}H^{\mp}$ becomes kinematically viable for {\bf C3} and Br($H \rightarrow W^{\pm}H^{\mp}$) $\sim$ Br($H \rightarrow ZA$). 
This leads to a smaller cross-section in the channel $gg \rightarrow H \rightarrow ZA$, hence a relaxed constraint on $m_A$ for {\bf C3} compared to the case {\bf C4} even in tree-level 2HDM. 

Till now we have discussed the changes in the excluded region on the $m_A - m_H$ plane at a  low value of $\tan \b$. 
 At large $\tan \beta$,\ie $\tan \b \gtrsim 10$, exotic decays such as $H \rightarrow ZA, A \rightarrow ZH$ become negligible even in the hierarchical scenarios and $gg/b\bar{b} \rightarrow H/A \rightarrow \tau\bar{\tau}$ leads to the only relevant constraint. 
 This can be read off eq.~\eqref{coupmult}.  In such cases, for $\cos (\b-\a) = 0$, the constraints on $m_H$ or $m_A$ are altered at the most by $\sim 5$~GeV.

\begin{figure}[h!]
 \begin{center}
\subfigure[]{
 \includegraphics[width=2.8in,height=2.8in, angle=0]{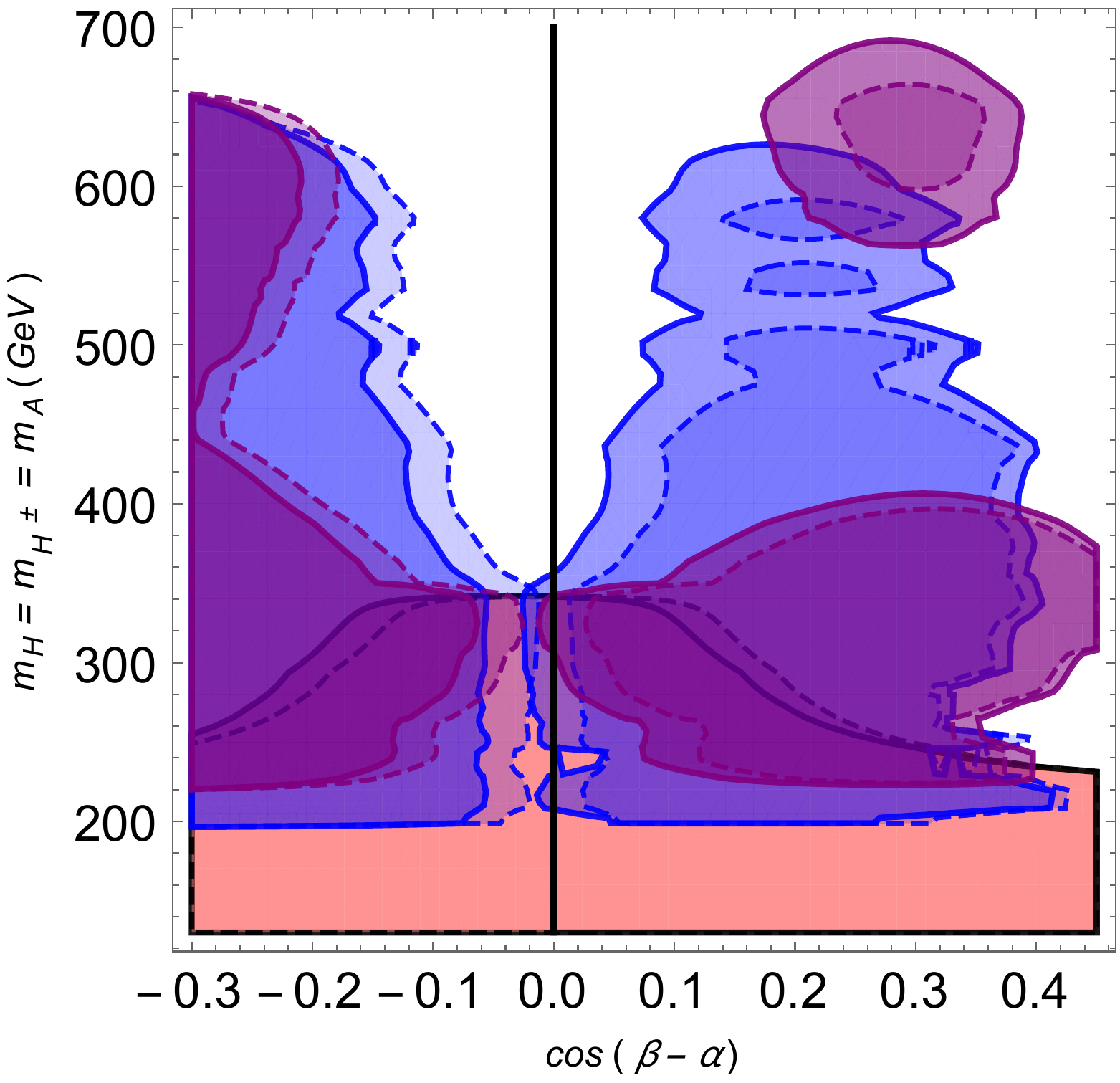}} 
 \hskip 15pt
 \subfigure[]{
 \includegraphics[width=2.8in,height=2.8in, angle=0]{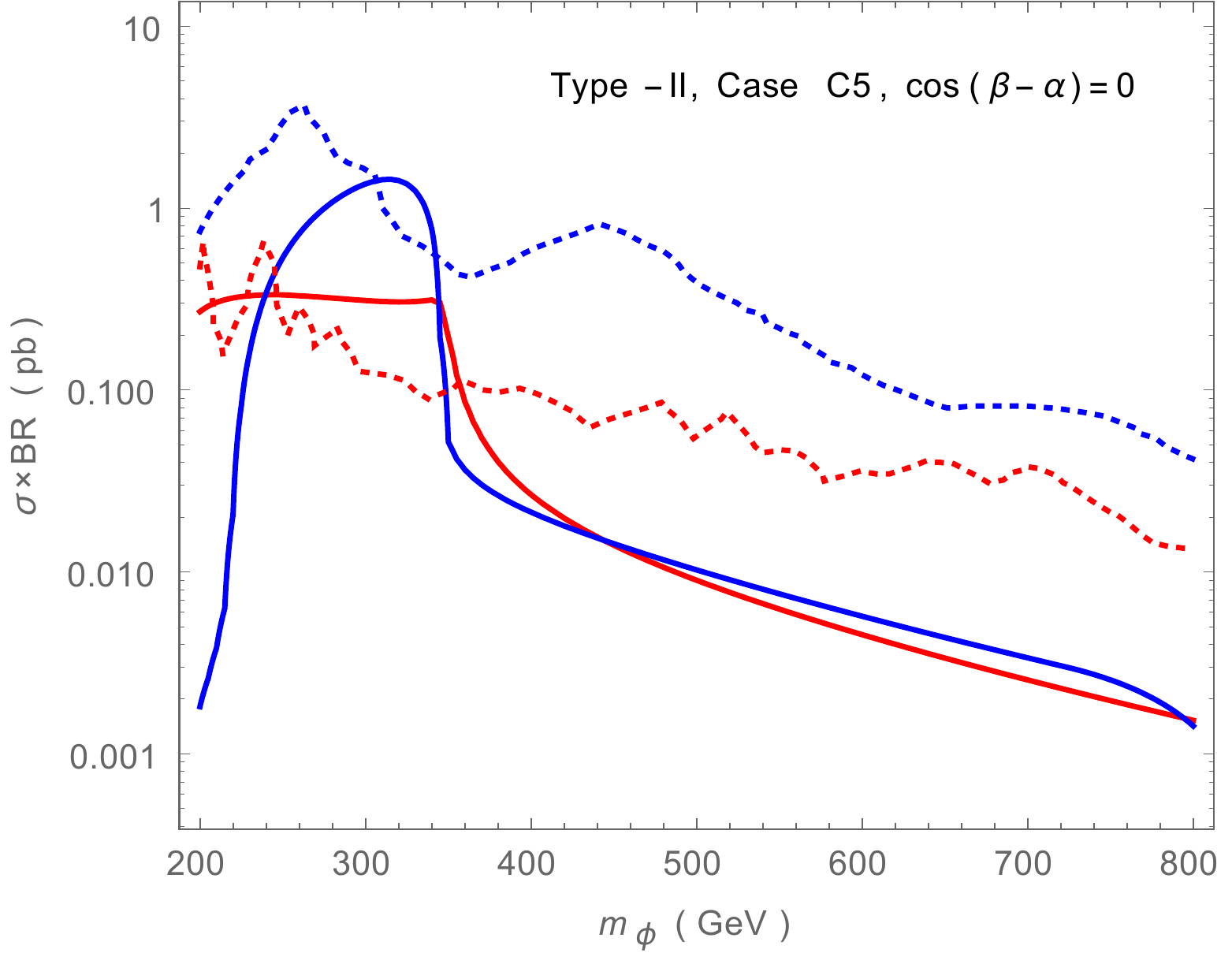}} 
 \caption{The effect of $\varphi^4 D^2$ type of operators for {\bf C5} in Type-II 2HDM. (a)~The  blue regions with the dashed and solid boundaries are ruled out from $H \rightarrow ZZ$~\cite{ATLAS:2017nxi}, the pink regions with the dashed and solid boundaries  are ruled out from $A \rightarrow \tau \tau$~\cite{ATLAS:2017mpg}  and the purple regions with the dashed and solid boundaries  are rules out from $A \rightarrow Zh$~\cite{TheATLAScollaboration:2016loc} in 2HDM and {\bf BP1} of 2HDMEFT respectively. (b)~The solid blue~(red) line represents $\sigma \times $Br for $gg \rightarrow H\rightarrow ZZ$~\cite{ATLAS:2017nxi} ($gg \rightarrow A \rightarrow Zh$~\cite{TheATLAScollaboration:2016loc}) at the alignment limit in presence of the 6-dim operators mentioned in {\bf BP1}. The experimental upper limit to the $\sigma \times $Br corresponding to the two processes are also shown as dotted lines in the same colour.}
 \label{fig:c5}
\end{center}
 \end{figure}

In fig.~\ref{fig:c5}~(a) we have depicted the constraints on the $\cos (\b-\a) - m_A$ plane due to the non-observation of $H$ and $A$ in the degenerate mass scenario for type-II 2HDM. As mentioned earlier, the exotic decay channels like $H (A) \rightarrow A(H) Z$, etc. are absent in such a case.
It can be seen from the eq.~(\ref{coupmult}) that couplings like $HVV$, $AZh$, etc., vanish at the exact alignment limit in 2HDM at the tree-level~\cite{Grzadkowski:2018ohf}.
 Thus the non-observation of $H \rightarrow ZZ$ or $A \rightarrow Zh$ cannot rule out $\cos (\b-\a) = 0$ irrespective of the value of $m_A$. 
This also implies that the discovery of a new scalar in the $VV$ or $Zh$ final states would rule out the exact alignment limit in the framework of a CP-conserving 2HDM. 
Though it is not the case if the 6-dim terms are also present, as those can lead to non-vanishing contribution to such decay channels.
It can be seen from fig.~\ref{fig:c5}~(a), in a 2HDM augmented with $\varphi^4D^2$ kind of operators, the bounds from $H \rightarrow ZZ$, $A \rightarrow \tau\bar{\tau}, Zh$ modify in comparison to 2HDM at the tree level. $H \rightarrow ZZ$ can become non-vanishing, ruling out a range of values of $m_H$ even at $\cos (\b-\a) = 0$.
There is an overall leftward shift in the region  ruled out by $A \rightarrow \tau \bar{\tau}$ which can be followed from eqs.~(\ref{coupmult}) and (\ref{hfieldred}).
For example, the values $m_H \sim 210 - 355$~GeV and $m_A \sim 300- 340$~GeV can be excluded from $H \rightarrow ZZ$ and $A \rightarrow Zh$ respectively at $95\%$~CL even in the alignment limit. 
In fig.~\ref{fig:c5}~(b) we have shown the cross-section of $H$ and $A$ {\it via} the gluon-fusion times the branching ratios in the channels $ZZ$ and $Zh$ at the alignment limit in presence of the higher dimensional operators. 
The value of $\sigma \times $Br for $H \rightarrow ZZ$ reaches $\sim 0.30$~pb in the range $m_H = 200 - 344$~GeV. For $A \rightarrow Zh$ it can reach up to $1.5$~pb in the range $m_A \sim 280 - 330$~GeV. 
For 2HDM at the tree-level, such processes would not at all exist in the alignment limit.  

It is to be noted that, according to eq.~(\ref{coupmult}), for a particular value of $|\cos (\b-\a)|$ and $\tan \b \gtrsim 1$, the $hb\bar{b}$ coupling multiplier is more pronounced in the negative direction of $\cos (\b-\a)$, compared to  the positive direction. 
This effect also propagates in the gluon-fusion cross-section of $H$.  
Thus the area  on the $\cos (\b-\a) - m_A$ plane ruled out by $gg \rightarrow H \rightarrow ZZ$ is larger on the negative $\cos (\b-\a)$ direction. The relevant branching ratios of the heavy scalars have been shown in fig.~\ref{fig:XsBr}. 

\begin{figure}[h!]
 \begin{center}

 \includegraphics[width=2.8in,height=2.8in, angle=0]{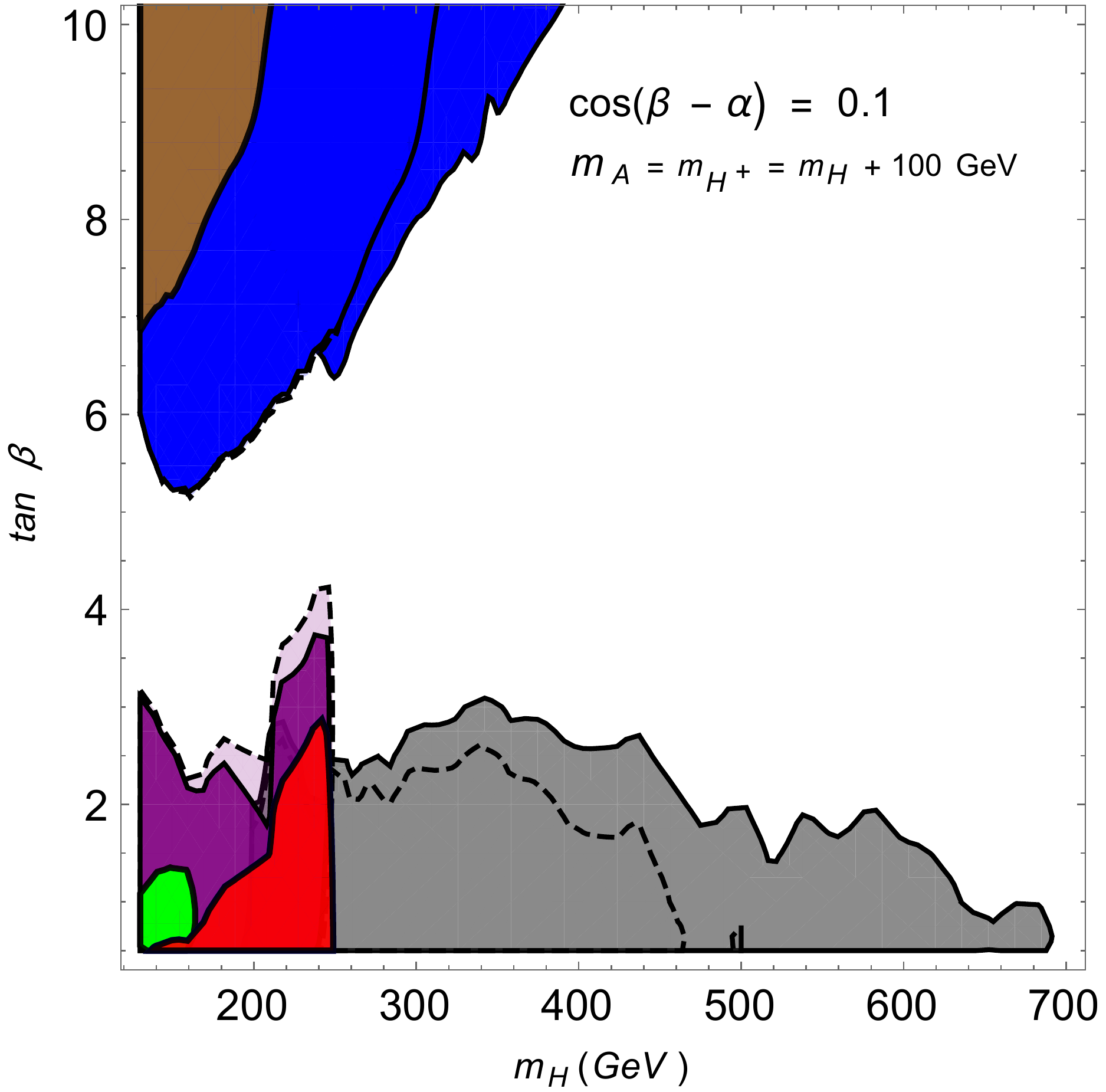} 
 \caption{The effect of $\varphi^4 D^2$ type of operators on the $m_H -\tan \b$ plane in type-II 2HDM. The regions are excluded from $gg \rightarrow H \rightarrow ZZ$~(grey)~\cite{ATLAS:2017nxi}, $gg \rightarrow A \rightarrow ZH$~(purple)~\cite{TheATLAScollaboration:2016loc}, $gg \rightarrow A \rightarrow Zh$~(red)~\cite{TheATLAScollaboration:2016loc}, $gg \rightarrow A \rightarrow \tau\bar{\tau}$~(green)~\cite{ATLAS:2017mpg}, $b\bar{b} \rightarrow A \rightarrow \tau\bar{\tau}$~(brown)~\cite{ATLAS:2017mpg}, $b\bar{b} \rightarrow H \rightarrow \tau\bar{\tau}$~(blue)~\cite{ATLAS:2017mpg,CMS:2015mca}. In case of 2HDMEFT~(dashed lines) the bounds from $A \rightarrow ZH$ and $H \rightarrow ZZ$ are different compared to 2HDM at tree-level~(solid lines).}
 \label{fig:mHtb}
\end{center}
 \end{figure}

In fig.~\ref{fig:mHtb} we have shown the change in the excluded region from various searches of the new scalars on the $m_H - \tan \b$ plane. 
Here, $\cos (\b-\a) = 0.1$ and $m_A = m_{H^{\pm}} = m_H + 100$~GeV. So, this is essentially a hierarchical mass scenario where the exotic decay channel $A \rightarrow ZH$ plays a significant role at lower values of $\tan \b$. 
 For type-II 2HDM both $hb\bar{b}$ and $h\tau\bar{\tau}$ couplings grow with increasing $\tan \b$.
Thus in fig.~\ref{fig:mHtb} higher values of $\tan \b$ are mostly ruled out from the measurements of $b\bar{b} \rightarrow H \rightarrow \tau\bar{\tau}$.
On introduction of the 6-dim terms the region excluded from $H \rightarrow ZZ$ increases, whereas that from $A \rightarrow ZH$ shrinks. For instance, in 2HDMEFT the values $m_H \approx 464 - 686$~GeV can be ruled out for $\tan \b \lesssim 1.5$. Moreover, the excluded region becomes larger in the direction of $\tan \b$ for $m_H \approx 254 - 464$~GeV. 

\begin{figure}[h!]
 \begin{center}
\subfigure[]{
 \includegraphics[width=2.8in,height=2.8in, angle=0]{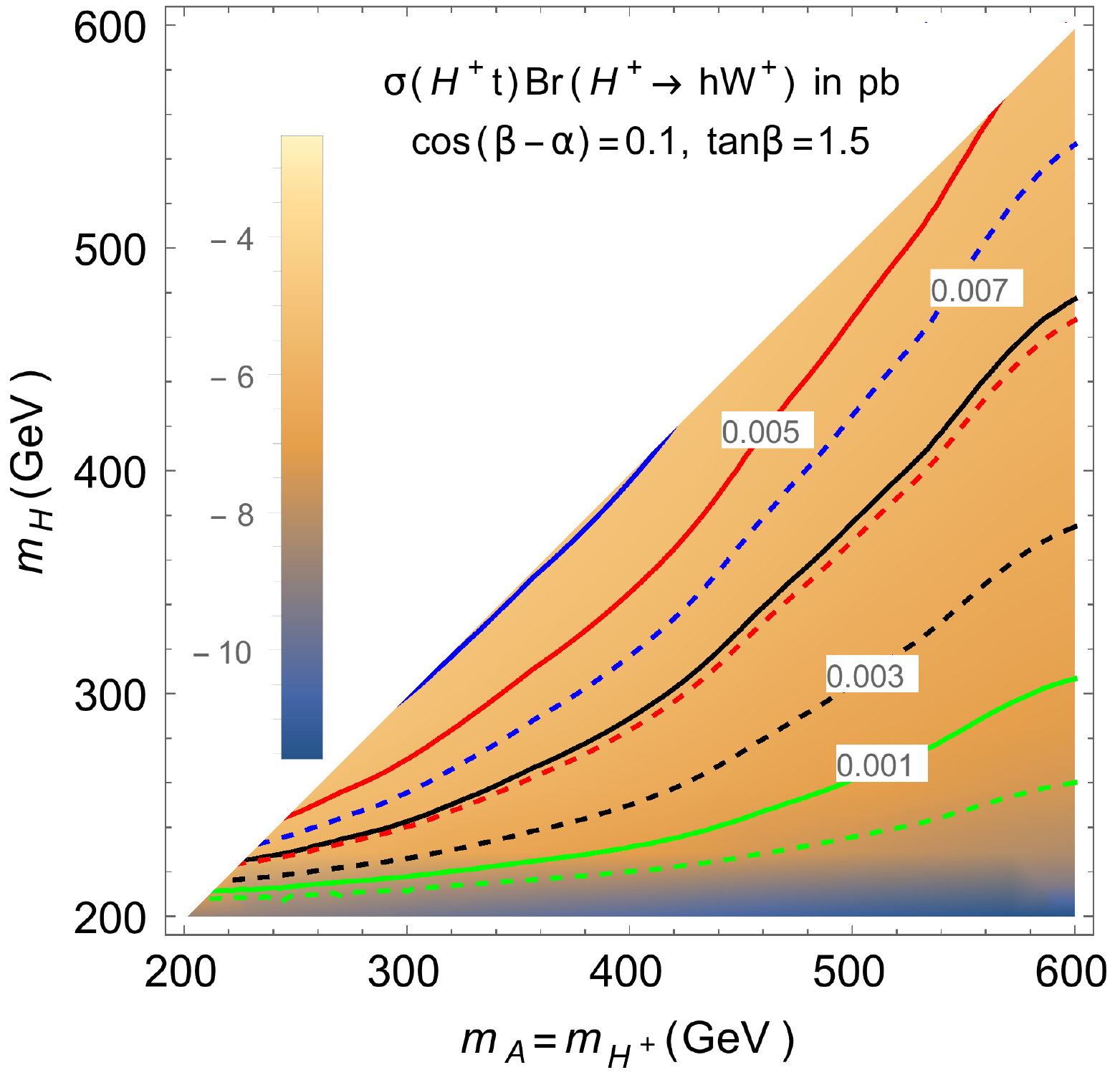}} 
 \hskip 15pt
 \subfigure[]{
 \includegraphics[width=2.8in,height=2.8in, angle=0]{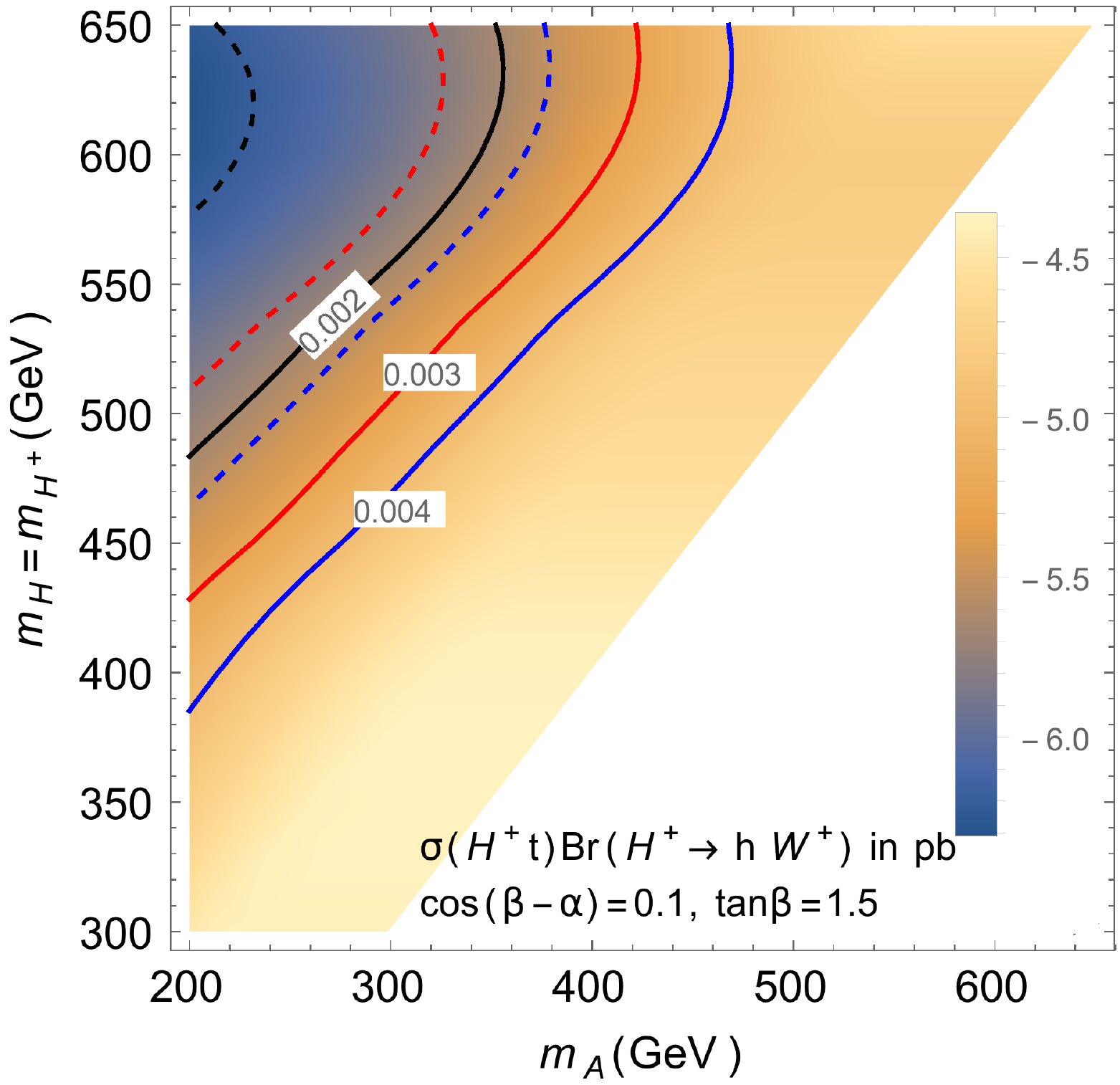}} 
 \subfigure[]{
 \includegraphics[width=2.8in,height=2.8in, angle=0]{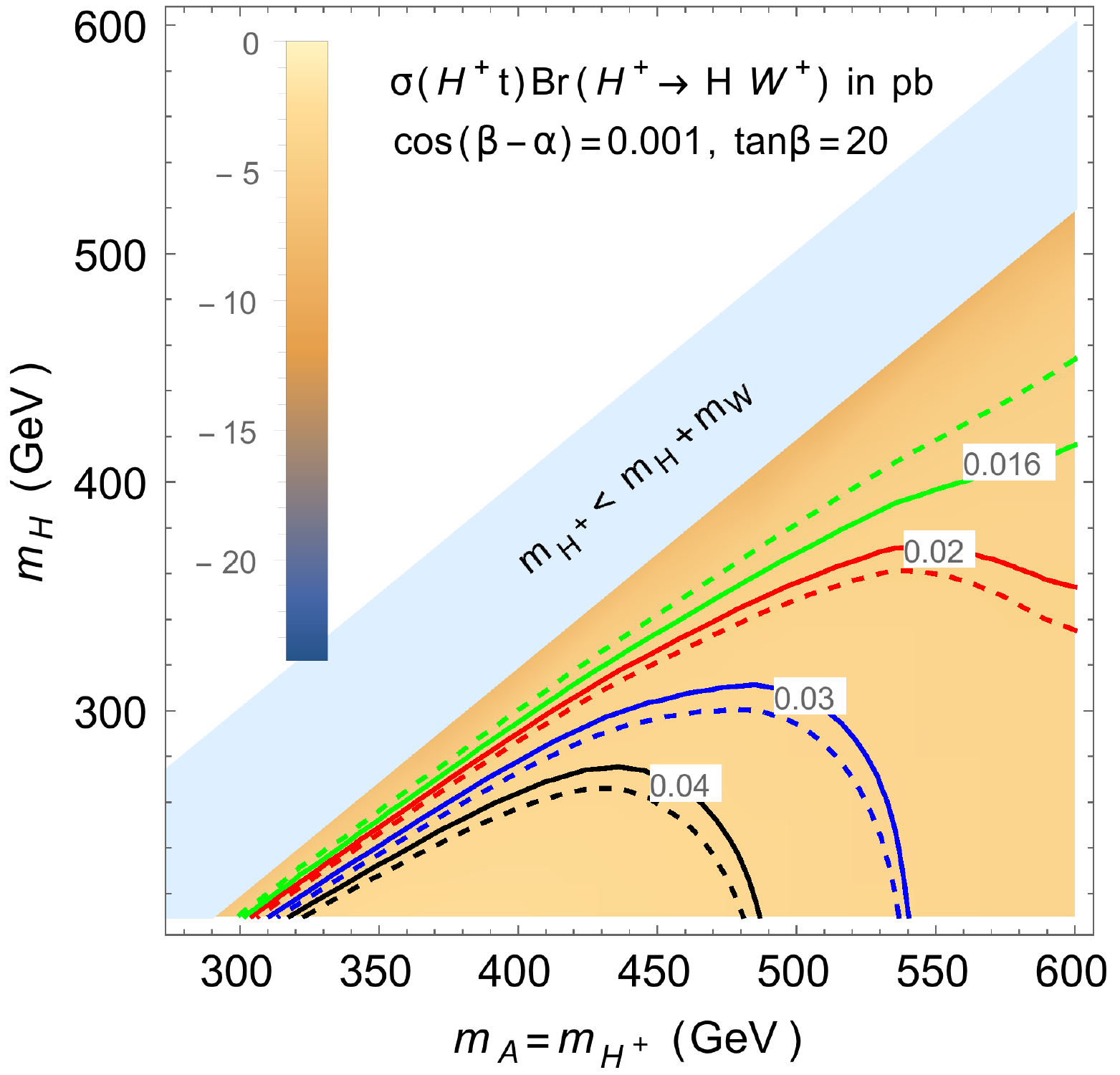}} 
 \hskip 15pt
 \subfigure[]{
 \includegraphics[width=2.8in,height=2.8in, angle=0]{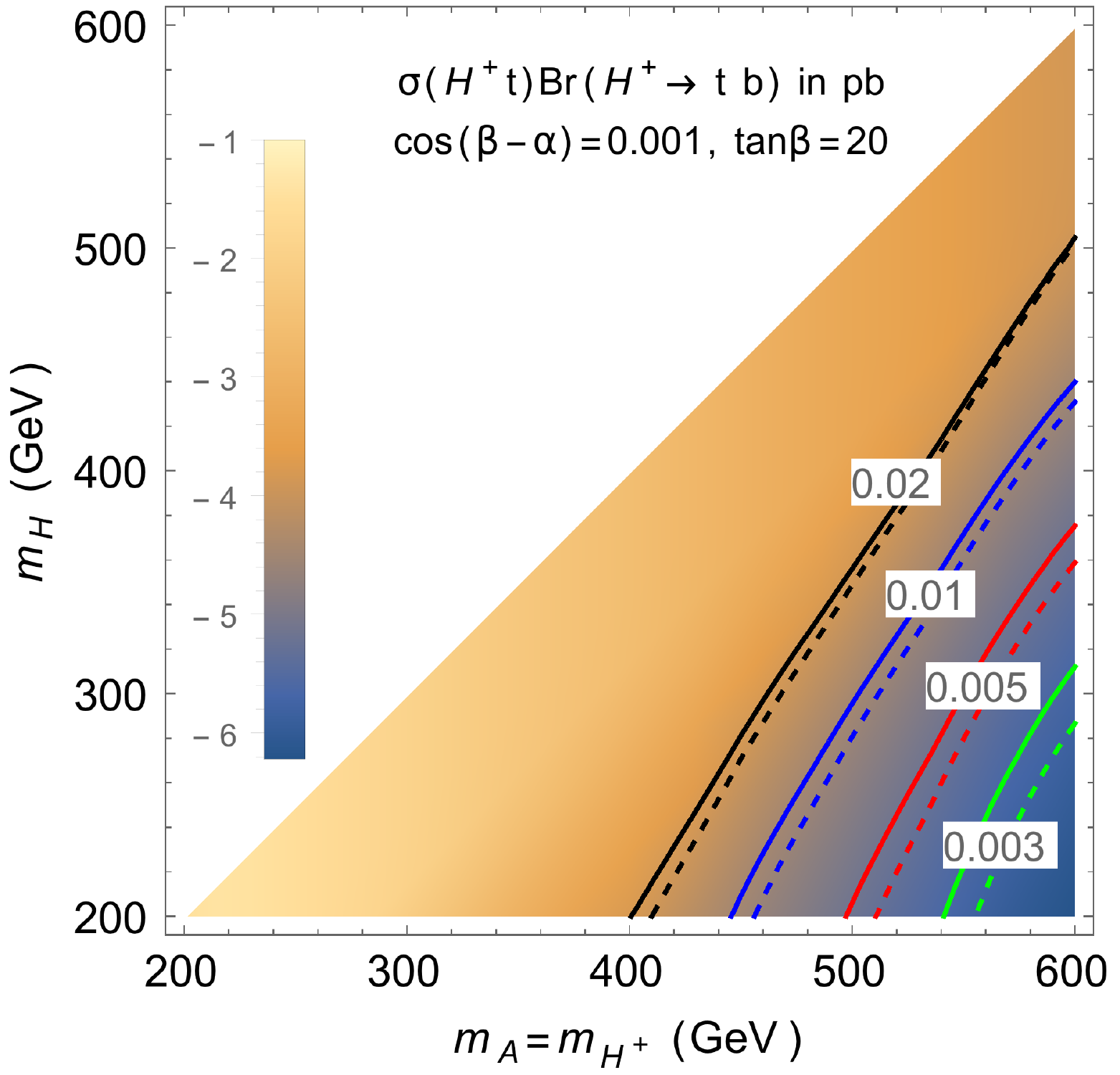}}  
 \caption{The effect of 6-dim terms in type-II 2HDM on (a) $\sigma(H^{\pm}t) \text{Br}( H^{\pm} \rightarrow h W^{\pm})$ for case~{\bf C1}, (b) $\sigma(H^{\pm}t) \text{Br}( H^{\pm} \rightarrow h W^{\pm})$ for case~{\bf C4}, (c)~$\sigma(H^{\pm}t) \text{Br}( H^{\pm} \rightarrow H W^{\pm})$ for case~{\bf C1} and (d)~$\sigma(H^{\pm}t) \text{Br}( H^{\pm} \rightarrow t \bar{b})$ for case {\bf C1}. The solid and dotted lines correspond to the same values of $\sigma \times $Br in 2HDM at tree-level and in {\bf BP1} of 2HDMEFT respectively. The density plots depict the values of the corresponding $\sigma \times $Br in the log-scale at LHC with $\sqrt{s} = 14$~TeV.} 
 \label{fig:HC}
\end{center}
 \end{figure}

So far we have considered the phenomenology of only the neutral scalars. Now we comment on a few effects of the 6-dim terms in 2HDMEFT on the decay modes of the charged scalars. We calculate the production cross-section of the charged scalar following ref.~\cite{Flechl:2014wfa} as it was recommended in ref.~\cite{Heinemeyer:2013tqa}.
For $m_{H^{\pm}} > m_t$, the key production channel of $H^{\pm}$ is through the process $ p p \rightarrow H^{\pm} t$. 
The $H^{\pm} t b$ coupling multiplier depends on the value of $\tan \b$ and the top and bottom quark masses. 
Following the $\tan \b$-dependence of $\sigma (H^{\pm} t)$ we rescale $\sigma (H^{\pm} t)$ at $\tan \b = 30$ with the appropriate numerical factor to obtain the cross-sections at $\tan \b = 1.5$ based on fig.~3 and 10 of ref.~\cite{Flechl:2014wfa}. 
The traditional search channels of a charged Higgs boson consider the decays $H^{+} \rightarrow \bar{\tau} \nu_{\tau},t\bar{b}$,~\cite{Aaboud:2016dig,CMS:2016szv,ATLAS:2016qiq} etc.
Pertaining to different mass spectra of 2HDM, the cascade decay channels with other Higgses as intermediate states can be interesting the bounds on charged Higgs mass can be relaxed significantly~\cite{Coleppa:2014cca,Arhrib:2016wpw}.
In fig.~\ref{fig:HC} we have shown the contours of $\sigma(H^{\pm}t) \text{Br}( H^{\pm} \rightarrow h W^{\pm})$ in {\bf BP1} of 2HDMEFT for the 2HDM mass spectrum corresponding to cases {\bf C1} and {\bf C4} as discussed earlier. 
Br($H^{\pm} \rightarrow H W^{\pm}$) can change for the mass spectrum {\bf C1} and can be followed from fig.~\ref{fig:HC}(c).
The decay channels consisting of SM fermions, such as $H^{+} \rightarrow t \bar{b}, \bar{\tau} \n_{\tau}, c \bar{s}$, etc., become quite important at higher values of $\tan \b$. 
In fig.~\ref{fig:HC}(d) we have shown the change in $\sigma \times$Br with the decay channel $H^{+} \rightarrow t \bar{b}$. Though the value of $\sigma \times$Br in this channel is around one order smaller compared to the current LHC bound on such a process. 

 The coupling multiplier $\kappa_{H^{\pm}tb} \propto (m_t \tan \b P_L + m_b \cot \b P_R)$, and thus it reaches its minimum around $\tan \b \sim 6-8$. 
 Thus the production cross-section of $H^{\pm}$ associated with a top quark becomes quite small for such values of $\tan \b$. 
 So we work only in scenarios when $\tan \b \sim 1$ and $\tan \b \sim \mathcal{O}(10)$. 
 We have not considered the case of $m_{H^{\pm}} \lesssim m_t$ when the key production mode of the charged scalar is $p p \rightarrow t \bar{t}$ with one of the tops in the final state decaying through $t \rightarrow bW^{+}$ and another one {\it via} $\bar{t} \rightarrow \bar{b}H^{-}$.
 For these values of $m_{H^{\pm}}$, bosonic decay channels of $H^{\pm}$, such as $H^{\pm} \rightarrow h W^{\pm}, H W^{\pm}$ are kinematically forbidden, unless one considers $m_H < m_h$, which is not the case for us.

As mentioned earlier,  the coupling $H^{\pm} h W^{\mp}$ vanishes at the alignment limit in 2HDM. But in presence of the 6-dim terms with the Wilson coefficients as in {\bf BP1} of 2HDMEFT, the channel $H^{\pm} \rightarrow h W^{\pm}$ can become significant following the eqs.~(\ref{hfieldred}). For the mass spectrum  in case {\bf C3},  the cross-section in the channel $\sigma(H^{\pm}t) \text{Br}( H^{\pm} \rightarrow h W^{\pm})$ can go up to $\sim 25$ fb for $m_{H^{\pm}} \sim 250$~GeV at LHC with $\sqrt{s} = 14$~TeV. The dependence of cross-section in this channel on $m_{H^{\pm}}$ for all four hierarchical mass spectra can be followed from Appendix~\ref{HphW}.
 

The couplings of $h(125)$ will be even more precisely measured in the future experiments.
For instance, the coupling multipliers $\kappa_{h\gamma \gamma}$ and $\kappa_{hWW}$ are to be measured with an accuracy of $\sim 5-7\%$ and $\sim 4-6\%$ respectively at HL-LHC with  luminosity $\sim 3$ \text{ab}$^{-1}$~\cite{ATLASHL}.
It can push a 2HDM, especially the ones with type-II, -III and -IV Yukawa couplings, further close to $\cos (\b-\a) = 0$.
However, the contribution of dim-6 terms to the signal strengths of $h(125)$ do not decrease with the same scale. 
As it was also discussed in ref.~\cite{Karmakar:2018scg}, even at the exact limit $\cos (\b-\a) = 0$, the effects of the dim-6 terms in masking the true alignment limit can be rather significant. 
Thus even at the limit when the couplings of $h(125)$ are exactly at par with the SM expectations, the heavier scalars are not decoupled from the rest of the particle spectrum. 
This remarkable feature can also be interpreted as the violation of the sum rules involving the couplings of the CP-even Higgses in 2HDM extended with dim-6 operators. 
In this paper, we have demonstrated several cases, where the cross-sections of certain decay channels of the heavier scalars are significant in the presence of 6-dim terms at $\cos (\b-\a) = 0$. 
This leads to an interesting possibility of detecting the heavier scalars in these channels at HL-LHC, even if $h(125)$ exactly resembles the SM Higgs.

\section{Summary and discussions}
\label{summary}
In the context of the searches for new scalars at LHC, it is an interesting possibility that the exotic scalars in a 2HDM exist below the TeV scale, pertaining to the so-called `alignment-without-decoupling' scenario.
A study in 2HDMEFT becomes relevant in this case.
Such an approach is appealing because it allows us to study the constraints in the 2HDM parameter space while remaining agnostic about any new physics beyond 2HDM.
In this paper, we have confined our discussion to the bosonic operators of 2HDMEFT. 
The changes in the constraints on the masses of the exotic scalars of 2HDM are studied in the presence of 6-dim operators of type $\varphi^4 D^2$, because the other bosonic operators are quite constrained from electroweak precision tests.   

We consider both degenerate and hierarchical mass spectra of the new scalars in this purpose and show the changes in the constraints for all four Yukawa types.
The theoretical constraints, such as stability, perturbativity, and unitarity, as well as the measurement of the oblique parameters, restrict the mass differences of such scalars. 
In light of that one can narrow down four types of the mass spectrum in the hierarchical case. 
We notice that in a couple of such cases, dubbed as {\bf C1} and {\bf C2} in the text, the constraints on the $m_A -m_H$ plane can be significantly relaxed in the presence of certain 6-dim operators of type $\varphi^4 D^2$.  
For example, in case {\bf C2} with $\cos (\b-\a) = 0$, $\tan \b = 1.5$ and type-I Yukawa coupling, the upper limit on $m_H$ reduces to $\sim 196~$GeV in {\bf BP1} of 2HDMEFT from $\sim 300~$GeV which is the case for 2HDM at tree-level.
Such changes are always more pronounced for type-I and -IV 2HDM compared to type-II and -III.

At $\cos (\b-\a) = 0$, processes such as $H \rightarrow ZZ$, $A \rightarrow Zh$, etc. vanish for 2HDM at the tree-level, which is not the case if dim-6 operators are present.
A non-zero value for Br($H \rightarrow WW$) reduces the value of Br($H \rightarrow b\bar{b}$), which brings down the cross-section for the process $pp \rightarrow A \rightarrow ZH(b\bar{b})$, thus relaxing the constraint on $m_H$. 
Such changes are not significant at higher values of $\tan \beta$ irrespective of the mass spectrum under consideration. This happens because the  SM fermionic decay modes of the heavier scalars dominate for higher values of $\tan \b$ and appearance of new bosonic decay channels cannot change the key decay channels involving SM fermions significantly. 
For the degenerate case we notice that, for our chosen benchmark scenario, the region excluded from the non-observation of $H$ and $A$ becomes larger in 2HDMEFT compared to 2HDM at the tree-level. 
It is also seen, as it was discussed above, a certain mass range for $m_H (= m_A)$ is ruled out even for $\cos (\b-\a) = 0$ from processes like $H \rightarrow WW$, $A \rightarrow Zh$, which usually vanish in 2HDM at tree-level.
We have also shown in fig.~\ref{fig:HC} the change in $\sigma \times $Br for various decay channels of the charged scalar in 2HDMEFT compared to 2HDM at the tree-level at LHC with $\sqrt{s} = 14~$TeV. 

The key reason for the change in the constraints on 2HDM parameter space upon including dim-6 operators of type $\varphi^4 D^2$ lies in the redefinition of the CP-even Higgs fields, $h$, and $H$.
This way the coupling multipliers involving the CP-even scalars are rescaled compared to 2HDM at the tree-level and lead to a change in the branching ratios of all the processes  which involve $h$ and $H$. 
It leads to the departure of the `true' alignment limit from its tree-level 2HDM counterpart,\ie $\cos (\b-\a) = 0$.
As the projected accuracy of the $h(125)$ coupling measurement at a future version of LHC, such as HL-LHC is at the level $\lesssim 5-6\%$, 2HDMs might get further pushed to the alignment limit.
Thus, in the presence of dim-6 operators, even if the couplings of $h(125)$ turn out to be completely aligned with the SM Higgs, the heavier scalars in 2HDM with masses $\lesssim~$TeV still do not decouple from the SM sector,\ie their discovery might still be viable. 
As mentioned earlier, some cascade-type decay channels of the heavier scalars vanish at the alignment limit in the tree-level 2HDM. 
It implies that the discovery of a new scalar in such a channel would perhaps rule out the alignment limit in a CP-conserving 2HDM.
But if dim-6 operators are present, even if a new scalar is discovered in such channels, it will no longer rule out the alignment limit.

In case of the discovery of the new Higgs(es), the verification of the sum rules involving their coupling multipliers can provide useful information about the nature of the extended Higgs sector.   
In 2HDMEFT, the redefinition of the CP-even Higgs fields due to $\varphi^4 D^2$ operators also imply that the sum rules involving these scalars are modified in a certain way.
We have discussed how the measurement of sum rules can help distinguishing between various options beyond a CP-even 2HDM.

If new scalars are discovered at the LHC in near future, the correlation of their signal strengths in different channels will be important to determine the exact nature of the underlying scalar sector. 
In this context, 2HDMEFT can be an efficient framework in quantifying the departure from tree-level 2HDM in various channels, providing an opportunity to narrow down the possible UV-complete scenarios.

\section{Acknowledgements}
S.K. acknowledges T.~Stefaniak for help regarding \texttt{2HDMC}. This work is supported by the Department of Science and Technology, India {\it via} SERB grant EMR/2014/001177 and DST-DAAD grant INT/FRG/DAAD/P-22/2018.

\vfill

\appendix

\section{Field redefinition}
\label{xydef}
The redefinition of the physical CP-even neutral scalars in 2HDMEFT compared to 2HDM at the tree-level is given by eq.~(\ref{hfieldred}) along with,
\bea
\label{hVVscalefac}
x_1 &=&\frac{v^2}{f^2} \Big( c_{H1} c_{\b}^2  s_{\a}^2 + c_{H2} c_{\a}^2 s_{\b}^2 + \frac{1}{8} c_{H1H2} s_{2\a} s_{2\b} + c_{H12} (c_{\a}^2 c_{\b}^2 + s_{\a}^2 s_{\b}^2 - \frac{1}{4} s_{2\a} s_{2\b})\nn\\
&&\hspace{50pt}+ c_{H1H12} c_{\b} s_{\a} (s_{\a} s_{\b} - \frac{1}{2} c_{\a} c_{\b}) + c_{H2H12} c_{\a} s_{\b} (c_{\a} c_{\b} - \frac{1}{2} s_{\a} s_{\b}) \Big) ,\nn\\
x_2 &=&\frac{v^2}{f^2} \Big( c_{H1} c_{\b}^2 c_{\a}^2 + c_{H2} s_{\a}^2 s_{\b}^2 + \frac{1}{8} c_{H1H2} s_{2\a} s_{2\b} + c_{H12} (s_{\a}^2 c_{\b}^2 + c_{\a}^2 s_{\b}^2 - \frac{1}{4} s_{2\a} s_{2\b})\nn\\
&&\hspace{50pt}+ c_{H1H12} c_{\b} c_{\a} (c_{\a} s_{\b} - \frac{1}{2} s_{\a} c_{\b}) + c_{H2H12} s_{\a} s_{\b} (s_{\a} c_{\b} - \frac{1}{2} c_{\a} s_{\b}) \Big),\nn\\
y &=& \frac{v^2}{f^2} \Big( \frac{1}{2} c_{H1} s_{2\a} c_{\b}^2 -\frac{1}{2} c_{H2}  s_{2\a} s_{\b}^2 - \frac{1}{8} c_{H1H2} c_{2\a} s_{2\b} - \frac{1}{2} c_{H12} (  c_{2\b}  s_{2\a} + \frac{1}{2} c_{2\a}  s_{2\b}) \nn\\
&&\hspace{50pt}+ \frac{1}{4} c_{H1H12} ( s_{2\a} s_{2\b} - c_{2\a} c_{\b}^2 ) - \frac{1}{4} c_{H2H12} ( s_{2\a} s_{2\b} + c_{2\a} s_{\b}^2 )\Big).
\eea


\section{Branching ratios in 2HDM vs. 2HDMEFT}
\label{subsection:brratio}
\begin{figure}[h!]
 \begin{center}
\subfigure[]{
 \includegraphics[width=2.2in,height=2.2in, angle=0]{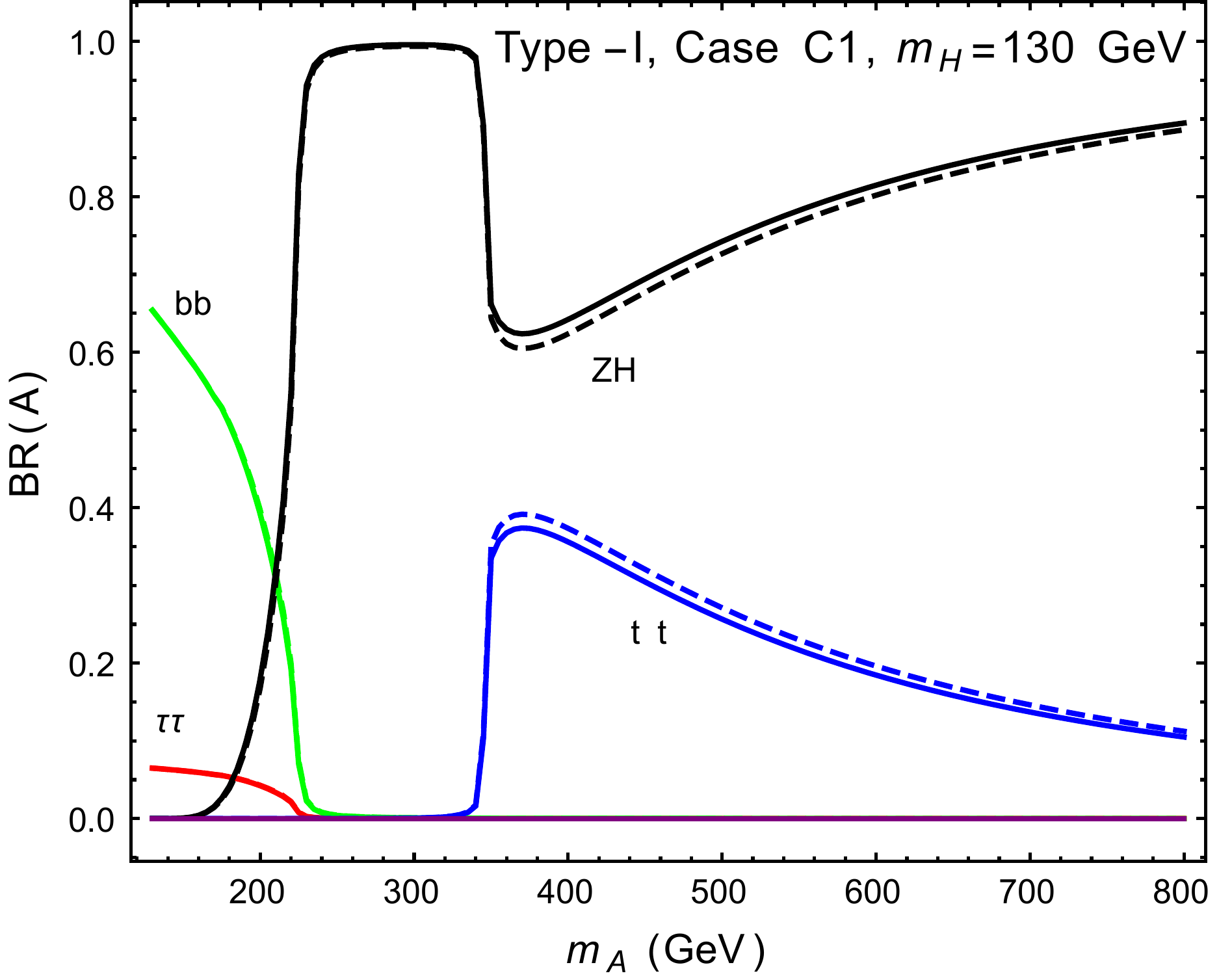}} 
 \hspace{15pt}
 \subfigure[]{
 \includegraphics[width=2.2in,height=2.2in, angle=0]{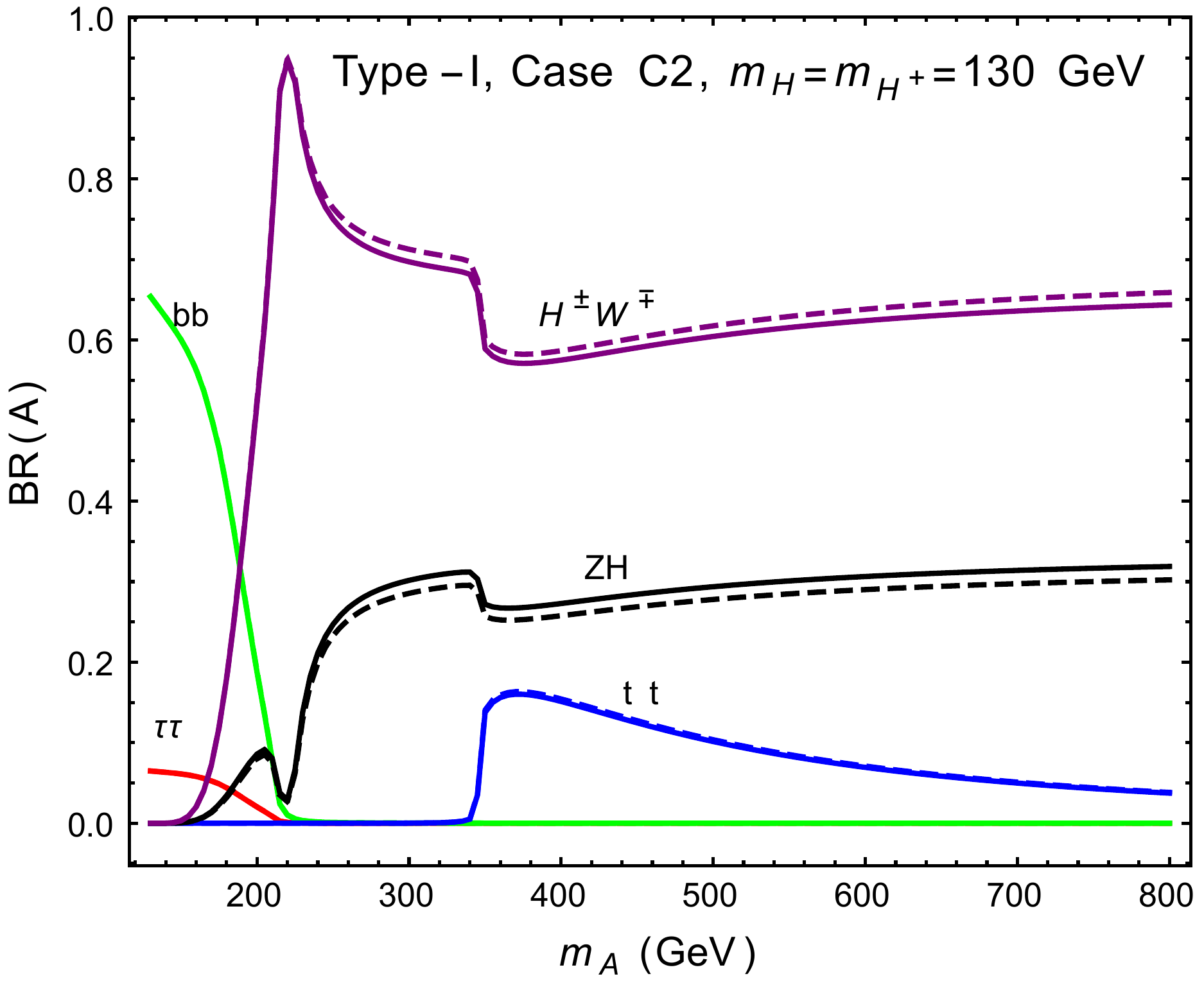}} 
 \\
 \subfigure[]{
 \includegraphics[width=2.2in,height=2.2in, angle=0]{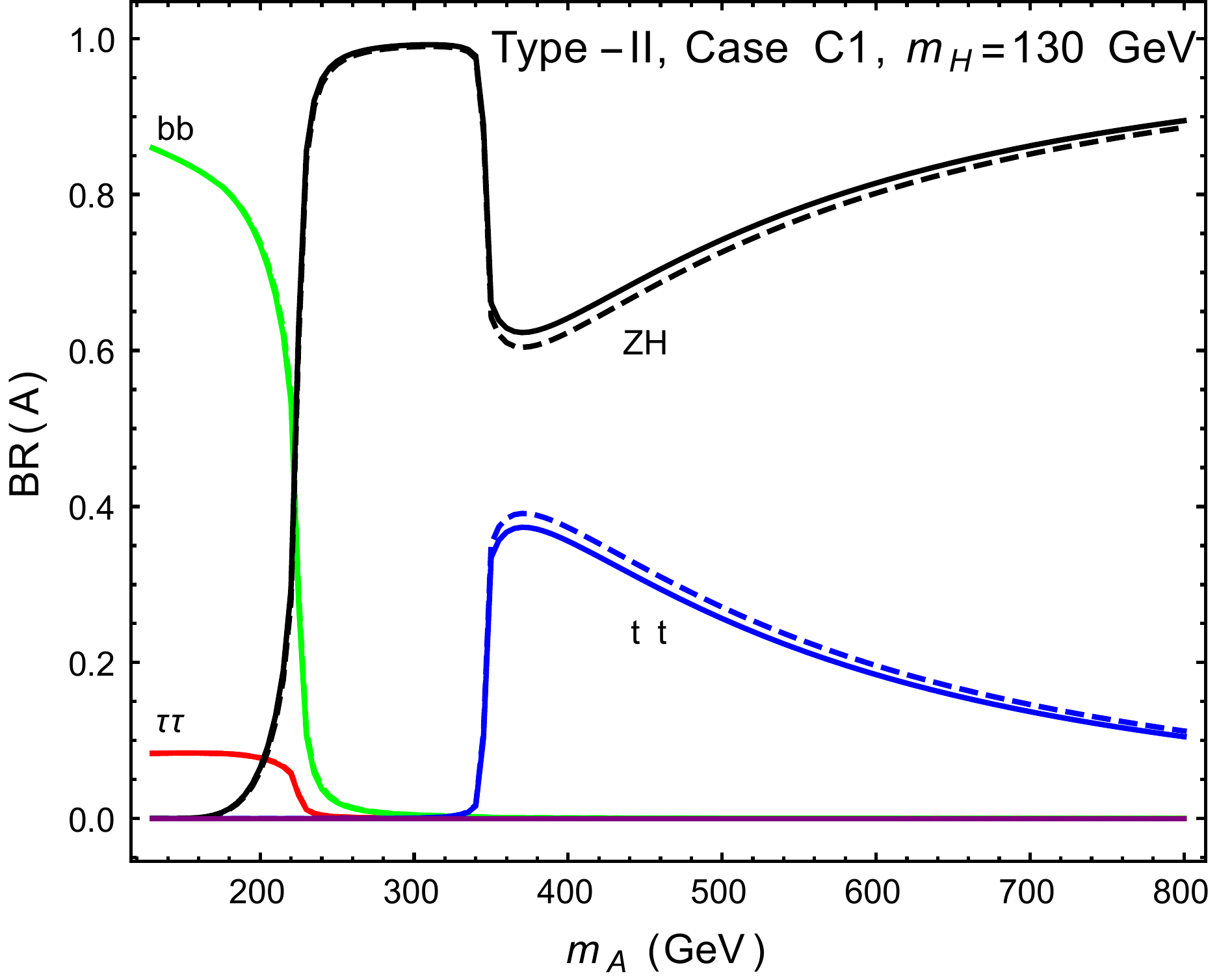}} 
  \hspace{15pt}
 \subfigure[]{
 \includegraphics[width=2.2in,height=2.2in, angle=0]{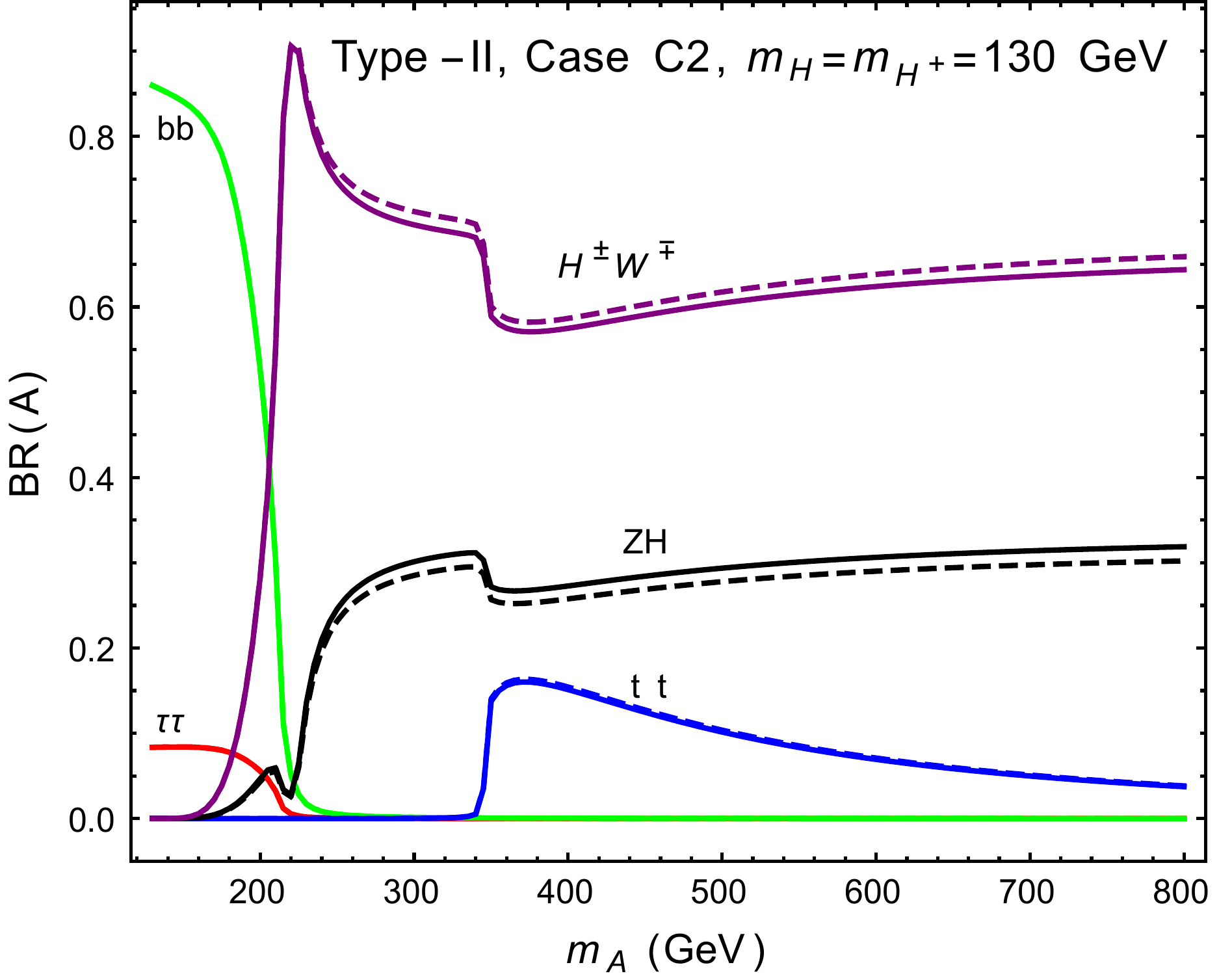}} 
  \caption{The branching ratios of  $A$ in various channels for the hierarchical cases for $\cos (\b-\a) = 0$, $\tan \b =1.5$. Dashed and solid lines represent the case of 2HDM at tree-level and BP1 of 2HDMEFT respectively.}
 \label{brratiosA}
 \end{center}
 \end{figure}
 
 \begin{figure}[h!]
 \begin{center}
 \subfigure[]{
 \includegraphics[width=2.3in,height=2.3in, angle=0]{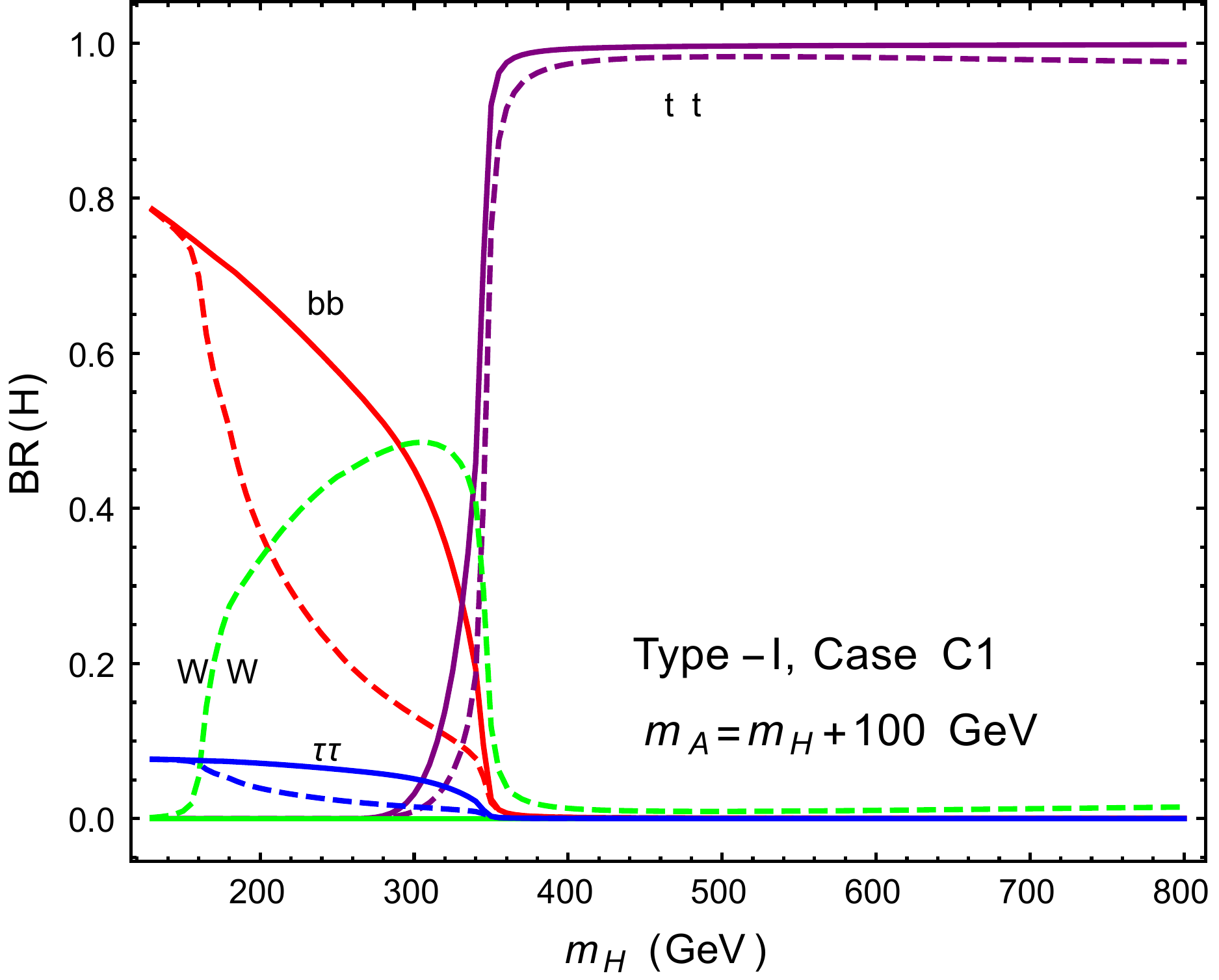}} 
\hspace{10pt} 
  \subfigure[]{
 \includegraphics[width=2.3in,height=2.3in, angle=0]{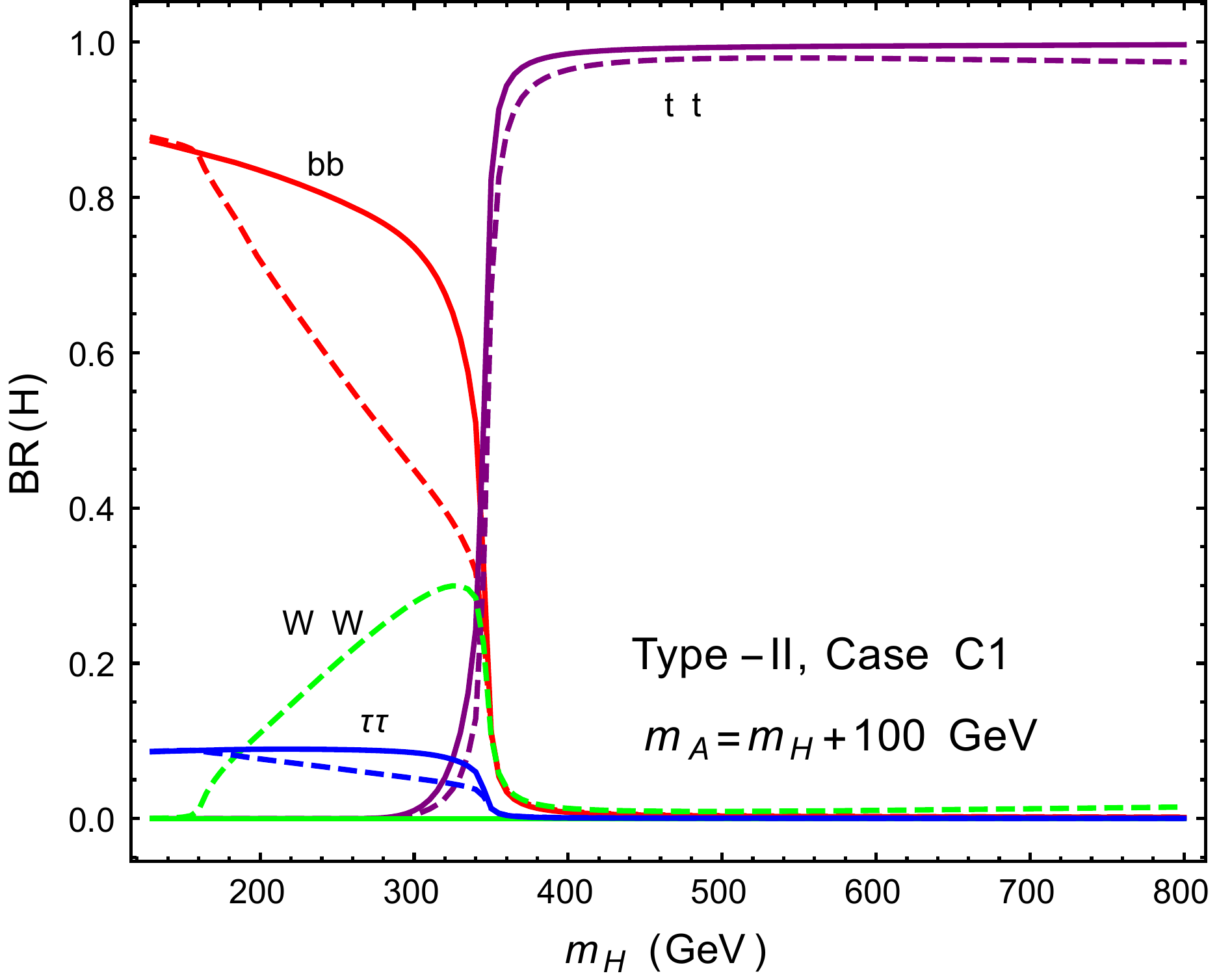}} \\
   \caption{The branching ratios of $H$ in various channels for the hierarchical case {\bf C1} for $\cos (\b-\a) = 0$, $\tan \b =1.5$.  Dashed and solid lines represent the case of 2HDM at tree-level and BP1 of 2HDMEFT respectively.}
 \label{brratiosH}
 \end{center}
 \end{figure}

\begin{figure}[h!]
 \begin{center}
\subfigure[]{
 \includegraphics[width=2.0in,height=2.0in, angle=0]{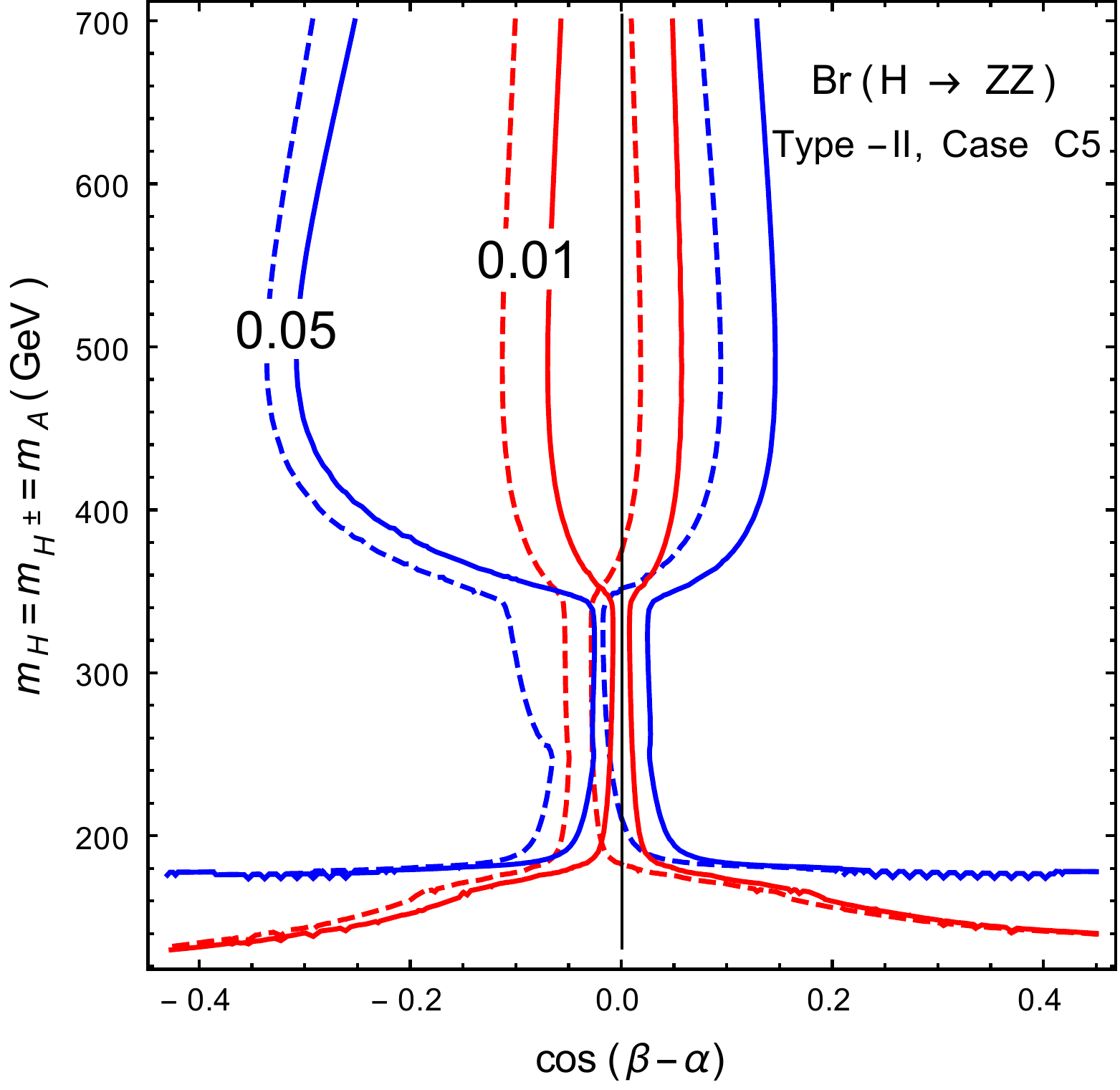}} 
 \subfigure[]{
 \includegraphics[width=2.0in,height=2.0in, angle=0]{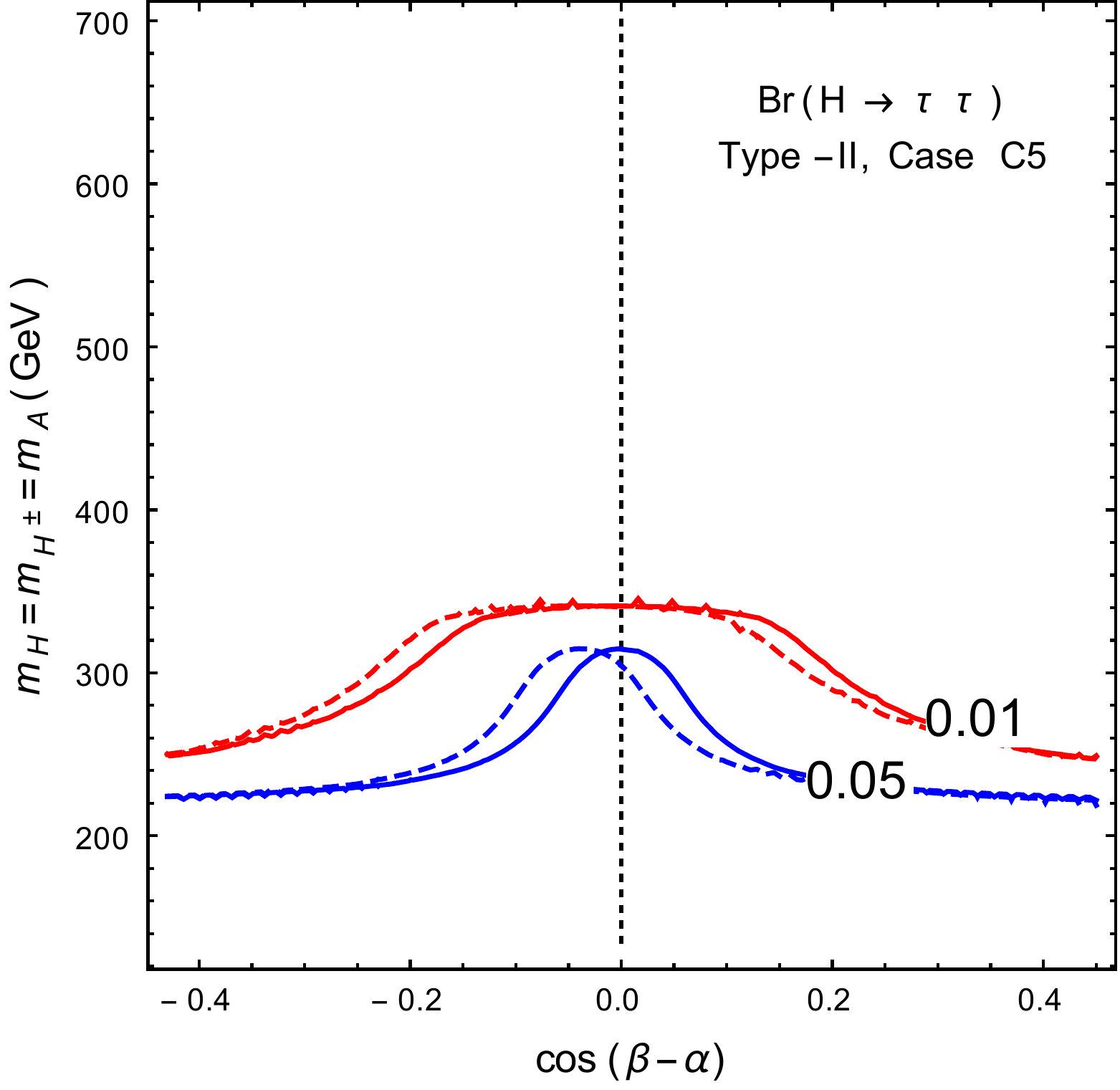}} 
 \subfigure[]{
 \includegraphics[width=2.0in,height=2.0in, angle=0]{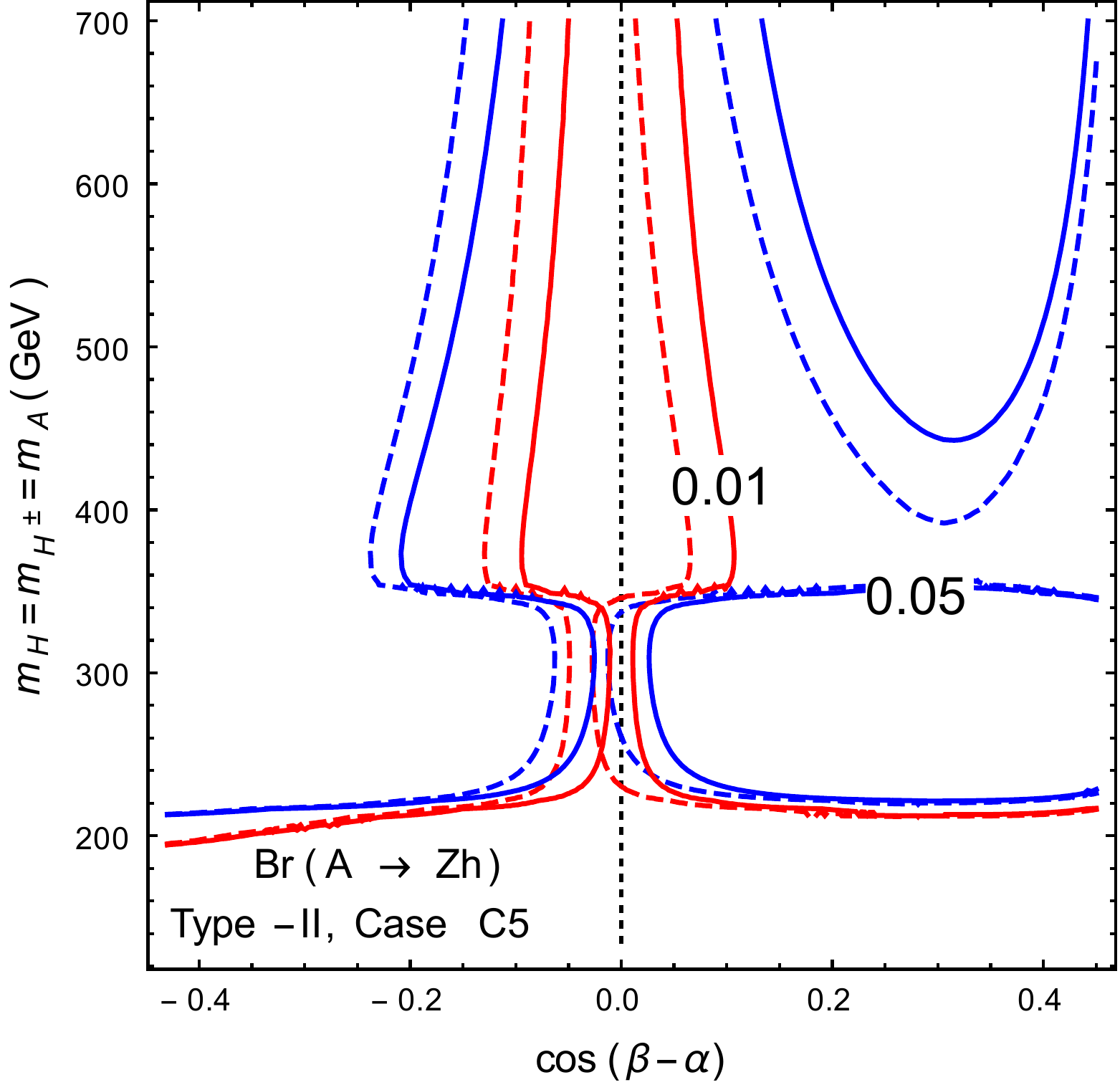}}  
 \caption{The branching ratios of $H$ and $A$ in various channels which constrain the parameter space on the $\cos(\b-\a) - m_A$ plane in the degenerate case {\bf C5} for Type-II 2HDM. In all three figures, the solid and dashed lines represent the corresponding branching ratios in 2HDM at the tree-level and {\bf BP1} of 2HDMEFT respectively.}
 \label{fig:XsBr}
\end{center}
\end{figure}

The key branching ratios of $H$ and $A$ have been shown in figs.~\ref{brratiosA} and \ref{brratiosH} with type-I and -II Yukawas for mass spectra {\bf C1} and {\bf C2}. When $m_{A(H)} \gtrsim 2 m_t$, the channel $A(H) \rightarrow t\bar{t}$ becomes viable in all the cases.
While showing the branching ratios for $A$ we assume, $m_H = m_{H^{\pm}} = 130$~GeV and while showing the same for $H$ we take, $m_A = m_H + 100$~GeV.
For both these cases we take, $\cos (\b-\a) = 0$ and $\tan \b = 1.5$.  
 Br($A \rightarrow t\bar{t}$) attains values up to $\sim 0.4$  and $\sim 0.19$ for cases {\bf C1} and {\bf C2} respectively.
In {\bf C2}, Br($A \rightarrow H^{\pm}W^{\mp}) \gtrsim 0.8$ for both type-I and -II 2HDM, whereas this process is absent for the case {\bf C1}. 
This leads to a much lower values of Br($A \rightarrow ZH$) in {\bf C2} compared to {\bf C1}, which can be seen from figs.~\ref{brratiosA}(a) and (b).
Hence the non-observation of $A \rightarrow ZH$ rules out a larger region of parameter space in the case {\bf C1} compared to {\bf C2} even in tree-level 2HDM, which can be seen from figs.~\ref{fig:c1} and \ref{fig:c2}.  
From figs.~\ref{brratiosA}(a) and (c) it can be seen that Br($A \rightarrow \tau\bar{\tau}$) is larger for type-II 2HDM compared to type-I.  
Thus the non-observation of $A \rightarrow \tau\bar{\tau}$ rules out a larger region for type-II 2HDM, as it can be seen from, for example, figs.~\ref{fig:c1}(a) and (b). 
Fig.~\ref{brratiosH}(a) shows that Br($H \rightarrow WW$) can attain values up to $\sim 0.5$ for type-I 2HDM for case {\bf C1}  at $\cos (\b -\a) = 0$ in presence of 6-dim term, whereas it is vanishing in tree-level 2HDM. 
In contrary, as it can be seen from fig.~\ref{brratiosH} (b), the corresponding value for type-II 2HDM can only go up to $\sim 0.3$.  
The branching ratios of $H$ for case {\bf C2} are exactly the same as in the case {\bf C1}. 
Such non-zero branching ratios of $H \rightarrow WW$ leads to a low value of Br($H \rightarrow b\bar{b}$) and eventually a lower value of $A \rightarrow ZH(b\bar{b})$ compared to 2HDM at the tree-level, which explains the relaxed constraints on $m_H$ as illustrated in fig.~\ref{fig:c1} and \ref{fig:c2} on the $m_A -m_H$ plane.   
Such branching ratios in type-III and -IV 2HDM can also be followed from figs.~\ref{brratiosA} and \ref{brratiosH} along with eq.~\eqref{coupmult}, exploiting the patterns of couplings across Yukawa types. It can be seen that the constraints for type-I Yukawa resemble that for type-IV, whereas the constraints on type-II 2HDM are similar to that in type-III case.

\section{$H^{\pm} \rightarrow h W^{\pm}$ cross-section at $\cos (\b-\a) = 0$}
\label{HphW}
\begin{figure}[h!]
 \begin{center}
\subfigure[]{
 \includegraphics[width=2.2in,height=2.0in, angle=0]{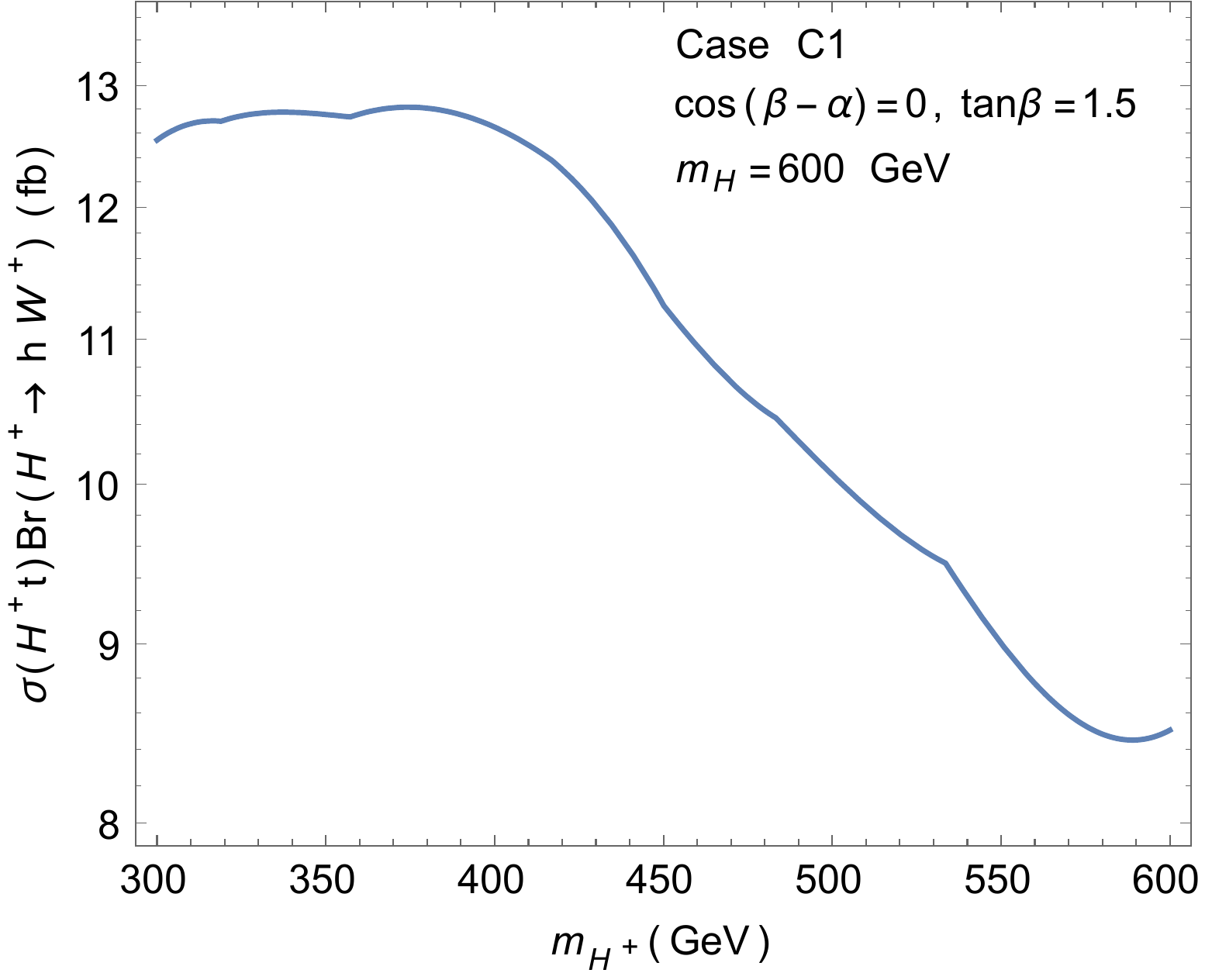}} 
 \hskip 15pt
 \subfigure[]{
 \includegraphics[width=2.2in,height=2.0in, angle=0]{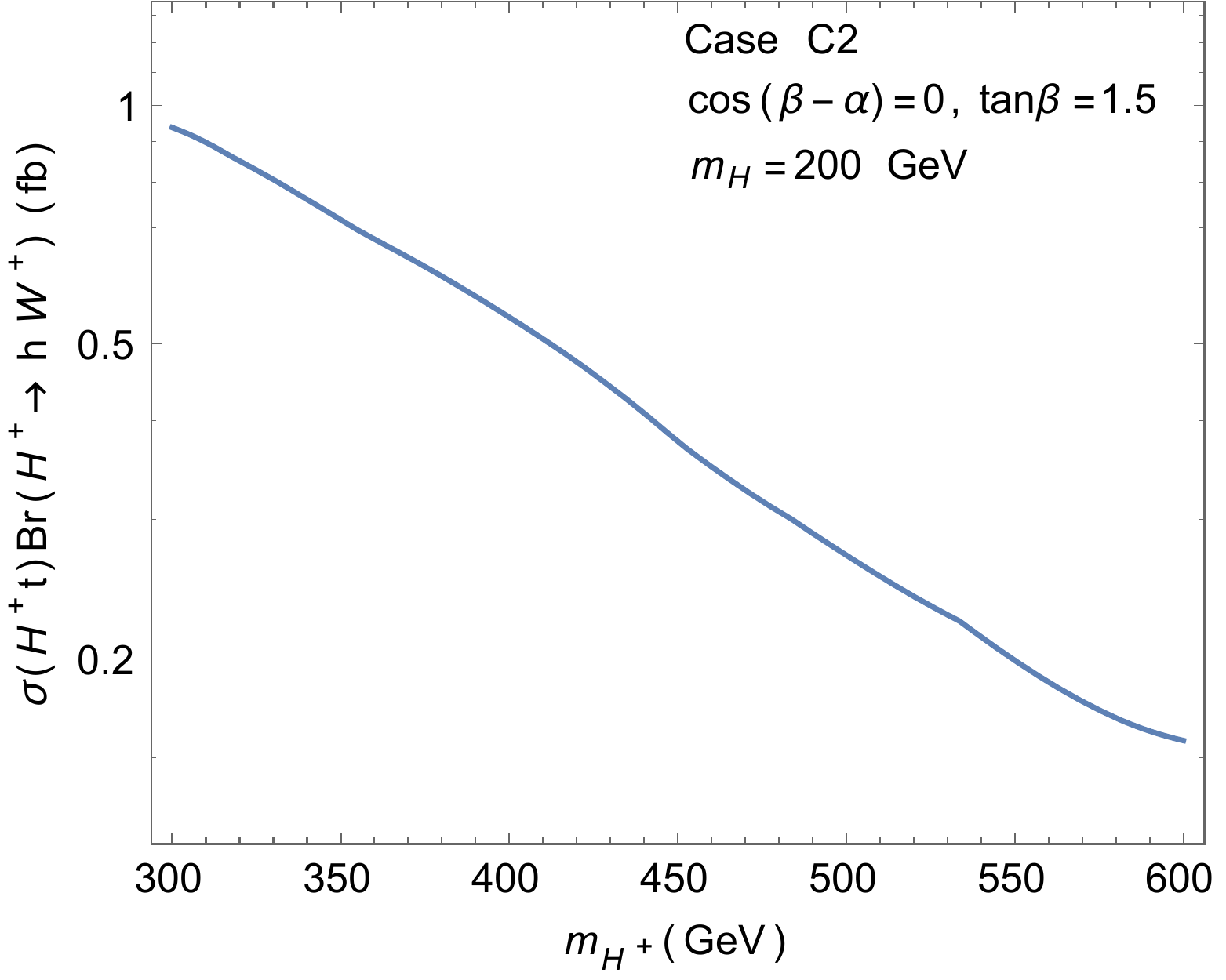} }
 \subfigure[]{
 \includegraphics[width=2.2in,height=2.0in, angle=0]{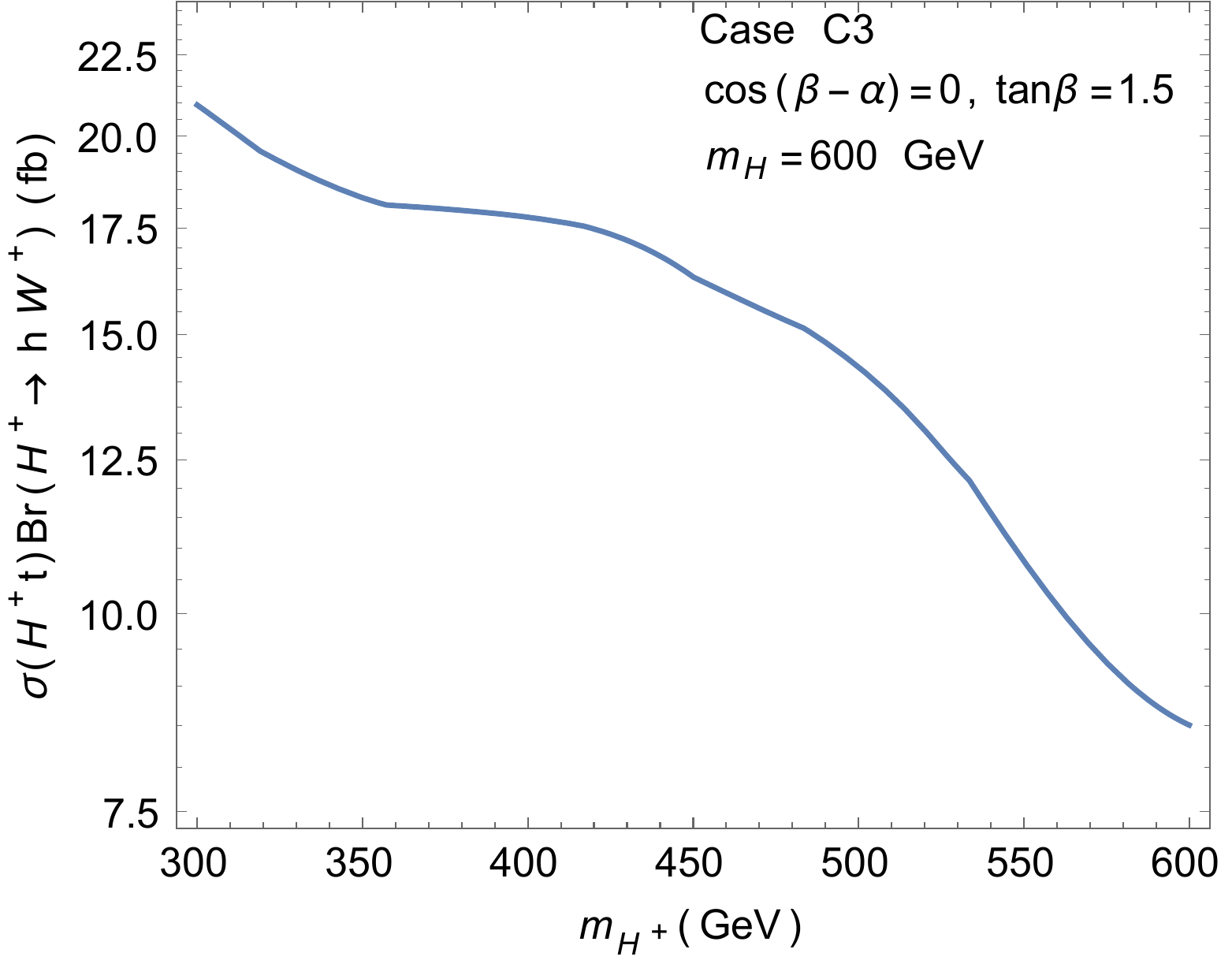}} 
 \hskip 15pt
 \subfigure[]{
 \includegraphics[width=2.2in,height=2.0in, angle=0]{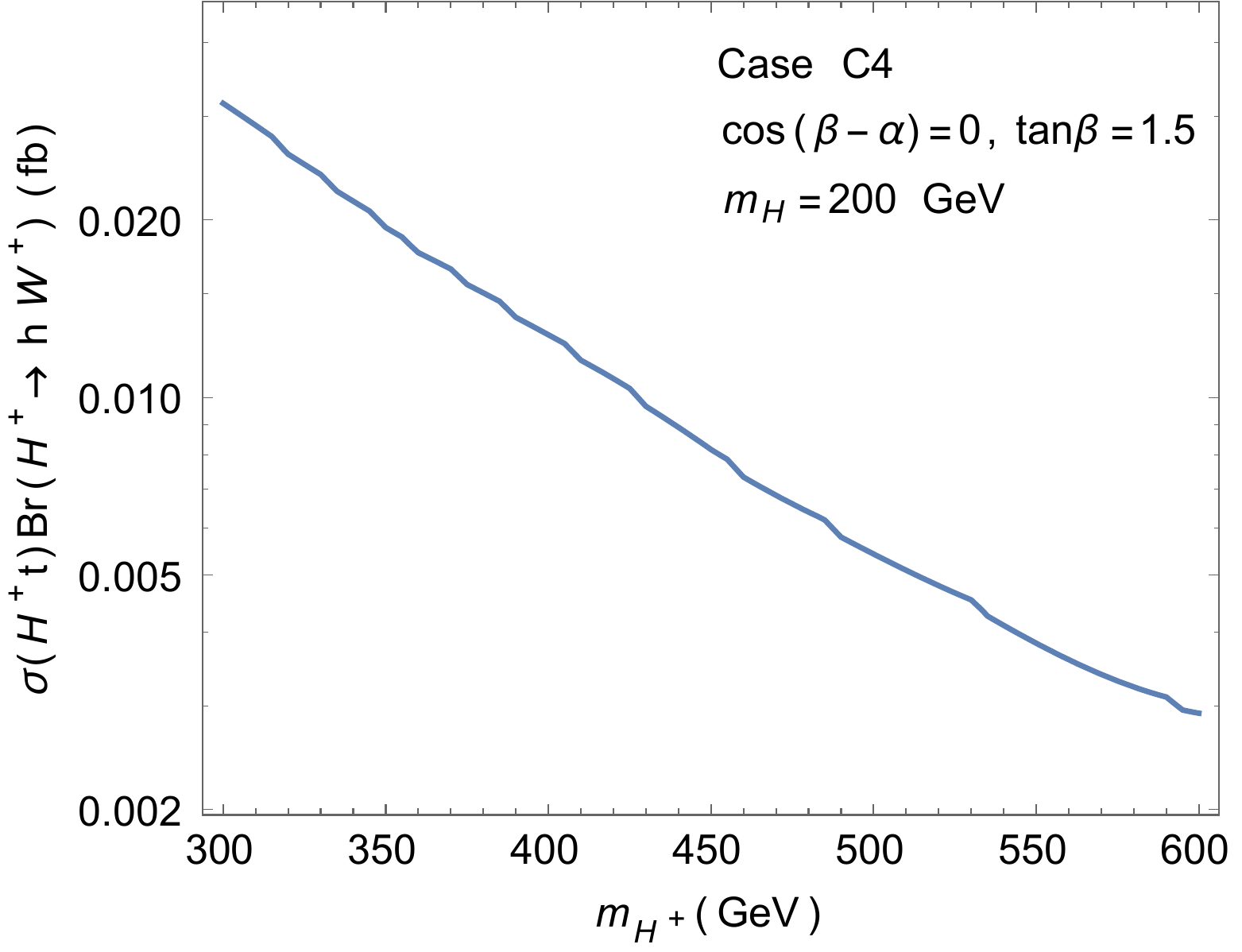}}   
 \caption{$\sigma(H^{\pm}t) \text{Br}( H^{\pm} \rightarrow h W^{\pm})$  in the alignment limit, in presence of the 6-dim terms at LHC with $\sqrt{s} = 14$~TeV in type-II 2HDM. }
 \label{fig:HC2}
\end{center}
 \end{figure}

The cross-sections in the channel $pp \rightarrow H^{\pm} \rightarrow h W^{\pm}$ for different hierarchical mass spectra and type-II Yukawa couplings have been presented in fig.~\ref{fig:HC2}. For $\cos (\b-\a) = 0$, in 2HDM at the tree-level, such a process in absent. But in present of the 6-dim operators, the cross-section for this process can go up to $\sim 25~$fb depending on the mass spectrum of heavy scalars in 2HDM. 
It can be noticed that the cross-section in this channel attains significantly higher values for the cases {\bf C1}, {\bf C2} and {\bf C3} compared to {\bf C4}.  
For {\bf C4}, the channel $H^{\pm} \rightarrow A W^{\pm}$ becomes kinematically accessible, thus lowering the value of Br$(H^{\pm} \rightarrow h W^{\pm})$, leading to lesser cross-section compared to the other cases.

\end{document}